\documentclass[twocolumn]{aastex61}
\pdfoutput=1

\begin{document}

\title{Tidal Interactions and Mergers in Intermediate Redshift EDisCS Clusters}

\author{Sinan Deger}
\affiliation{The University of Kansas, Department of Physics and Astronomy, Malott Room 1082, 1251 Wescoe Hall Drive, Lawrence, KS, 66045, USA; sinan.deger@ku.edu}

\author{Gregory Rudnick}
\affiliation{The University of Kansas, Department of Physics and Astronomy, Malott Room 1082, 1251 Wescoe Hall Drive, Lawrence, KS, 66045, USA}

\author{Kshitija Kelkar}
\affiliation{Raman Research Institute, Bangalore, India}

\author{Alfonso Arag\'{o}n-Salamanca}
\affiliation{School of Physics and Astronomy, The University of Nottingham, University Park, Nottingham NG7 2RD, UK}

\author{Vandana Desai}
\affiliation{Spitzer Science Center, California Institute of Technology, MS 220-6, Pasadena, CA 91125, USA}

\author{Jennifer M. Lotz}
\affiliation{Space Telescope Science Institute, 3700 San Martin Drive, Baltimore, MD 21218, USA}

\author{Pascale Jablonka}
\affiliation{Institut de Physique, Laboratoire d'astrophysique, Ecole Polytechnique F\'{e}d\'{e}rale de Lausanne (EPFL), Observatoire de Sauverny, CH-1290 Versoix, Switzerland}
\affiliation{GEPI, Observatoire de Paris, Universit\'{e} PSL, CNRS, 5 Place Jules Janssen, 92190 Meudon, France}

\author{John Moustakas}
\affiliation{Department of Physics \& Astronomy, Siena College, 515 Loudon Road, Loudonville, NY 12211, USA}

\author{Dennis Zaritsky}
\affiliation{University of Arizona, 933 N. Cherry Ave, Tucson, AZ 85721, USA}

\begin{abstract}
We study the fraction of tidal interactions and mergers with well identified observability timescales ($f_{\rm TIM}$) in group, cluster, and accompanying field galaxies and its dependence on redshift ($z$), cluster velocity dispersion ($\sigma$) and environment analyzing HST-ACS images and catalogs from the ESO Distant Cluster Survey (EDisCS). Our sample consists of 11 clusters, 7 groups, and accompanying field galaxies at $0.4 \leq z \leq 0.8$. We derive $f_{\rm TIM}$ using both a visual classification of galaxy morphologies and an automated method, the $G-M_{20}$ method. We calibrate this method using the visual classifications that were performed on a subset of our sample. We find marginal evidence for a trend between $f_{\rm TIM}$ and $z$, in that higher $z$ values correspond to higher $f_{\rm TIM}$. However, we also cannot rule out the null hypothesis of no correlation at higher than 68\% confidence. No trend is present between $f_{\rm TIM}$ and $\sigma$. We find that $f_{\rm TIM}$ shows suggestive peaks in groups, and tentatively in clusters at $R > 0.5\times R_{200}$, implying that $f_{\rm TIM}$ gets boosted in these intermediate density environments. However, our analysis of the local densities of our cluster sample does not reveal a trend between $f_{\rm TIM}$ and density, except for a potential enhancement at the very highest densities. We also perform an analysis of projected radius-velocity phase space for our cluster members. Our results reveal that tidal interactions and mergers (TIM), and undisturbed galaxies only have a 6\% probability of having been drawn from the same parent population in their velocity distribution and 37\% in radii, in agreement with the modest differences obtained in $f_{\rm TIM}$ at the clusters.

\end{abstract}

\section{Introduction}
\label{Sec:Introduction}

Past few decades have witnessed the shaping of the question of \textquotedblleft nature versus nurture" in galactic evolution. This question addresses whether the properties of galaxy populations we observe today are the result of intrinsic mechanisms, or the result of their environments and the interactions they underwent. It is highly likely that both of these play a role, but it is still unclear if either one is the dominant factor in shaping galactic evolution. An important observation that helped shape this scheme is the so-called morphology-density relation. The fractions of galaxies with "early-type" morphology, or galaxies that are classified as ellipticals (E's) and lenticulars (S0's) are found to peak in dense environments, whereas the fractions of spiral and irregular (Irr) galaxies show a comparable decrease \citep{Dressler80, Dressler97}. This comparable change implies that the increase in early-types has been in expense of transforming late-type galaxies. The fraction of early types depends both on global environment \citep{Dressler80, Dressler97, Fasano00, Blanton09, Just10, Vulcani10}, and local environment \citep{Dressler80, Postman05, Wilman09, Tasca09}. Likewise the fraction of passive galaxies is also lower in denser global \citep{Lewis02, Gomez03, Balogh04, Hogg04, Poggianti06, Gerke07} and local environments \citep{Lewis02, Gomez03, Balogh04, Kauffmann04, Poggianti08}. Though the main culprit for these observations eludes identification as of yet, there are studies attempting to pinpoint the exact mechanics at play.

Multiple processes have been proposed as candidates to explain this observation. One such process is ram pressure stripping, which occurs when the hot intracluster medium acts as a source of drag for galaxies moving through it, which can strip the cold gas within the galaxies \citep{Gunn72, Quilis00}. Another mechanism, referred to as either \textquotedblleft strangulation" or \textquotedblleft starvation", occurs when the hot gas reservoir bound to a galaxy is stripped when the galaxy falls into a dense environment such as a cluster. After losing access to this reservoir to replenish its gas content, the galaxy will consume whatever fuel it has left for star formation and will gradually show lower and lower star formation rate (SFR) as it runs out of fuel \citep{Larson80, Bekki02}. Both of these processes result in the depletion of gas in galaxies and may result in the presence of passive disks \citep{Bundy10, Cantale16b}, potentially also with larger bulges \citep{Kawata08}. Due to the high velocity dispersions of cluster environments, encounters between member galaxies occur at high speeds. Changes to the internal energy of galaxies after such encounters make them more and more susceptible to disruptions by later encounters with other members or by the tidal interactions with the cluster potential, either of which is capable of alterations to morphology. The cumulative effect of these high speed encounters is called \textquotedblleft galaxy harassment" \citep{Richstone76, Farouki82, Moore98}. Finally, the process of the central galaxy of a halo accreting satellite galaxies that lost their momentum due to dynamical friction is called \textquotedblleft galactic cannibalism". The most massive central galaxies of halos almost invariably have elliptical morhologies, possibly due to many such events \citep{Ostriker75, White76, Hausman78}. Even though these processes underline the importance of environment, environmental factors may not represent the entire picture. Examples of transition galaxies, such as E+A galaxies can be found in the field \citep{Zabludoff96}, demonstrating that a dense environment is not a necessary condition.

We focus on another candidate process in this paper, namely galaxy mergers and galaxy-galaxy interactions. While related to cannibalism, in the context of this paper tidal interactions and mergers are those events that occur between satellite galaxies. Mergers are a likely suspect in explaining the observed transformation in morphology, as merger events are usually violent events that trigger drastic change. \cite{Toomre72} proposed that elliptical galaxies can be the outcome of the merging of two disk galaxies. This morphological transformation via mergers has been subsequently demonstrated in many simulations since then \citep{Barnes96, Naab03, Lotz08b}. While multiple papers argued that it is possible for galaxies to retain their disk after major merger events, and even potentially have star formation present on the disk \citep{Springel05, Robertson06}. Simulations of major mergers \citep{Cox06} and observations of local gas-rich mergers indicate that merger events are capable of putting galaxies in states of intense star formation called starbursts, where galaxies have much higher star formation rates (SFRs) compared to their normal production \citep{Larson78}. This has a clearly observable effect on galaxy morphology.

Furthermore, \cite{Christlein04} show that models that generate early-type S0 galaxies by fading the disks of late-type galaxies fail to generate the bulge and disk luminosities they studied, and that bulge enhancement models are in good agreement with their clusters. They further conclude that their results are in favor of galaxy interactions and mergers, which can play a role in bulge enhancement. \cite{Johnston14} on the other hand finds it is possible to fade disks and grow bulges through centrally-concentrated star formation. \cite{Wilman09} also emphasizes the importance of bulge growth, and proposes minor mergers as a favored mechanism to explain S0 production. They also find that the fraction of S0's in their sample is much higher in groups compared to the field, and they propose galaxy groups to be the prominent environment in S0 production. This may be expected as groups have moderately high densities and low velocity dispersions, and are thus conducive sites for mergers. \cite{Just10} also finds that S0 type galaxies, which are likely products of mergers, are evolving in number faster in galaxy groups than in clusters. This suggests that the galaxies that will later on fall into a cluster are preprocessed in these moderate density environments, which host conditions favorable for merger-based morphological transformation. \citep{Zabludoff98, Fujita04, Cortese06, Dressler13, Abramson13, Vijayaraghavan13, Man16}.

In this paper we analyze galaxy merger events using the ESO Distant Cluster Survey (EDisCS) sample of cluster, group, and field galaxies with Hubble Space Telescope (HST) imaging at a redshift range of $0.4 \leq z \leq 0.8$ \citep{White05}. To understand the role galaxy mergers play in galaxy evolution it is vital to study them in different environments. Our sample allows us to study mergers at multiple environments of varying density, such as galaxy groups, clusters, and the accompanying field. We will study the effects of mergers on the star formation of galaxies in an upcoming paper, in this paper we present our merger detection, and our analysis of merger fractions. This study is complementary to those of \cite{Desai07} and \cite{Just10} who studied the morphological fractions in the EDisCS systems. We on the other hand are directly exploring the mergers that potentially drove this transformation.

Galaxy interactions have a variety of visible effects on morphologies of galaxies. Detection of these alterations has been a prime tool in the identification and study of galaxy interactions and mergers. The morphological detection of mergers is enabled by the asymmetries and distortions in the structures of galaxies that result from gravitational interactions, and from the compression and heating of the gas that results from hydrodynamical effects. Visual identification of these morphological disturbances is therefore common practice in galaxy interaction research. Such methods are subjective, and also are not immune to misclassification, as not every visually asymmetric/distorted galaxy is the result of interactions. Likewise, some signatures of merging, such as diffuse tidal tails are hard to identify long after the merger has occurred, causing incompleteness in some merger classifications. Recent years have seen extensive improvement in automated methods that is based on quantifying these distortions \citep{Abraham96, Conselice03, Lotz04, Hoyos12, Freeman13}. Automated methods have multiple advantages over visual classification in that they are easily reproducible, are generally faster compared to visual identification methods, especially for large sample sizes, and can easily be run on large simulation suites to assess the detection efficiency \citep{Lotz10}. A shortcoming these methods suffer from is that they are susceptible to both missing asymmetric features (incompleteness), and contamination due to noisy measurements which becomes especially prevalent at low signal-to-noise ratios. They therefore require careful calibration, as accurate merger detection is key to measuring the prevalence of galaxy mergers and their role in galaxy evolution. That is why we decided to use an automated method which we calibrate using a visually classified subsample. The automated method we use for this paper is the $G-M_{20}$ method (details in \cite{Lotz04}, brief explanation in \S\ref{Sec:MorphClass}).

We used a visual merger classification from \cite{Kelkar17} that was performed on the subset of our sample with spectroscopic redshifts. We then measured $G$ and $M_{20}$ values for our entire catalog, including those with photometric redshifts. Using our visually classified sample we calibrated a tidal interaction and merger (TIM) decision boundary on the $G-M_{20}$ space. We then calculated the fraction of tidal interactions and mergers ($f_{\rm TIM}$) of our clusters, groups, and field galaxies and analyzed the dependence of $f_{\rm TIM}$ on redshift, velocity dispersion, and both global and local environment. We used the local density measures derived by \cite{Poggianti08} for the spectroscopic cluster members of EDisCS for our analysis of the dependence of $f_{\rm TIM}$ on local density. Finally, we examined where tidal interactions and mergers lie with respect to undisturbed galaxies in projected radius-velocity phase space.

The paper consists of the following sections; In \S\ref{Sec:Sample} we present samples we used in our analysis. In \S\ref{Sec:MorphClass} we describe the two approaches taken in this paper for merger identification; namely visual classification and $G-M_{20}$ classification \citep{Abraham03, Lotz04} to obtain $f_{\rm TIM}$. We present our results for the variation of $f_{\rm TIM}$ with redshift, velocity dispersion, and global and local environment in \S\ref{Sec:Results}. We discuss the implications of these results in \S\ref{Sec:Discussion}. Finally, we summarize the main results of this paper in \S\ref{Sec:Summary}. Throughout the paper we assume $H_{0} = 70$ km $\rm s^{-1} \, Mpc ^{-1}$, and use AB magnitudes.

\section{Sample}
\label{Sec:Sample}

ESO Distant Cluster Survey (EDisCS, \cite{White05}) is a detailed photometric and spectroscopic survey of clusters, groups, and field galaxies, with structures drawn from the Las Campanas Distant Cluster Survey (LCDCS; \cite{Gonzalez01}). The EDisCS fields have either BVIK, BVIJK, or VRIJK photometry depending on the redshift estimate of the original cluster candidate. The sample was also observed with extensive FORS2 spectroscopy on the Very Large Telescope (ESO) \citep{Halliday04, MilvangJensen08}. To study the morphological content of the EDisCS sample, we used Hubble Space Telescope (HST) Advanced Camera for Surveys (ACS) imaging in the F814W filter (depths of 1 orbit at cluster outskirts, 5 orbits at cluster core) of 10 of the highest redshift clusters from \cite{Desai07}. We make use of these 10 fields with HST ACS images plus photometric and spectroscopic catalogs for the analysis presented in this paper.

\begin{deluxetable}{ccccc}
\tablecolumns{5}
\tablewidth{0pc}
\tablecaption{The EDisCS-HST Sample}
\tablehead{
\colhead{Structure Name} & \colhead{Redshift} & \colhead{$\sigma$} & \colhead{$N_{phot+spec}$} & \colhead{$N_{spec}$}}

\startdata
CL1040.7-1155 & 0.7043 & $418^{+55}_{-46}$ & 24  & 10 \\
CL1054.4-1146 & 0.6972 & $589^{+78}_{-70}$ &  71 &  24\\
CL1054.7-1245 & 0.7498 & $504^{+113}_{-65}$ & 57 &  16\\
CL1138.2-1133 & 0.4796 & $732^{+72}_{-76}$ & - &  13\\ 
CL1138.2-1133a & 0.4548 & $542^{+63}_{-71}$ & - &  7\\ 
CL1216.8-1201 & 0.7943 & $1018^{+73}_{-77}$ & 102  & 36 \\ 
CL1227.9-1138 & 0.6357 & $574^{+72}_{-75}$ & 54 &  12\\
CL1232.5-1250 & 0.5414 & $1080^{+119}_{-89}$ & 82 &  31\\
CL1354.2-1230 & 0.7620 & $648^{+105}_{-110}$ & 36 &  8\\
CL1354.2-1230a & 0.5952 & $433^{+95}_{-104}$ & - &  6\\
\hline
Clusters Total &      &     & 429 & 163\\
\hline
CL1037.9-1243 & 0.5783 & $319^{+53}_{-52}$ & -  & 7\\
CL1040.7-1155a & 0.6316 & $179^{+40}_{-26}$ & -  & 2\\
CL1040.7-1155b & 0.7798 & $259^{+91}_{-52}$ & -  & 2\\
CL1054.4-1146a & 0.6130 & $227^{+72}_{-28}$ & -  & 4\\
CL1054.7-1245a & 0.7305 & $182^{+58}_{-69}$ & -  & 7\\
CL1103.7-1245a & 0.6261 & $336^{+36}_{-40}$ & -  & 7\\
CL1103.7-1245b & 0.7031 & $252^{+65}_{-85}$ & -  & 5\\
\hline
Groups & - & - & - & 34\\
\hline
Field $0.4\leq z < 0.6$ & - & - &   85  &  22\\
Field $0.6\leq z < 0.8$ & - & - &  93  &  47\\
\hline
Field Total &      &     &  178 & 69\\
\enddata
\tablecomments{Column 1: Structure Name. Column 2: Cluster Redshift. Column 3: Cluster velocity dispersion. Column 4: Number of phot+spec members. Column 5: Number of spectroscopically confirmed members. Numbers are given after quality cuts described in \S\ref{Sec:Sample} are applied.}
\label{Table:Clusters}
\end{deluxetable}

Our sample consists of 11 galaxy clusters, 7 groups, and the accompanying field galaxies at $0.4 \leq z \leq 0.8$. Following \cite{Poggianti09} we define galaxy groups as structures with $\sigma < 400 \, \mathrm{km\, s^{-1}}$. The catalog of objects that have spectroscopic redshifts will be addressed as the \textquotedblleft spectroscopic sample" throughout this paper. Likewise, the catalog of objects that only have photometric redshifts \citep{Rudnick09, Pello09} will be referred to as the \textquotedblleft photometric sample". The other sample we use for our analysis consists of these two samples together, the spectroscopic sample plus galaxies identified as members or field galaxies using photometric redshifts from the EDisCS catalog, to which we will refer to as the \textquotedblleft phot+spec sample" throughout the paper.

\begin{figure*}
\centering
\includegraphics[width=.33\textwidth]{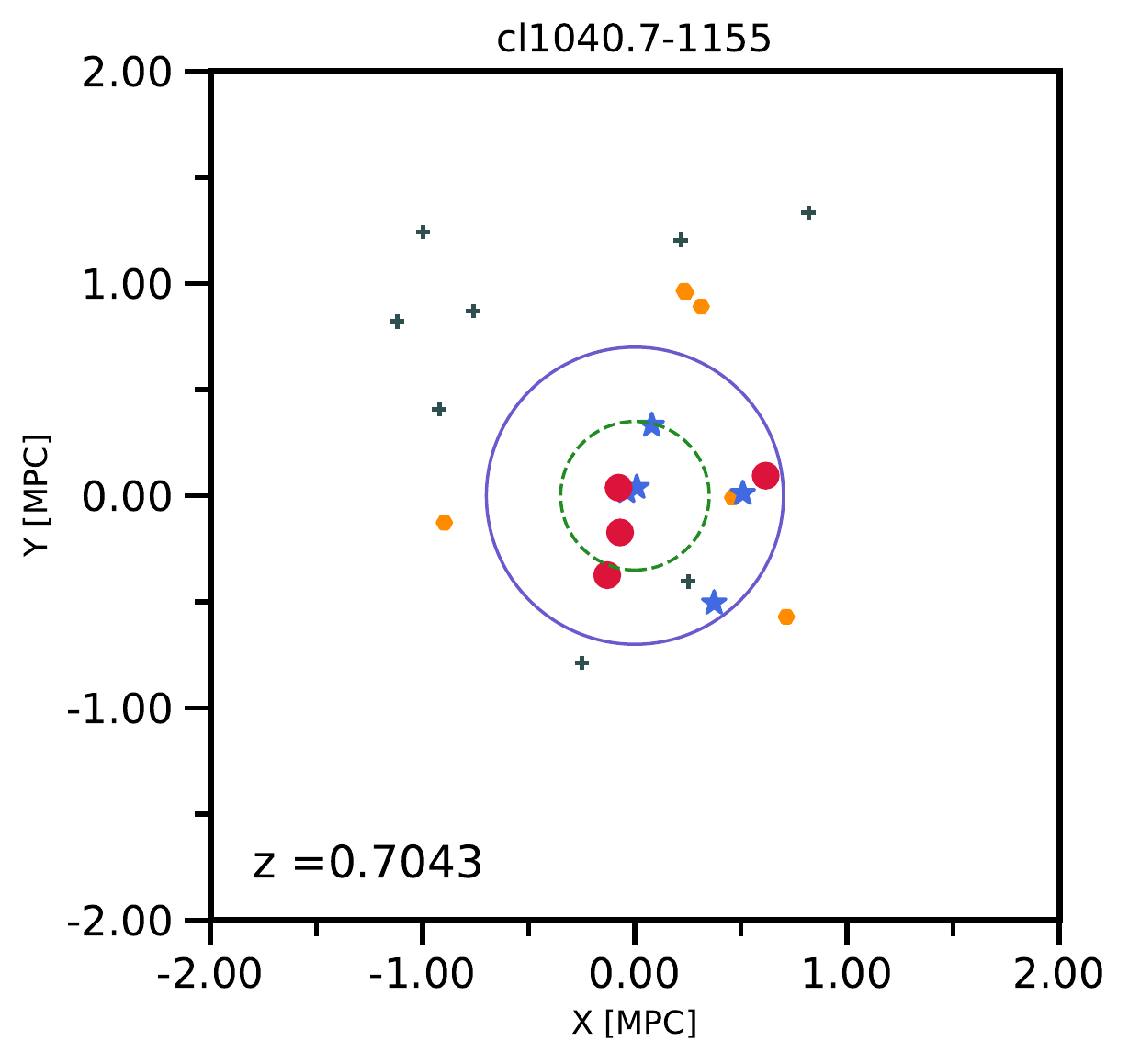}\hfill
\includegraphics[width=.33\textwidth]{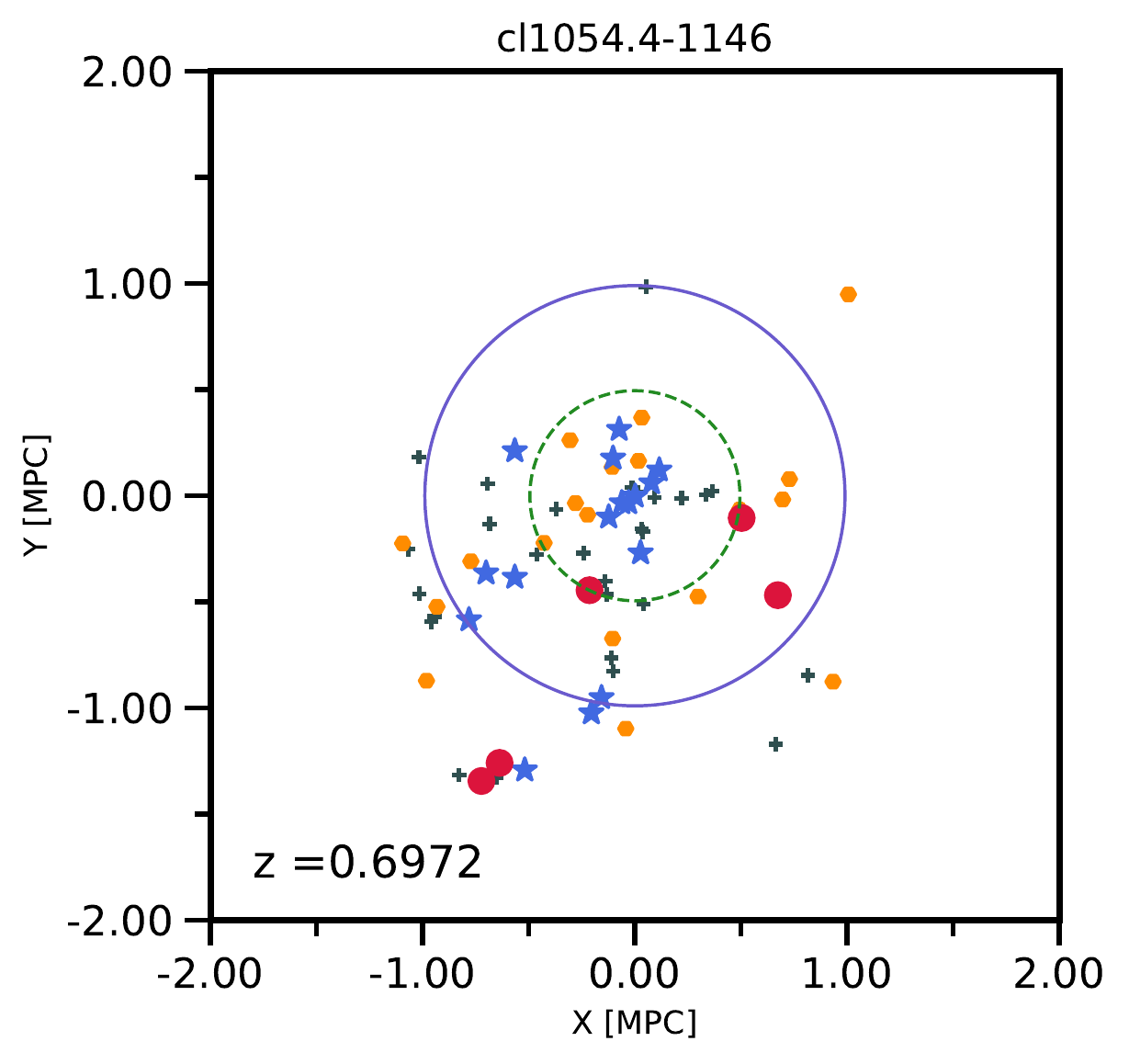}\hfill
\includegraphics[width=.33\textwidth]{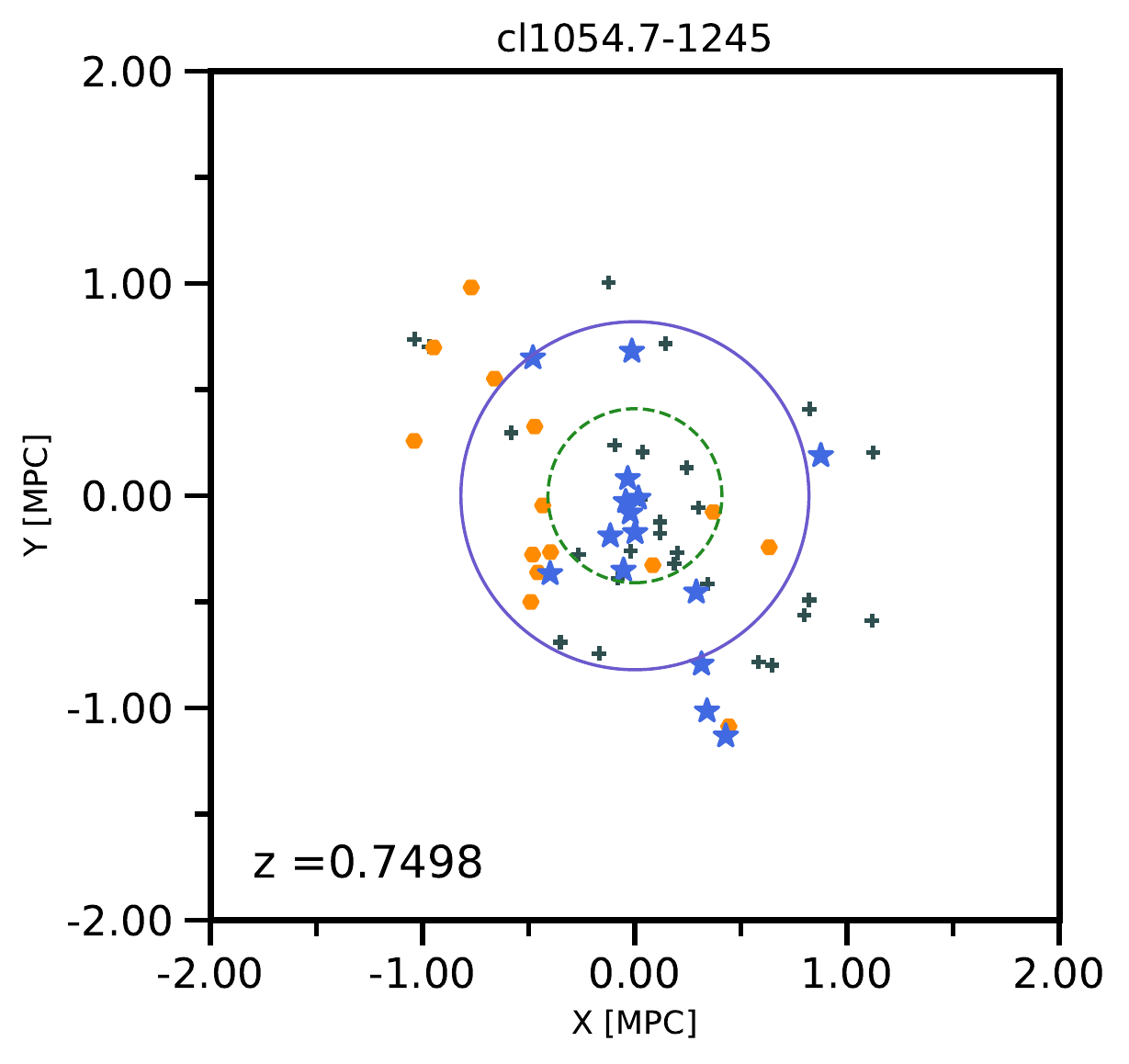}\\
\includegraphics[width=.33\textwidth]{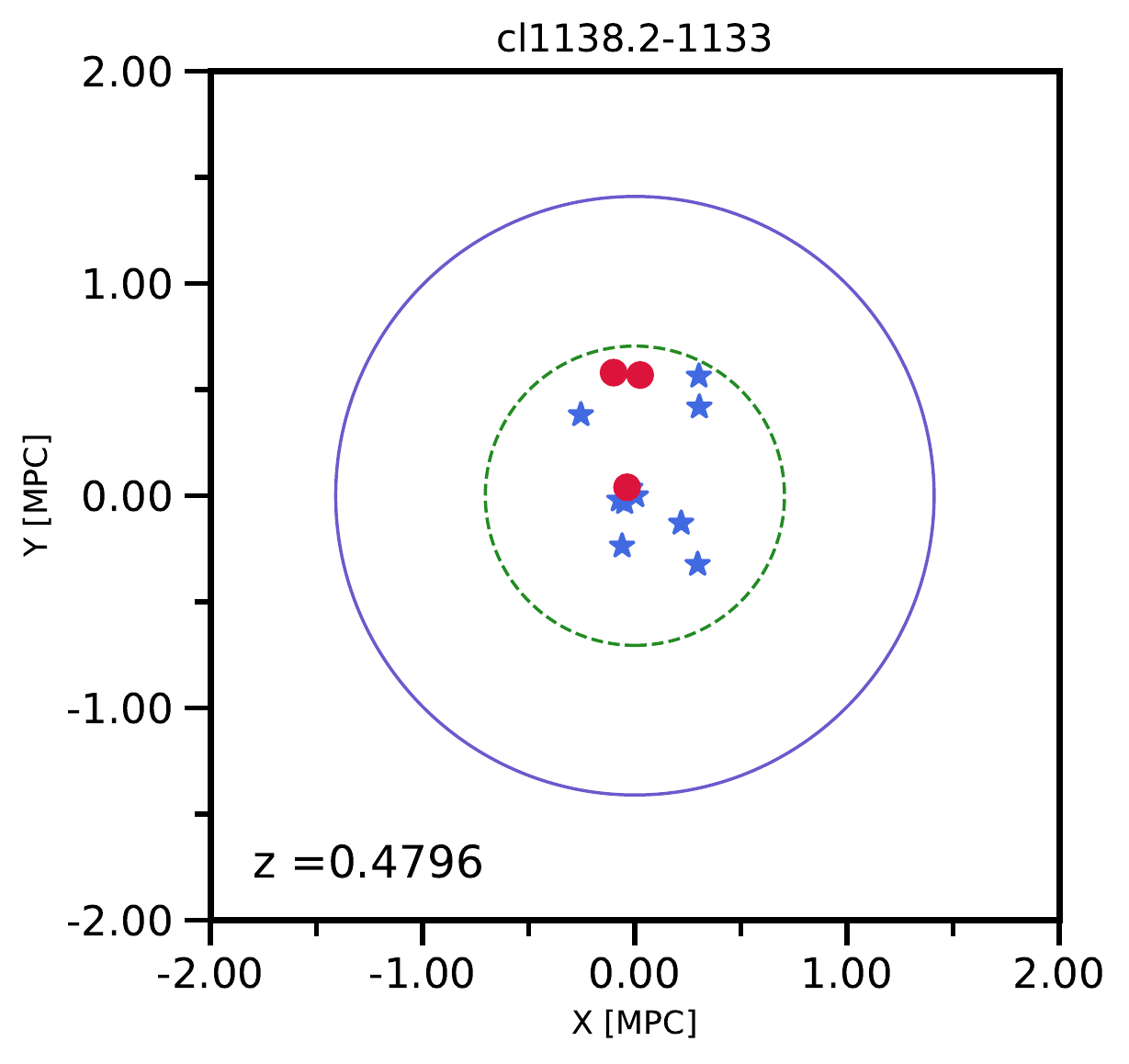}\hfill
\includegraphics[width=.33\textwidth]{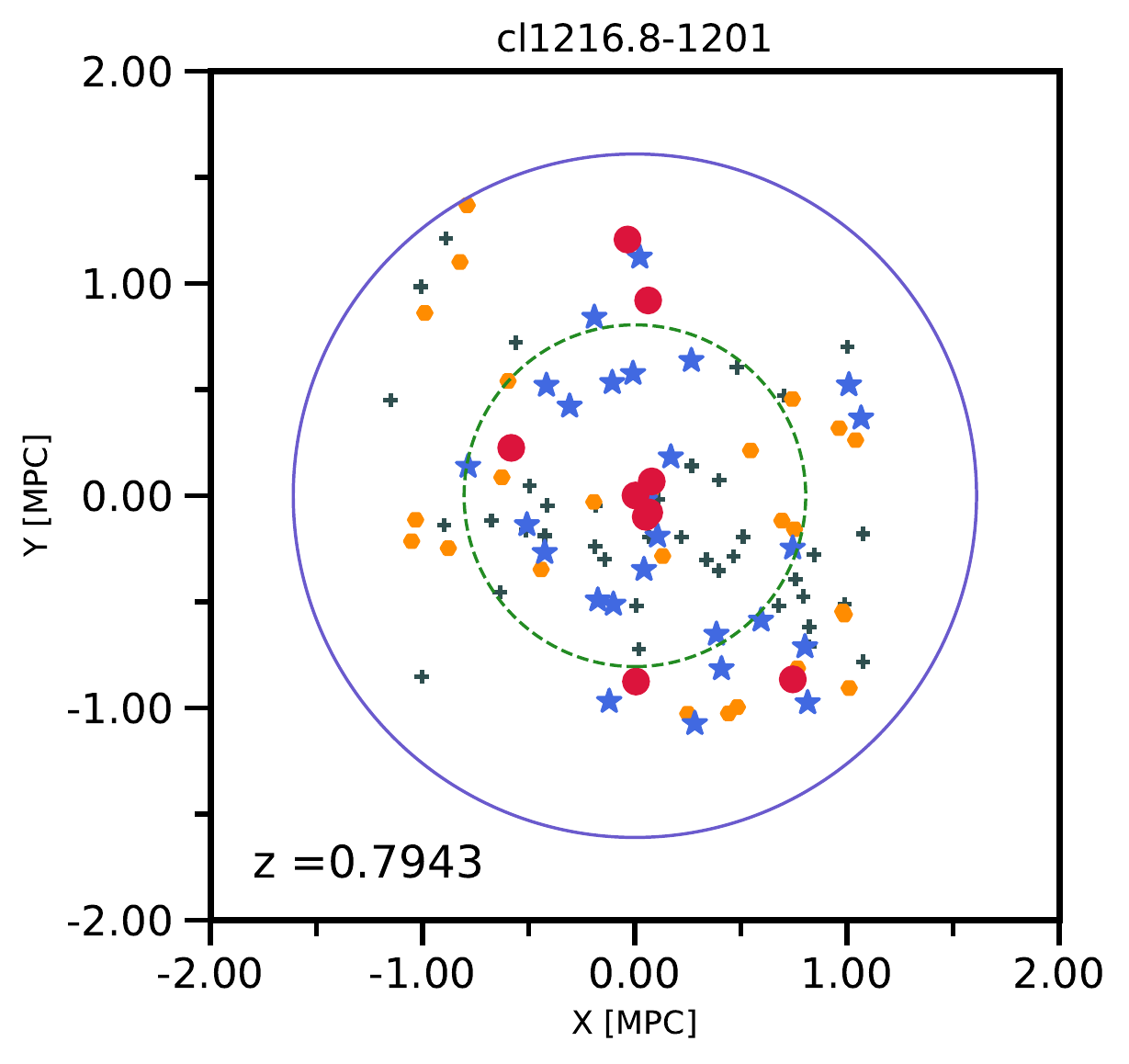}\hfill
\includegraphics[width=.33\textwidth]{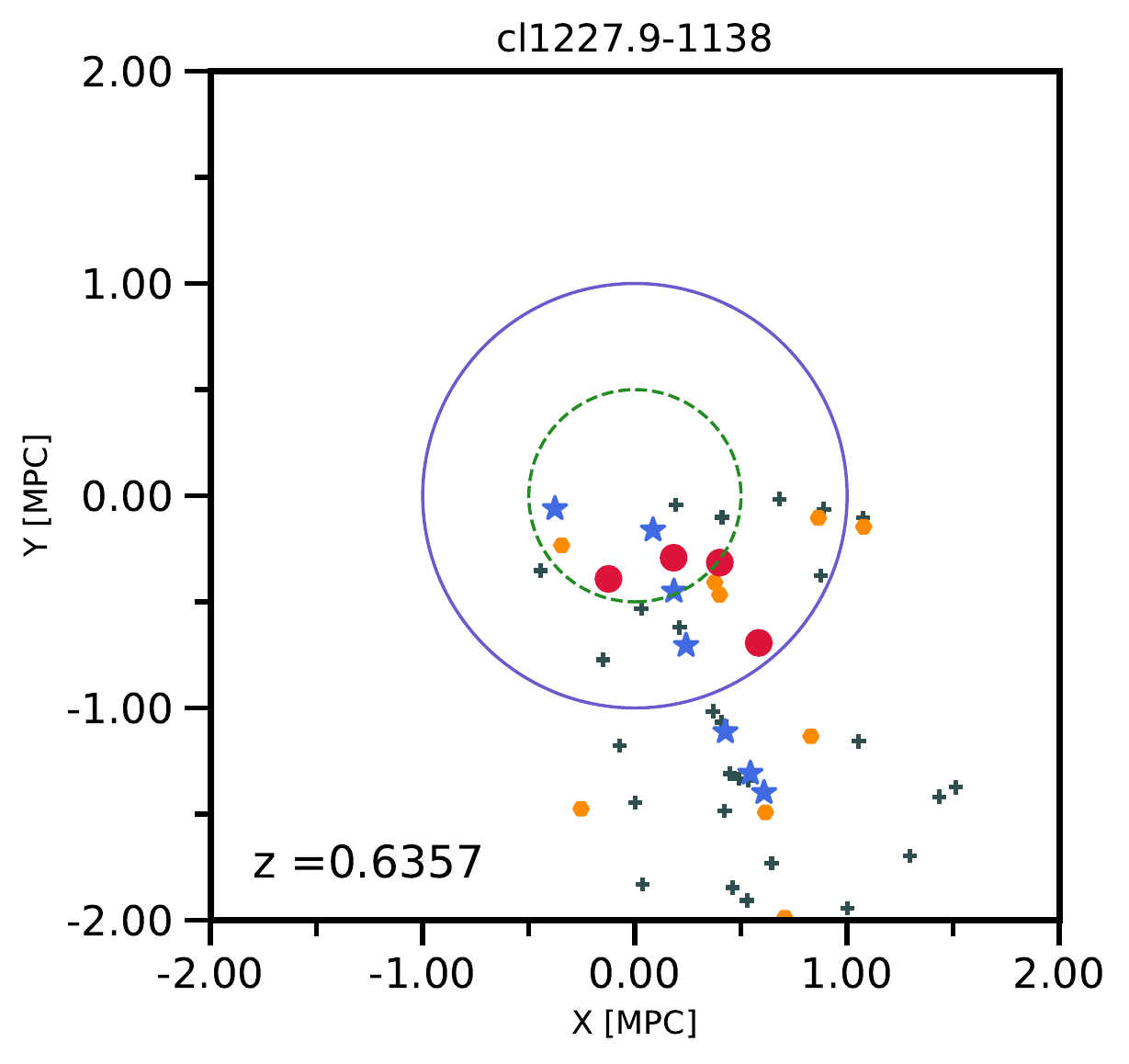}\\
\includegraphics[width=.33\textwidth]{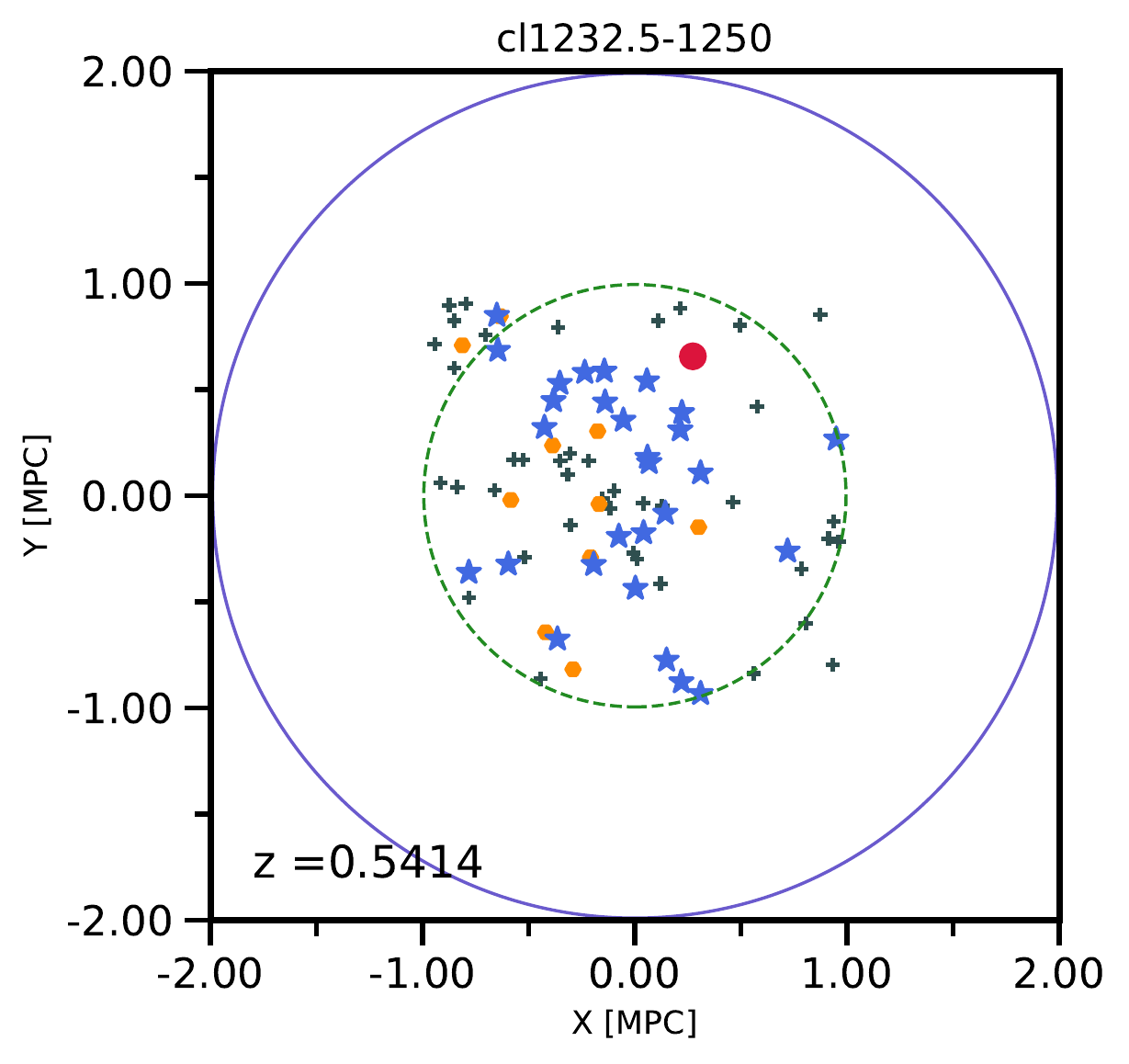}\hfill
\includegraphics[width=.33\textwidth]{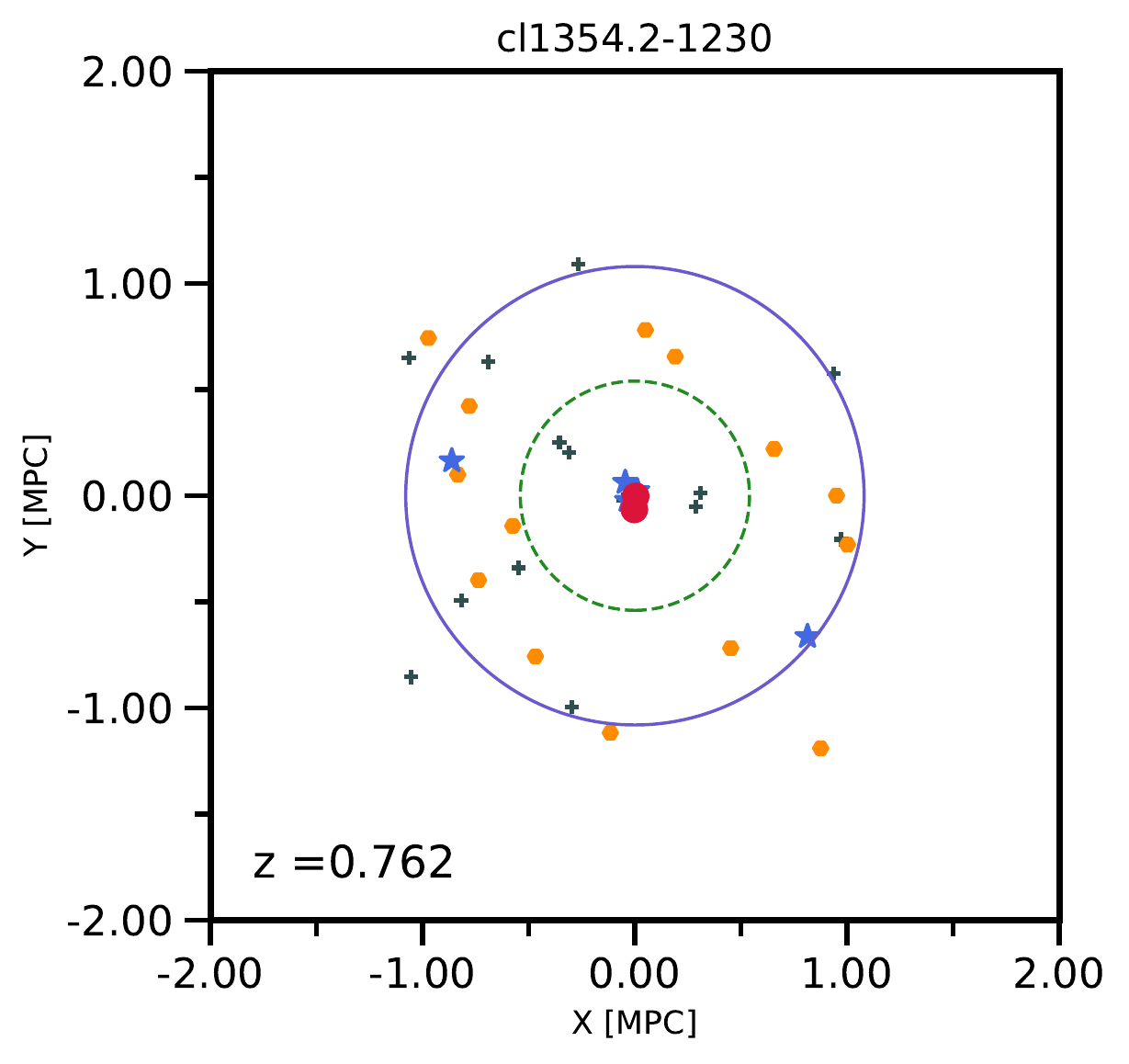}\hfill
\includegraphics[width=.33\textwidth]{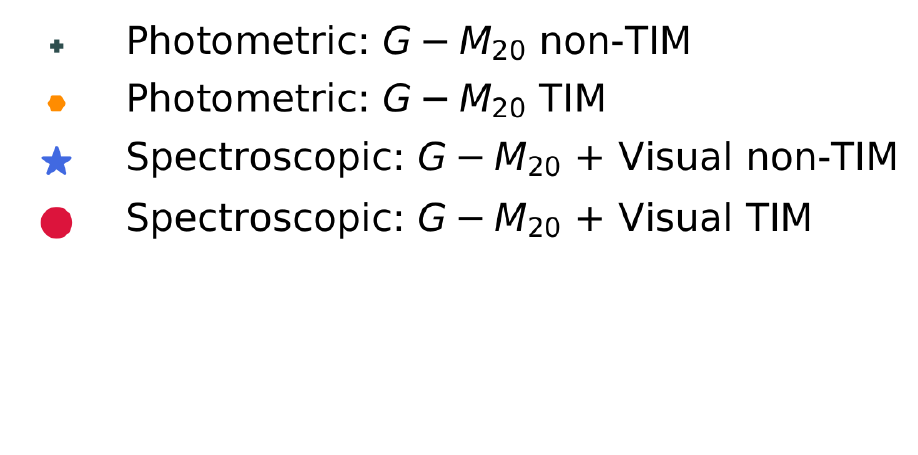}\hspace*{\fill}

\caption{X-Y plots for the cluster members in our sample. All plots are centered at the brightest cluster galaxy of the individual cluster. In every plot, red circles are visually classified TIM that also reside above our TIM selection line ($G-M_{20}$ + Visual TIM, see Figure~\ref{Fig:GM20}), blue stars are either visually classified undisturbed spectroscopic members or visually classified TIM below our line ($G-M_{20}$ + Visual non-TIM), orange points are photometric members above our line (Photometric $G-M_{20}$ TIM), and gray plus signs are photometric members below our line (Photometric $G-M_{20}$ non-TIM). The solid blue ring in each plot shows $R_{200}$ for each cluster, and the green dashed circle has a radius of $0.5\times R_{200}$. Cluster CL1138.2-1133 uses only its respective spectroscopic catalogs, as discussed in \S\ref{Sec:Sample}. Some clusters, such as CL1232.5-1250, do not have HST data that extends out to full $R_{200}$. We don't show CL1138.2-1133a and CL1354.2-1230a here, for reasons discussed in \S\ref{Sec:Sample}.}
\label{Fig:AllClust}
\end{figure*}

For galaxy groups we only use our spectroscopic sample. Our groups are poorer systems with a lower contrast against the background than the clusters, and precise redshift values are needed for clear identification of their members. The modest precision of even our good photometric redshifts would result in too high of a contamination from non-members if only using photometric redshifts to assign group membership.

In obtaining our results we chose to exclude certain structures from the analysis of our samples. CL1227.9-1138a is a poorer side structure in the same field as the targeted cluster CL1227.9-1138, with a much lower number of spectroscopic membership. It only had two spectroscopic members remaining after the application of our sample selection criteria. Due to this low sample size this structure has been excluded from our analysis.

CL1354.2-1230a is a cluster where we used the spectroscopic sample only. It has a small number of members and attempting to pick this structure using our photometric sample would have suffered from high contamination.

CL1138.2-1133 and CL1138.2-1133a are two clusters in the same field. Both these clusters are at $z < 0.5$, and therefore outside of the redshift interval where we have reliable photometric redshifts as our photometry does not extend shortward of the $4000\mathrm{\AA}$ break for those systems. Hence we only used these two clusters for our spectroscopic analysis.

CL1138.2-1133a and CL1354.2-1230a are too off-center in our spectroscopic observations to probe out to $0.5\times R_{200}$. Therefore we exclude them from any analysis that depends on the radial distribution or velocity distribution. We do include them in analyses that include the clusters as aggregates. We note that excluding these two systems does not affect any of our conclusions. CL1227.9-1138 has a brightest cluster galaxy (BCG) that is off-center compared to the rest of the members (see Figure~\ref{Fig:AllClust}), but since there is spectroscopic observations out to $R_{200}$ we included this cluster in any radial distribution analysis.

We choose field galaxies for our spectroscopic sample and photometric sample in a similar fashion. In each sample, we define our field galaxies to be within $\Delta z = 0.2$ of the cluster redshift, excluding galaxies that are cluster members. As described in \cite{MilvangJensen08}, galaxies within a $\Delta z=0.2$ slice around the cluster redshift form a magnitude limited sample that is unbiased by SED type. For the spectroscopic sample, this results in pure and complete field and cluster samples. For the photometric redshift sample, \cite{Pello09} showed that our photometric redshift cut is 90\% complete in selecting cluster members independent of SED type. The high membership completeness of our photometric redshift selection ensures that our our photometric field sample will have little contamination by cluster members. Due to the same reasoning as for our cluster galaxies, we also limit our field sample to $z > 0.5$. Hence in our phot+spec sample, field galaxies with $z < 0.5$ are coming from our spectroscopic sample only.

\subsection{Stellar Masses, Stellar Mass Completeness, and Final Galaxy Sample}

\begin{figure*}
\epsscale{1.15}
\plottwo{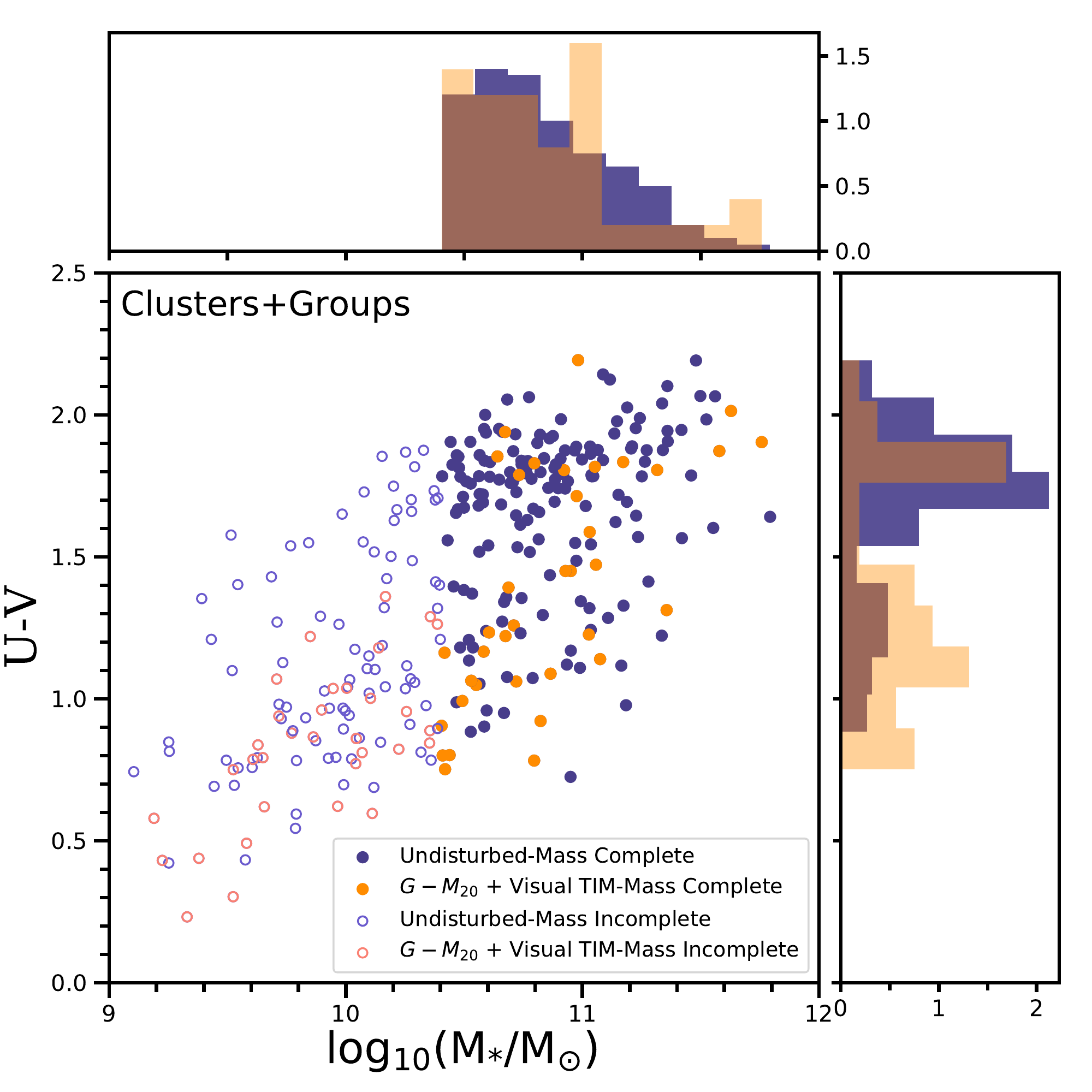}{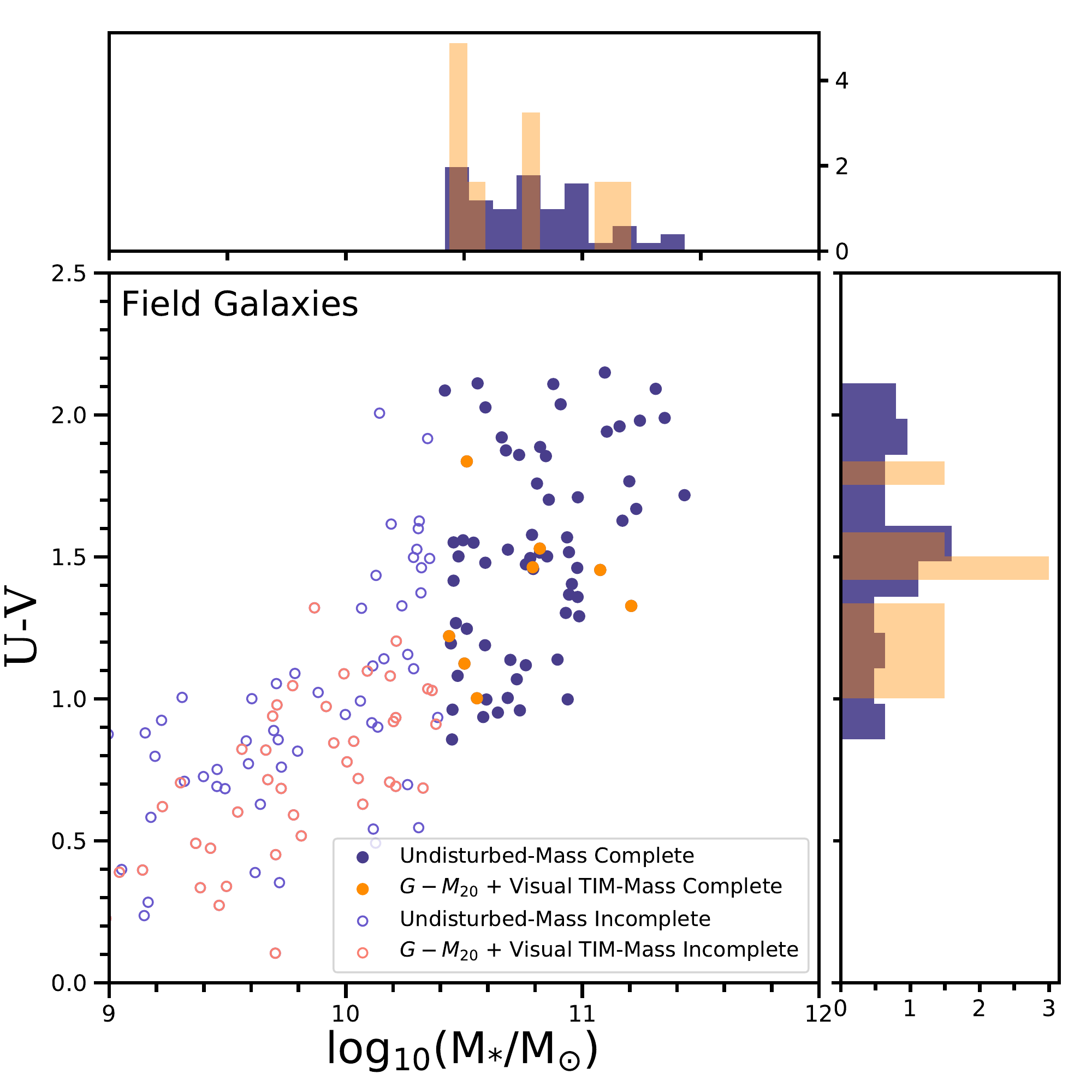}
\caption{U-V color versus stellar mass plots of our spectroscopic samples. All galaxies in \textit{both panels} have visual classifications. Galaxies shown in these plots have passed all quality cuts detailed in \S\ref{Sec:Sample}, except for the stellar mass completeness cut. In both panels, galaxies below our mass completeness limit of $\log_{10}(M_{*}/M_{\odot}) = 10.4$ are shown using open circles, galaxies above this threshold are shown in full circles. While we are complete below this limit for our photometric sample, we adopt the 10.4 limit to allow us to straightforwardly combine both samples. In both panels orange circles are visually classified TIM that are also classified as $G-M_{20}$ TIM, as explained in \S\ref{Sec:MorphClass}. Purple circles are galaxies visually classified as undisturbed or visual TIM that were not $G-M_{20}$ TIM. The normalized histograms for both panels show the number density of these classes, with colors being the same as the respective symbols. \textit{Left panel --} Our aggregate sample of spectroscopic cluster and group members. \textit{Right panel --} Our spectroscopic field galaxy sample.}
\label{Fig:UV-Mass}
\end{figure*}

We made use of the iSEDfit suite for the calculation of our stellar masses (for detailed information on iSEDfit see \cite{Moustakas13}). iSEDfit uses the redshift and observed photometry of galaxies to derive their stellar mass via a statistical likelihood analysis of a large ensemble of model SEDs. For our spectroscopic sample the masses were calculated at the galaxy spectroscopic redshifts. For our photometric cluster members masses were calculated with their redshifts fixed at the cluster redshift, where for the field galaxies masses were calculated at their photometric redshifts. We used a stellar mass cut of $\log_{10}(M_{*}/M_{\odot}) > 10.4$ to both our photometric and spectroscopic samples \citep{Rudnick17}. Above this limit we are mass complete. The $G-M_{20}$ code (more details on the $G-M_{20}$ method in \S\ref{Sec:MorphClass}) has a quality flag indicating whether the measurement can be trusted. Any objects that failed to pass this test was taken out of our sample as well. We mapped the distribution of objects for which a flag was raised across all our fields (17 objects total), and found via visual inspection that the distribution is spatially uniform. The rejection is not biased towards whether the object resides in a 1 orbit or 5 orbit depth region. After these quality cuts, there are a total of 163 cluster members in our spectroscopic sample, and 429 cluster members in our photometric+spectroscopic sample. Our samples sizes after these quality cuts is shown in Table~\ref{Table:Clusters}. Our spectroscopic cluster plus group sample, and all spectroscopic field galaxies are shown in a U-V color versus stellar mass plot in Figure~\ref{Fig:UV-Mass}. The galaxies in both panels are after all the quality cuts described above, except for the stellar mass cut. This plot also shows galaxies according to their visual class. The galaxies are split by their visual classification as determined by \cite{Kelkar17} and as described in detail \S\ref{Sec:MorphClass}.

\section{Morphological Classification}
\label{Sec:MorphClass}

We chose to make use of two different techniques to quantify the morphologically disturbed features in our galaxy sample, an automated method and visual classification of galaxies. These two methods have particular strengths that complement the intrinsic weaknesses of each other. Interactions between galaxies leave an imprint on the morphologies of the galaxies involved, and visually identifying these is a common method in merger analysis. This procedure invariably suffers from subjectivity, as visual morphological distortions a galaxy displays may have multiple causes. Automated methods are generally faster methods that carry the advantage of being reproducible. However, such methods can miss certain signatures of merger events and hence suffer from incompleteness. They also require careful calibration to increase completeness and to reduce contamination.

For our case, we use the visual classification method to calibrate our automated method of choice. The automated morphology analysis we use for this work uses $G$ \citep{Abraham03}, the Gini coefficient and $M_{20}$ \citep{Lotz04} as parameters. Briefly, $G$ is a measure of how the flux is distributed among the pixels of the target galaxy, and $M_{20}$ is defined as the normalized second-order moment of the brightest 20\% of the galaxy’s flux (further details in \cite{Lotz04}). This method, henceforth referred to as the $G-M_{20}$ method, is a nonparametric measure of morphology and hence does not assume any analytic functions for the light distribution of the measured object. This brings applicability of the method to irregular galaxies as well. The method has been shown to be effective especially at picking up bright double nuclei, which might be indicative of a merger event. \cite{Lotz04} showed that this method is able to detect morphological disturbances even at low signal-to-noise ratio (S/N). Furthermore, \cite{Lotz10} compares observability time scales at various baryonic mass ratios for different tests of morphology, namely $G-M_{20}$, $G-A$ and $A$ (asymmetry, \cite{Conselice03}). They conclude that the merger detection timescale of $G-M_{20}$ does not drop significantly even at baryonic mass ratios of around 10:1, and that it is, therefore, just as capable of detecting 9:1 mass ratio minor mergers as 1:1 major mergers \citep{Lotz10}. This favors the use of $G-M_{20}$ for detection of minor mergers. Hence another clear advantage the use of this method grants us is the well determined timescales of merger events, which we plan to use for future papers.

\begin{figure*}
\centering
\includegraphics[width=.33\textwidth]{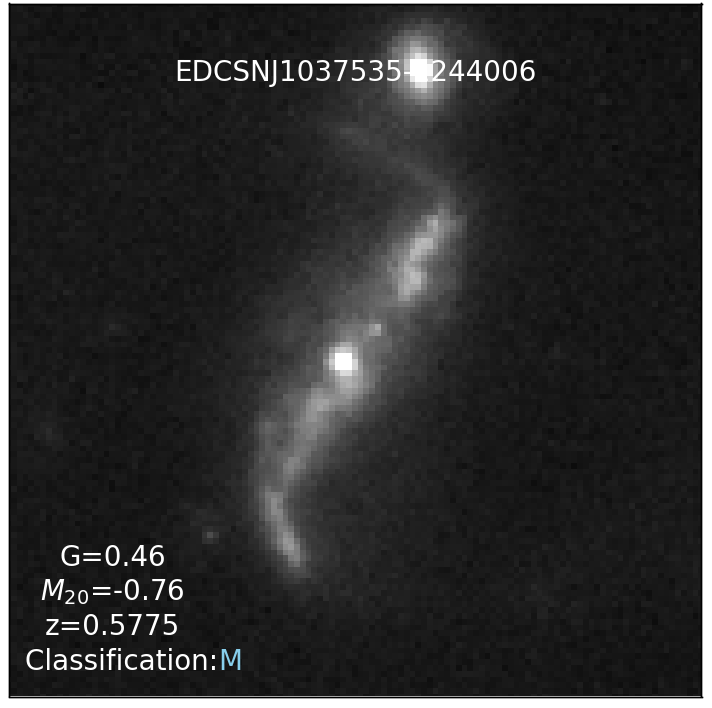}\hfill
\includegraphics[width=.33\textwidth]{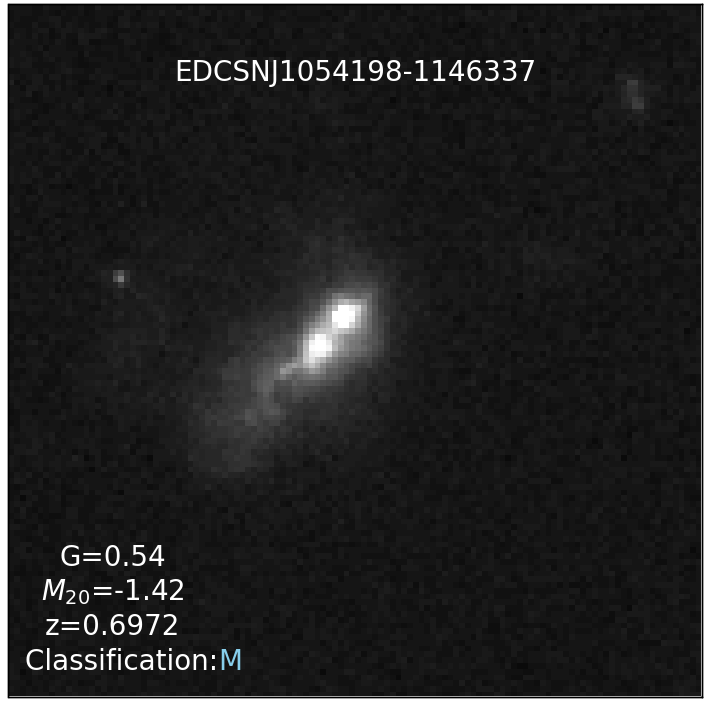}\hfill
\includegraphics[width=.33\textwidth]{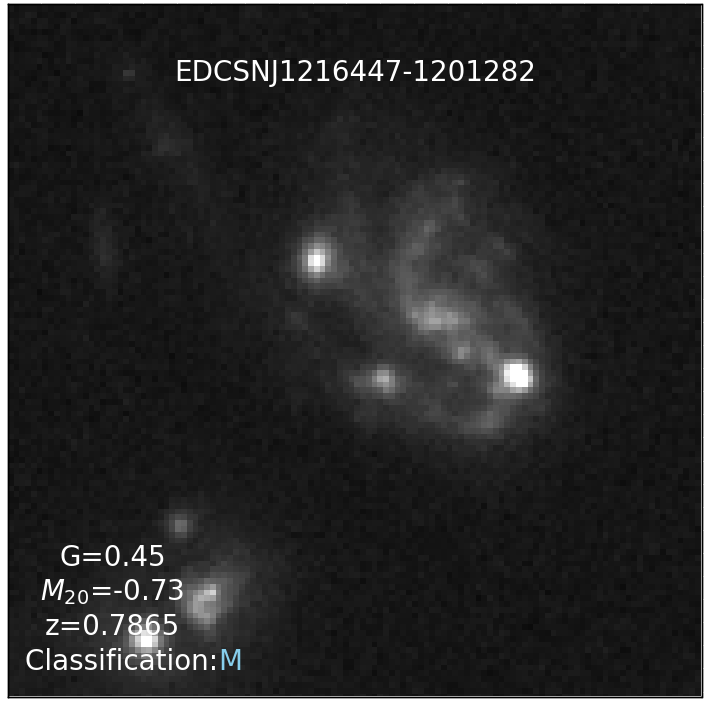}\\
\includegraphics[width=.33\textwidth]{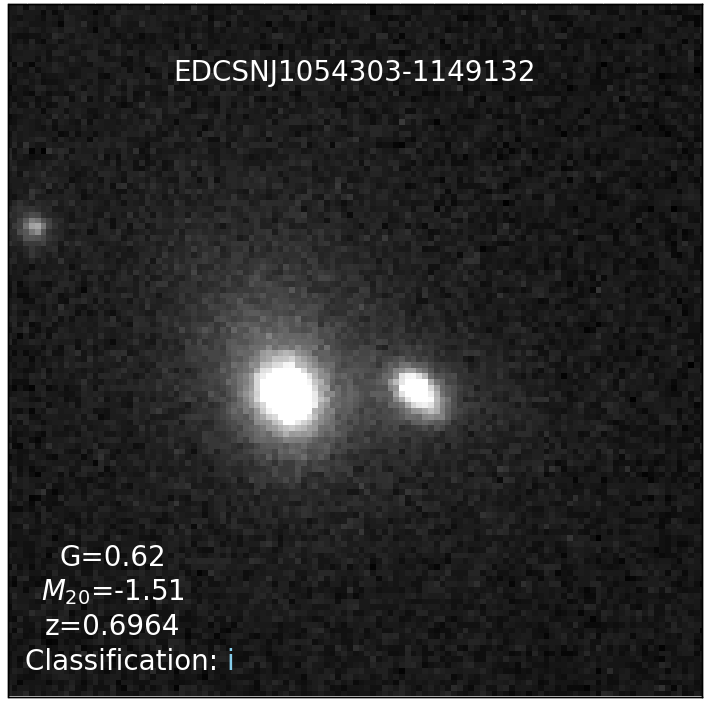}\hfill
\includegraphics[width=.33\textwidth]{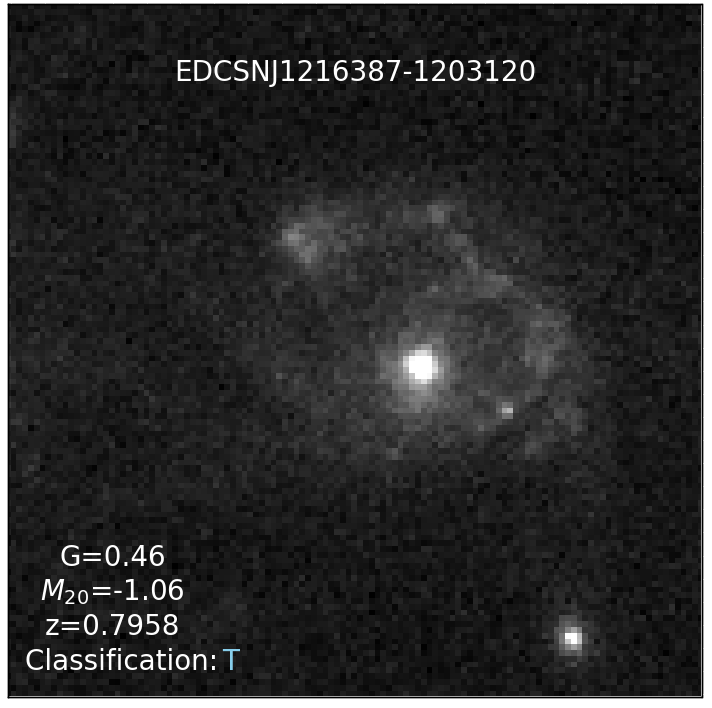}\hfill
\includegraphics[width=.33\textwidth]{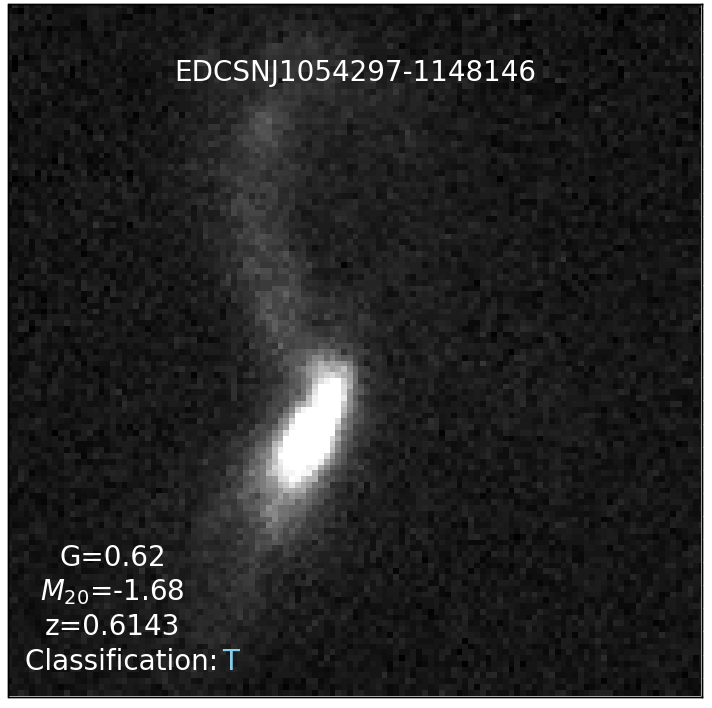}\\
\includegraphics[width=.33\textwidth]{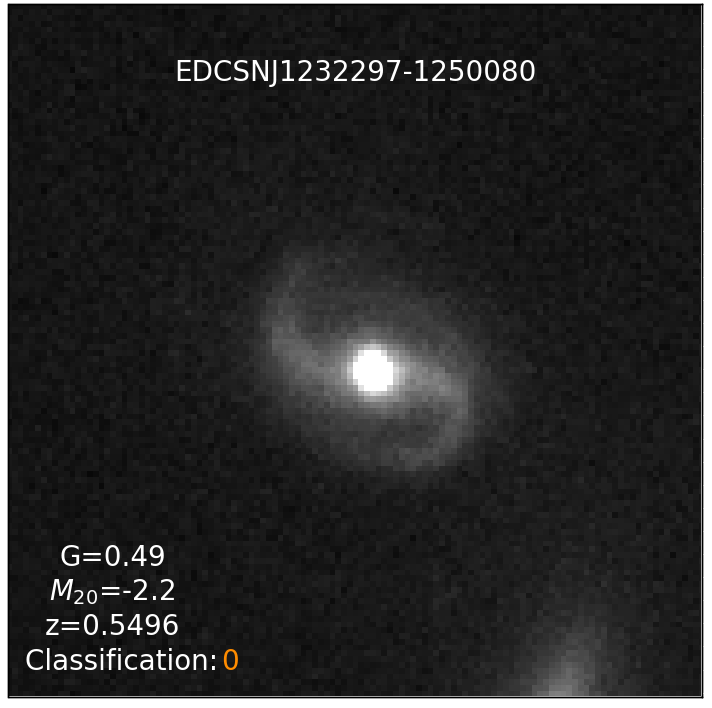}\hfill
\includegraphics[width=.33\textwidth]{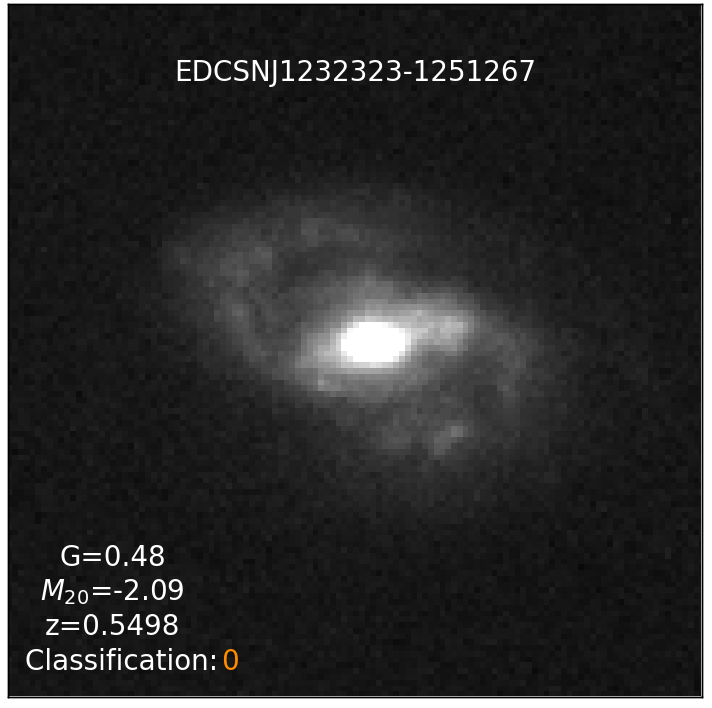}\hfill
\includegraphics[width=.33\textwidth]{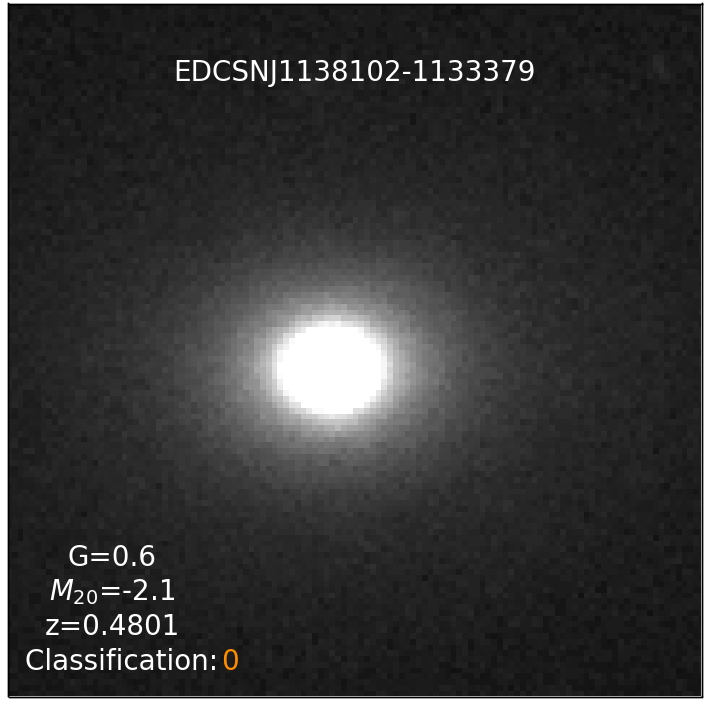}
\caption{Example postage stamps from the EDisCS-HST spectroscopic sample for which we performed a visual classification of morphology. Every panel shows a $6^{\prime\prime}\times6^{\prime\prime}$ region. Every panel shows the galaxy ID on top, then G, $M_{20}$, redshift, and its visual classification info at the bottom. M/m denote major/minor mergers, I/i denote strong/weak interactions, T/t denote strong/mild tidal features, and 0 denotes undisturbed galaxies \citep{Kelkar17}. Light blue color for the visual class is used to indicate that the object is a $G-M_{20}$ identified TIM (see Figure~\ref{Fig:GM20} for the line, and \S\ref{Sec:MorphClass} for its derivation), and orange color to indicate that it is below our line and hence is not identified as a $G-M_{20}$ TIM.}
\label{Fig:Classes}
\end{figure*}

In order to calibrate the completeness and contamination of our $G-M_{20}$ classification, we use the visual classification of \cite{Kelkar17} for galaxies from our sample with spectroscopic redshifts. In \cite{Kelkar17} three identifiers independently classified structural disturbances in order to control for variation between the identifiers. Every galaxy in our spectroscopic sample was classified into classes of minor/major mergers, strong/weak interaction and strong/mild tidal features and undisturbed (non-interacting galaxies), independent of morphology. A classification of merger or interaction required at least one visually nearby neighbor, whereas tidal features did not require any since tidal features can remain intact after the merger is complete. In Figure~\ref{Fig:Classes} we present examples of our visual classification scheme. Even with the best of efforts, no visual classification of morphology is foolproof. It is unfitting to appropriate every morphological asymmetry a galaxy displays to interactions with another galaxy. Regardless, classes other than undisturbed still have a higher probability of being the result of some form of galaxy-galaxy gravitational interaction or merger event. Therefore for the purposes of our merger analysis, all classes except for undisturbed are considered under one composite tag and will hence be referred to as ``tidal interactions and mergers", or TIM for short. After careful examination of the visual classifications of \cite{Kelkar17}, we reclassified three of the galaxies in our sample. These new classifications are given in Table 3.

\begin{figure*}
\epsscale{1.1}
\plottwo{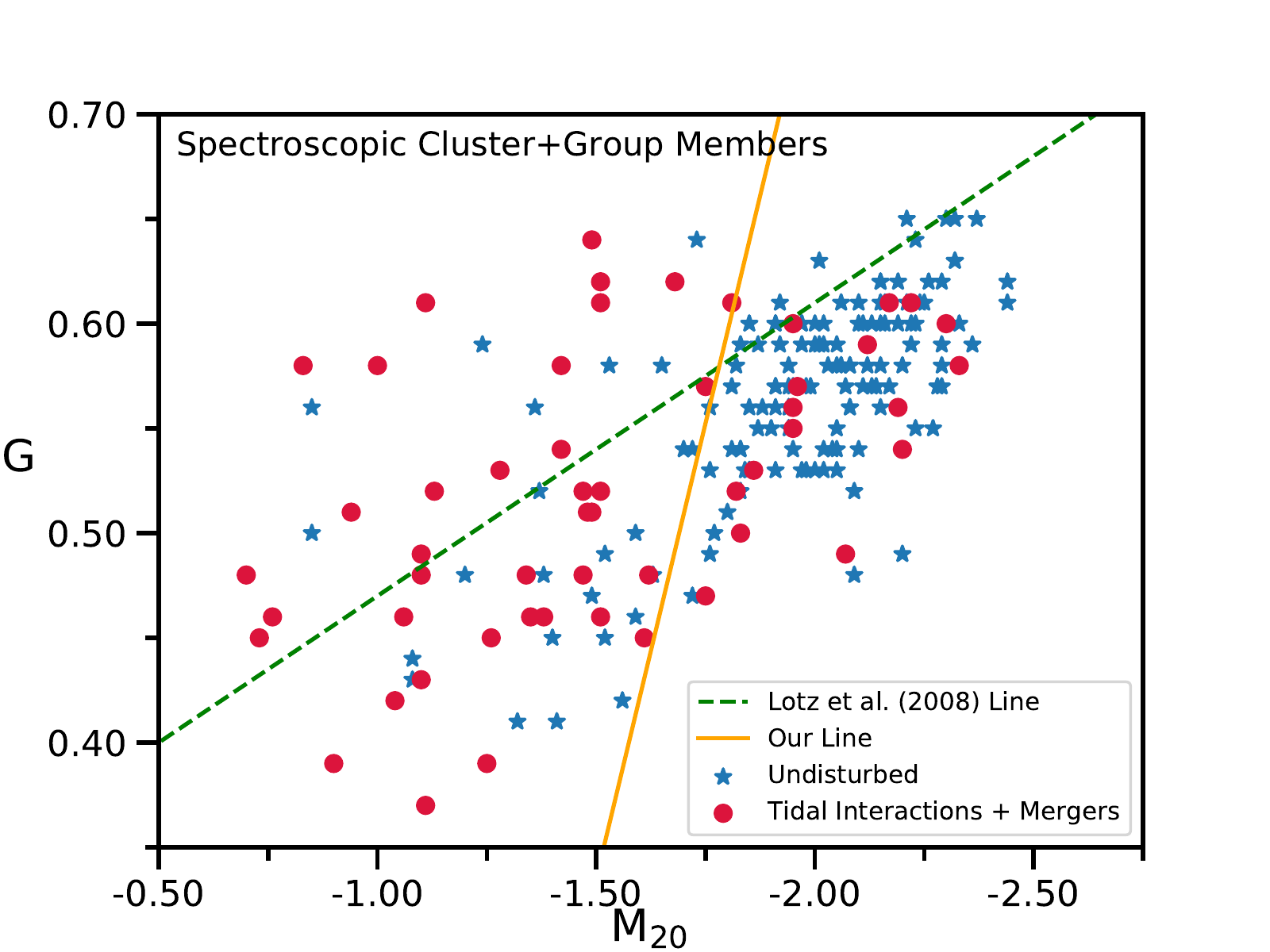}{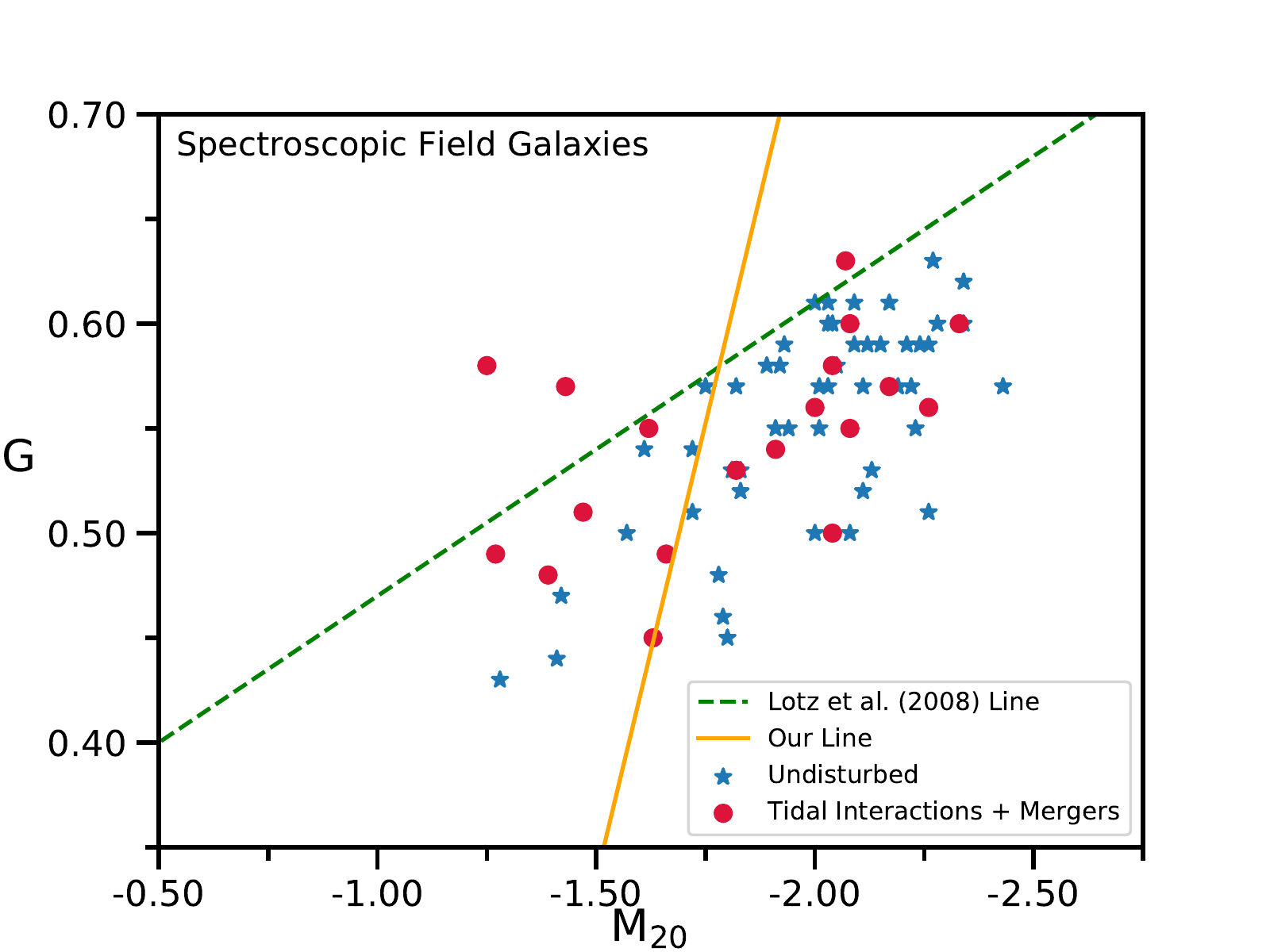}
\caption{The $G-M_{20}$ plots of our spectroscopic sample. All galaxies in \textit{both panels} have visual classifications. \textit{Left panel --} All spectroscopically confirmed cluster and group member galaxies are shown together. Red circles are galaxies that are visually classified as TIM as described in \S\ref{Sec:MorphClass}. Blue stars are galaxies that show no sign of interaction and are hence classified as undisturbed. The orange line corresponds to the highest purity value obtained through our calibration detailed in \S\ref{Sec:MorphClass}, whereas the green dotted line is the merger selection line from L08. \textit{Right panel --} Spectroscopically confirmed field galaxies. Symbols are the same as in the left panel. In both panels we see that our line picks many visually classified TIM that would have been left out by the L08 line.}
\label{Fig:GM20}
\end{figure*}

\begin{figure*}
\epsscale{1.1}
\plottwo{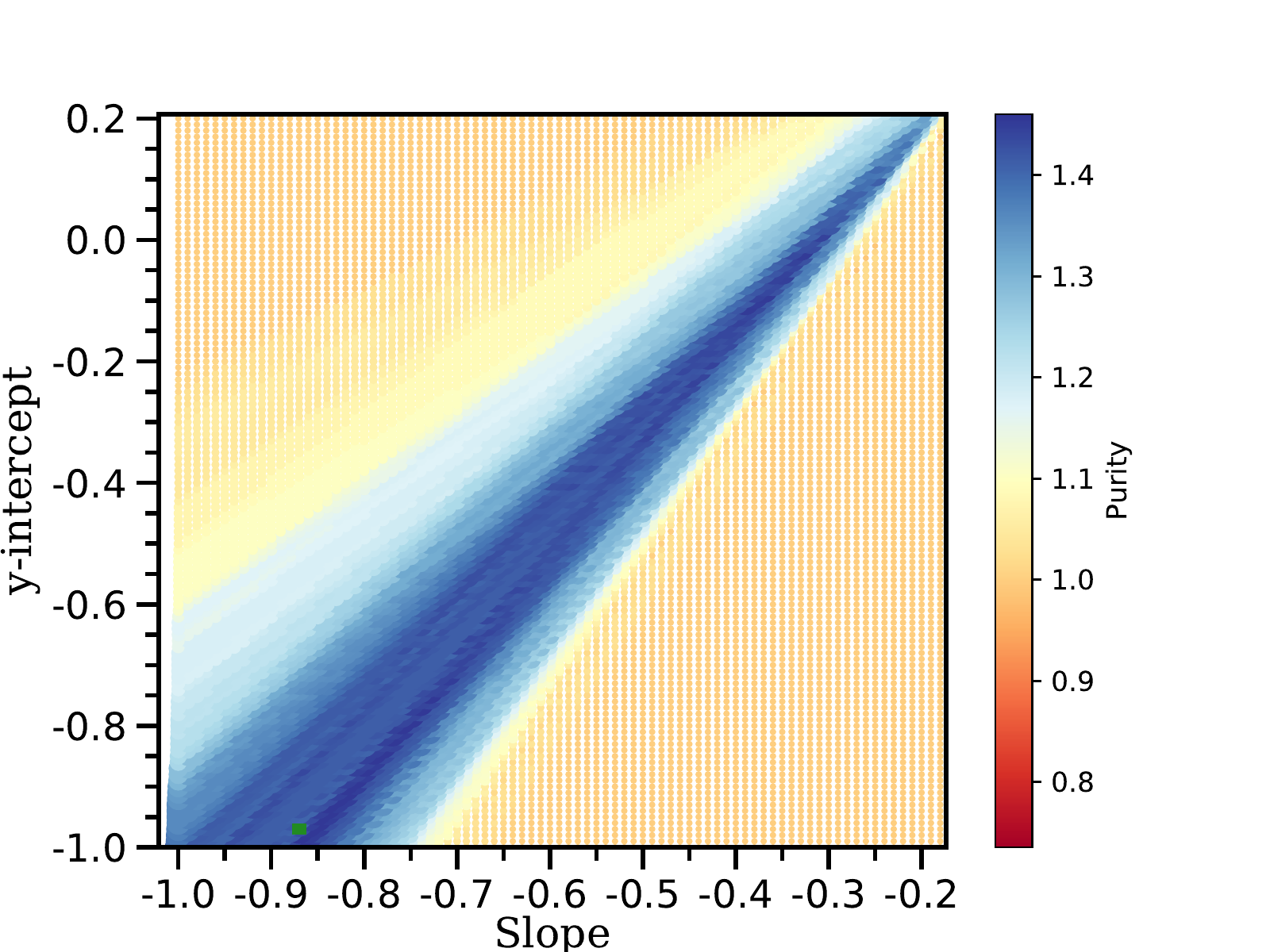}{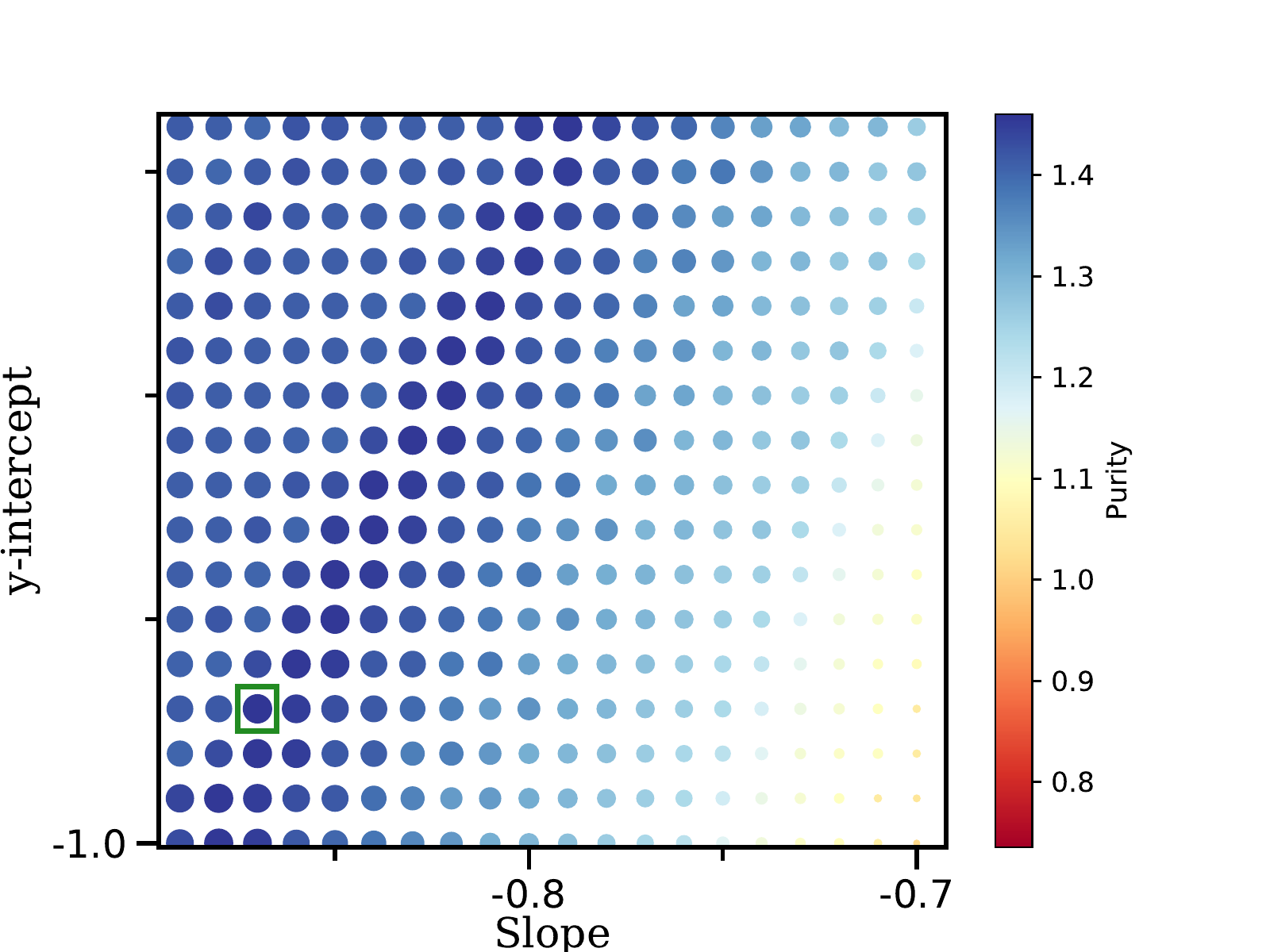}
\caption{Purity plot we used for the calibration of our line. The right panel is a zoom-in to a region of the left panel where our highest purity value resides (shown inside the green box). The plot has been obtained by calculating values of purity (as defined in \S\ref{Sec:MorphClass}) at different y-intercept and slope values. Larger and darker blue points represent higher purity results. We had only one result with the highest purity value of 1.46, which corresponded to -0.87 for the slope, and -0.97 for the y-intercept. Those values have been used for our merger selection line for all $G-M_{20}$ plots in this paper. We also tested purity values close to our highest value and using these did not change the results of our analysis in a significant way.}
\label{Fig:Purity}
\end{figure*}

\begin{figure*}
\centering
\includegraphics[width=.33\textwidth]{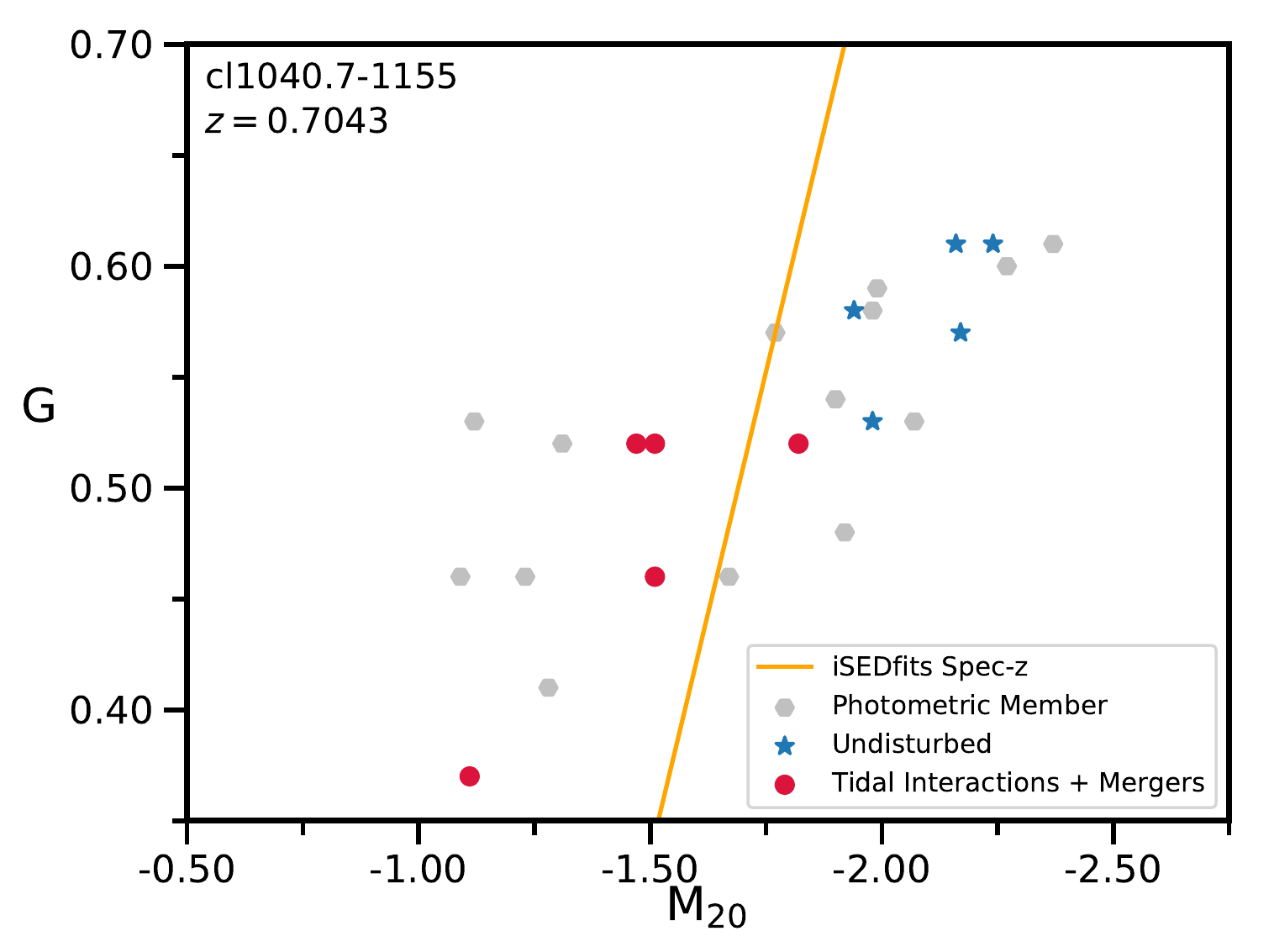}\hfill
\includegraphics[width=.33\textwidth]{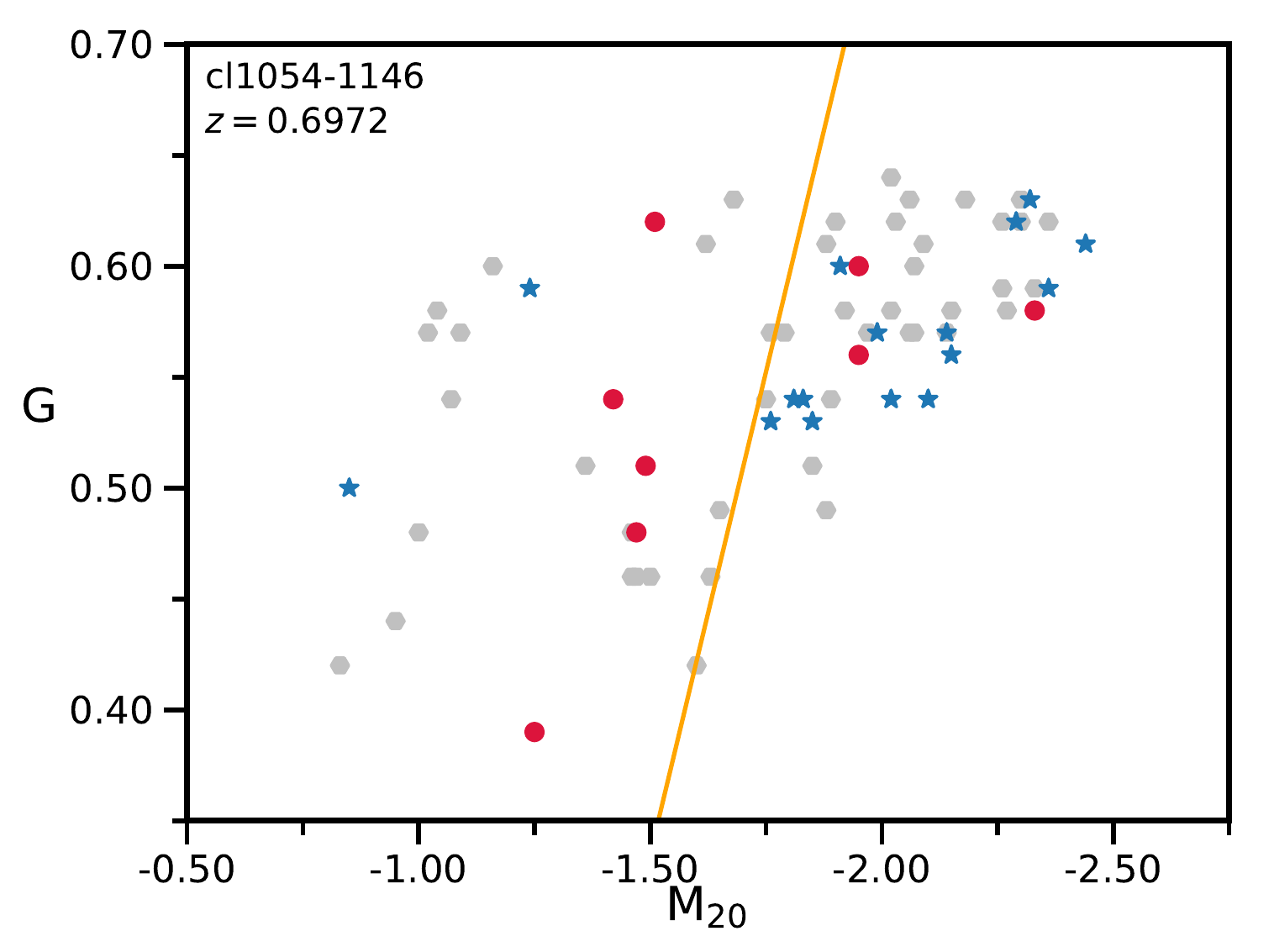}\hfill
\includegraphics[width=.33\textwidth]{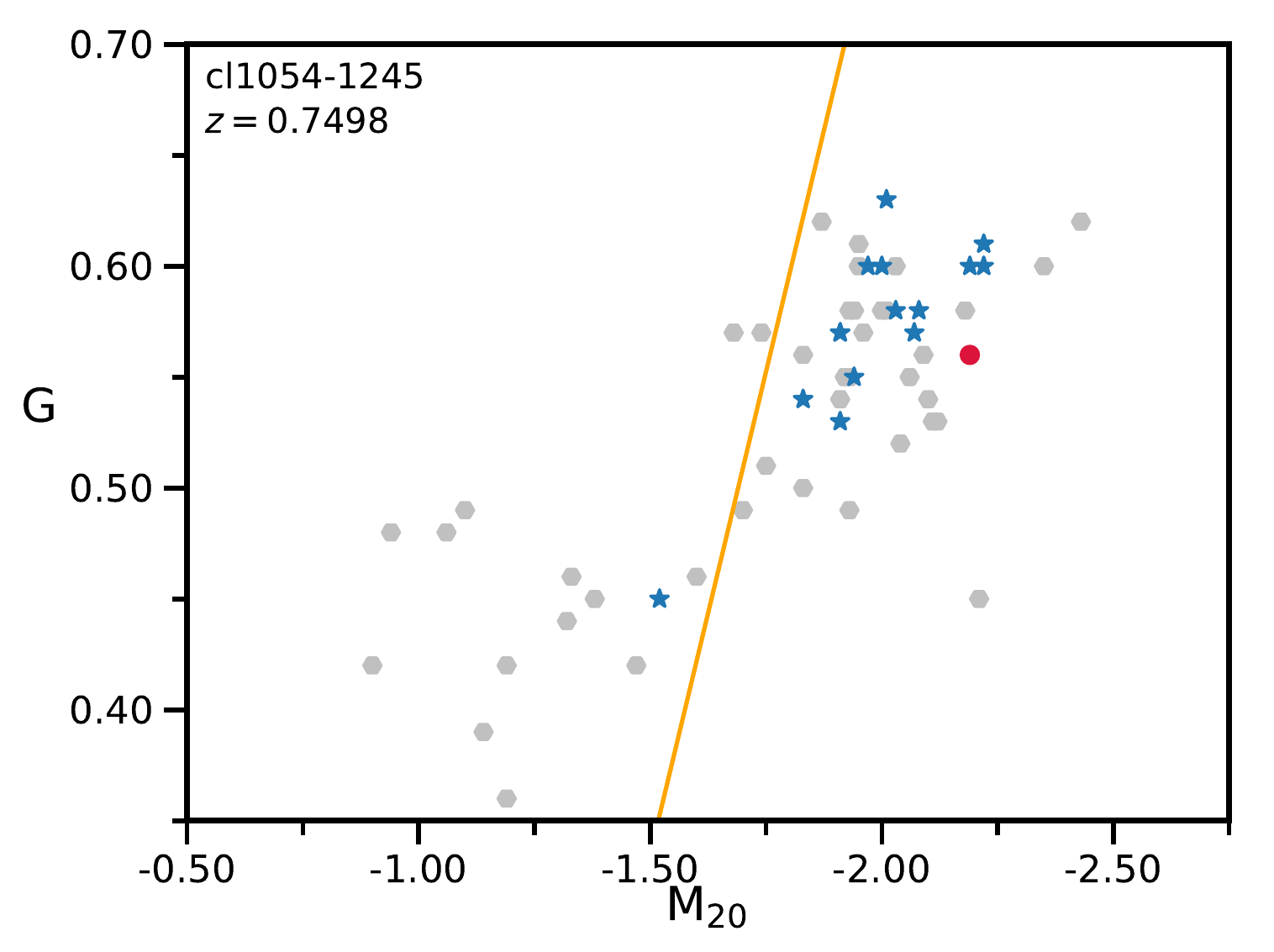}\\
\includegraphics[width=.33\textwidth]{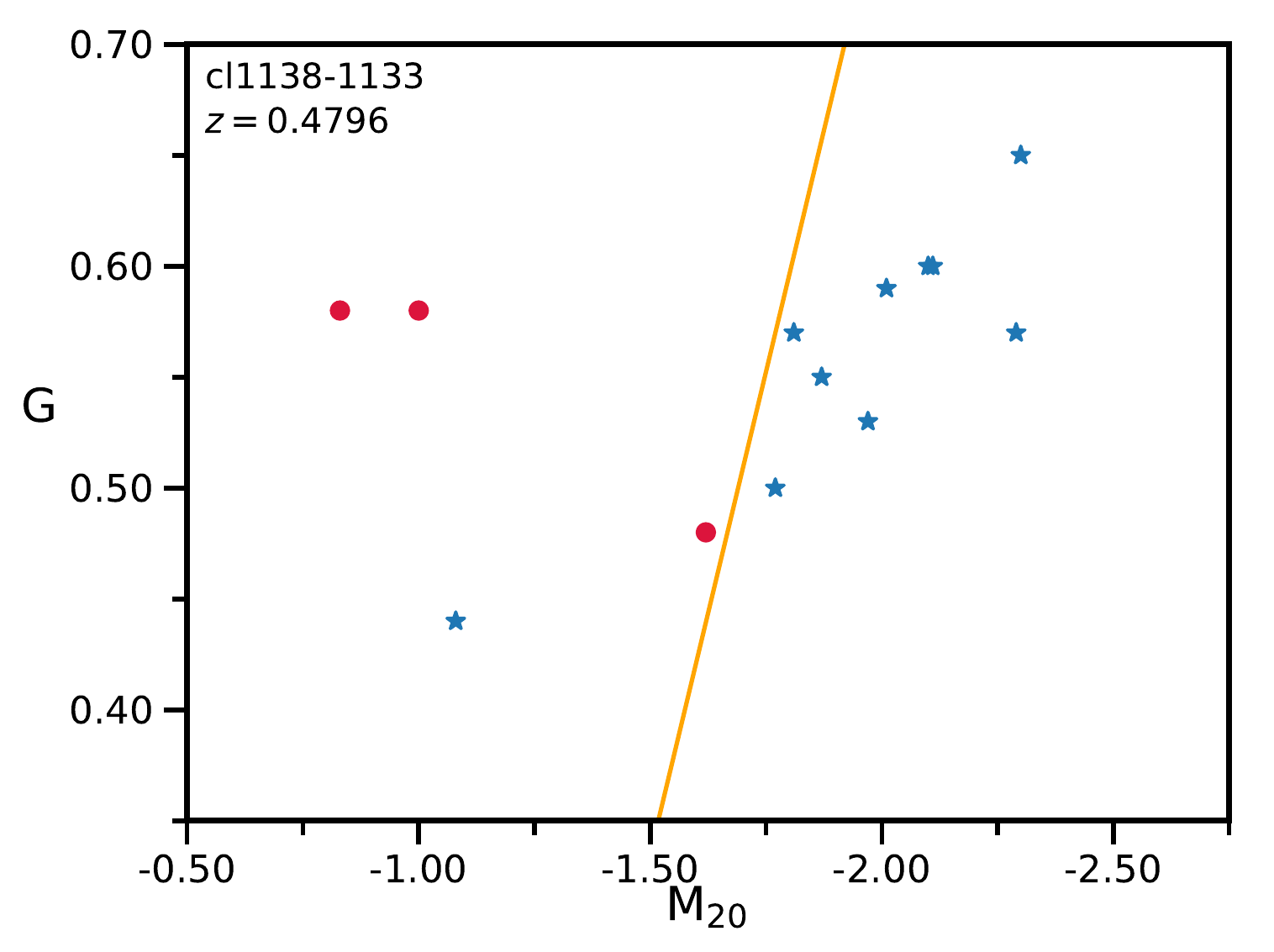}\hfill
\includegraphics[width=.33\textwidth]{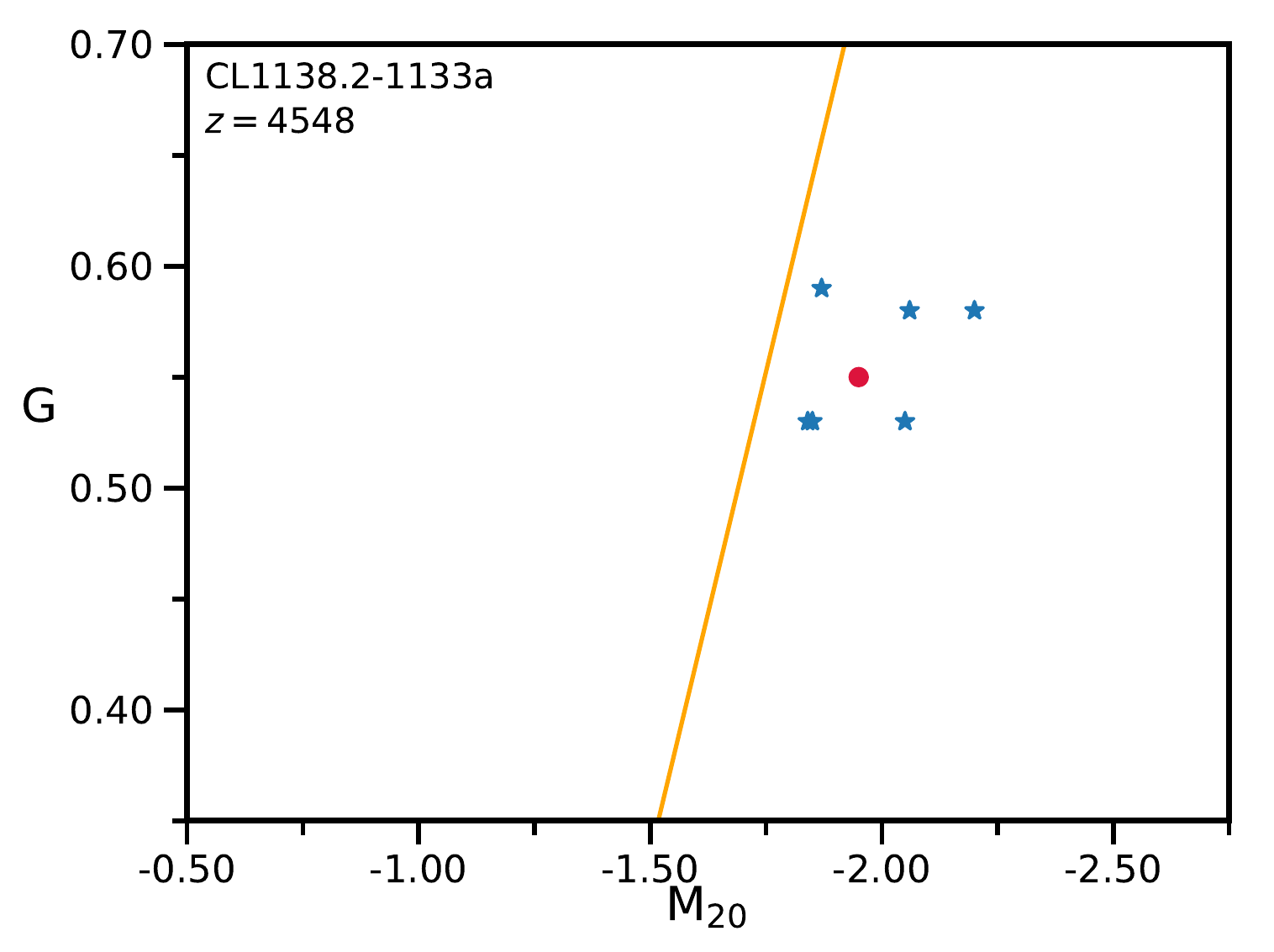}\hfill
\includegraphics[width=.33\textwidth]{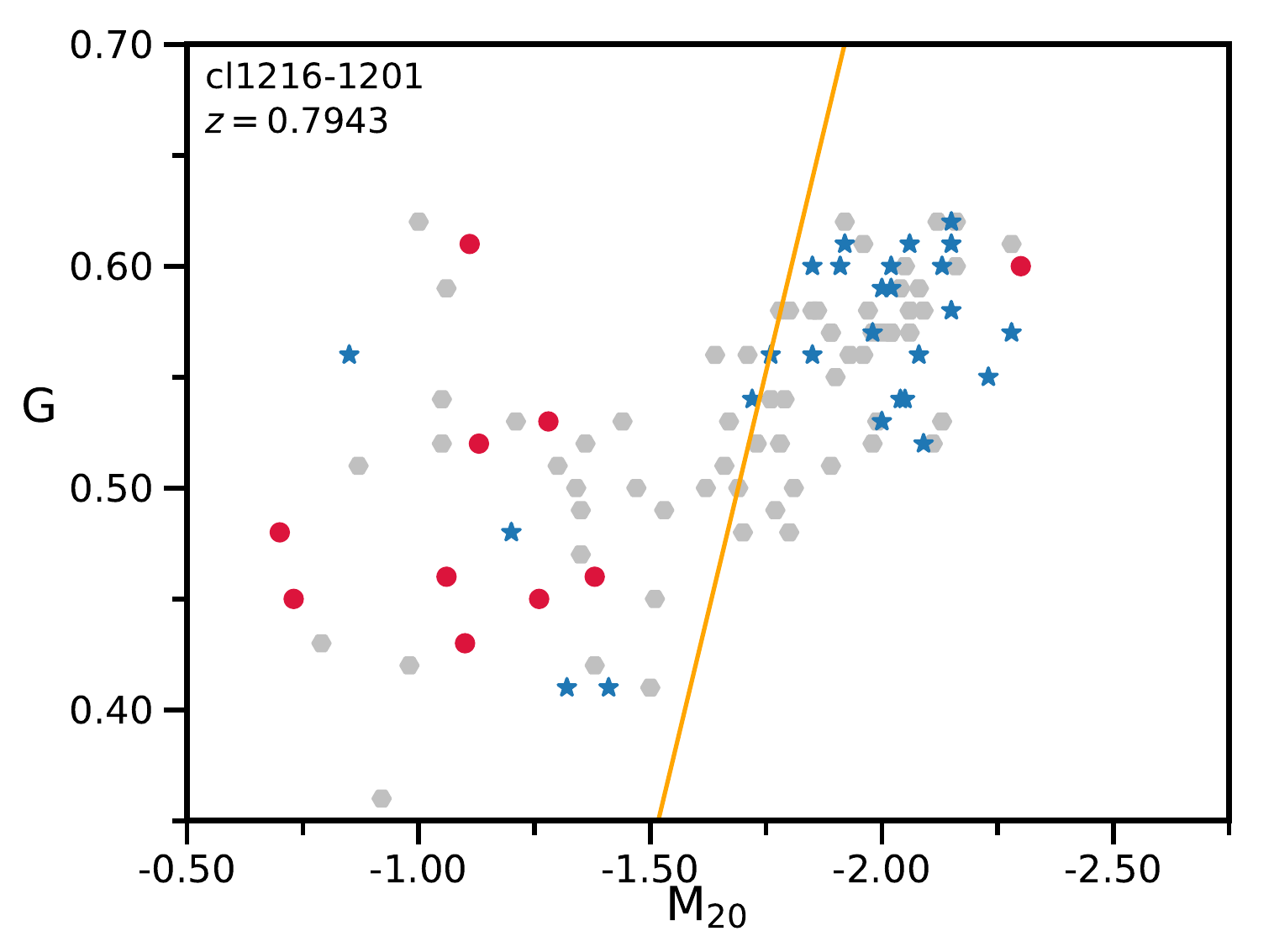}\\
\includegraphics[width=.33\textwidth]{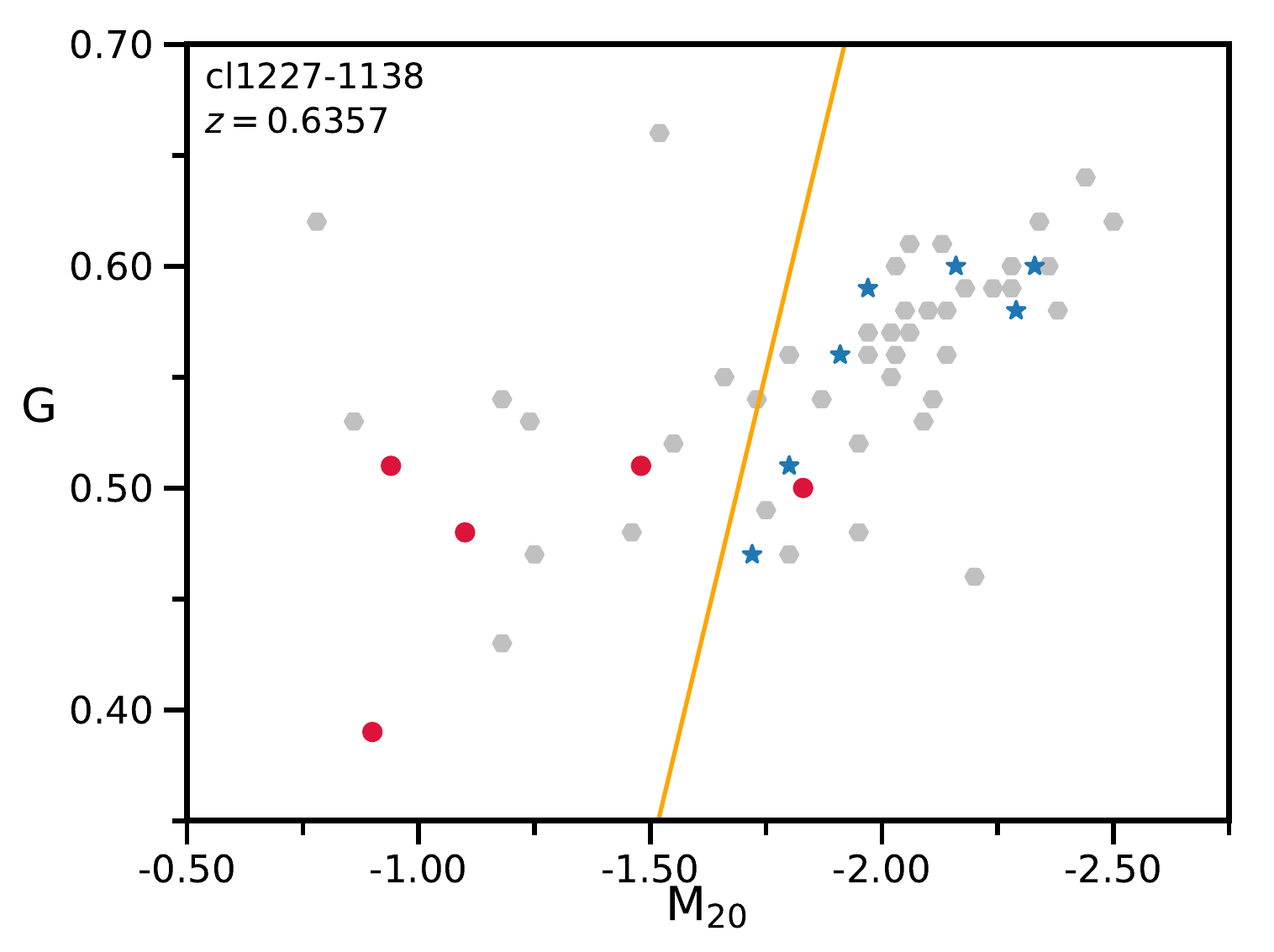}\hfill
\includegraphics[width=.33\textwidth]{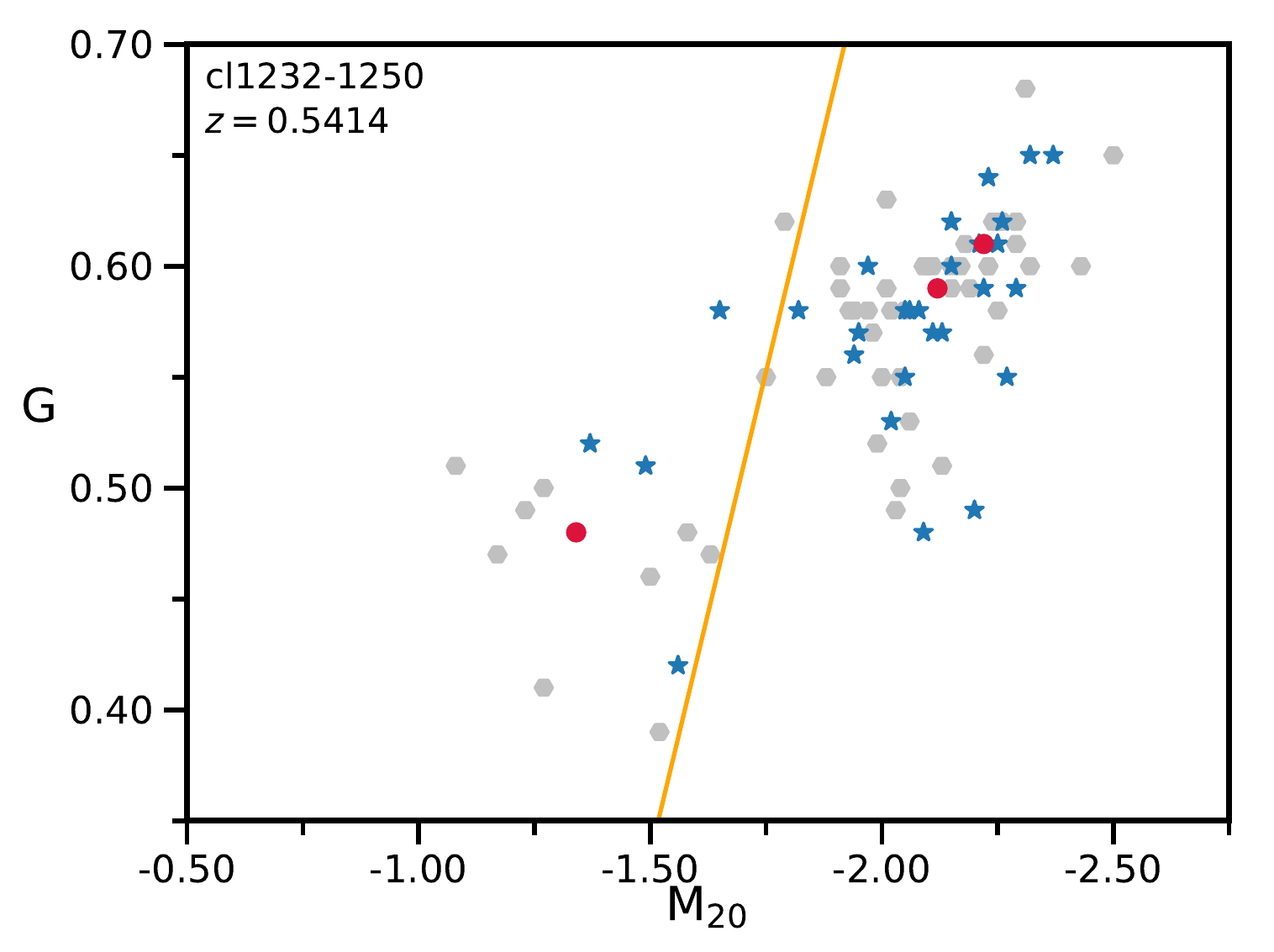}\hfill
\includegraphics[width=.33\textwidth]{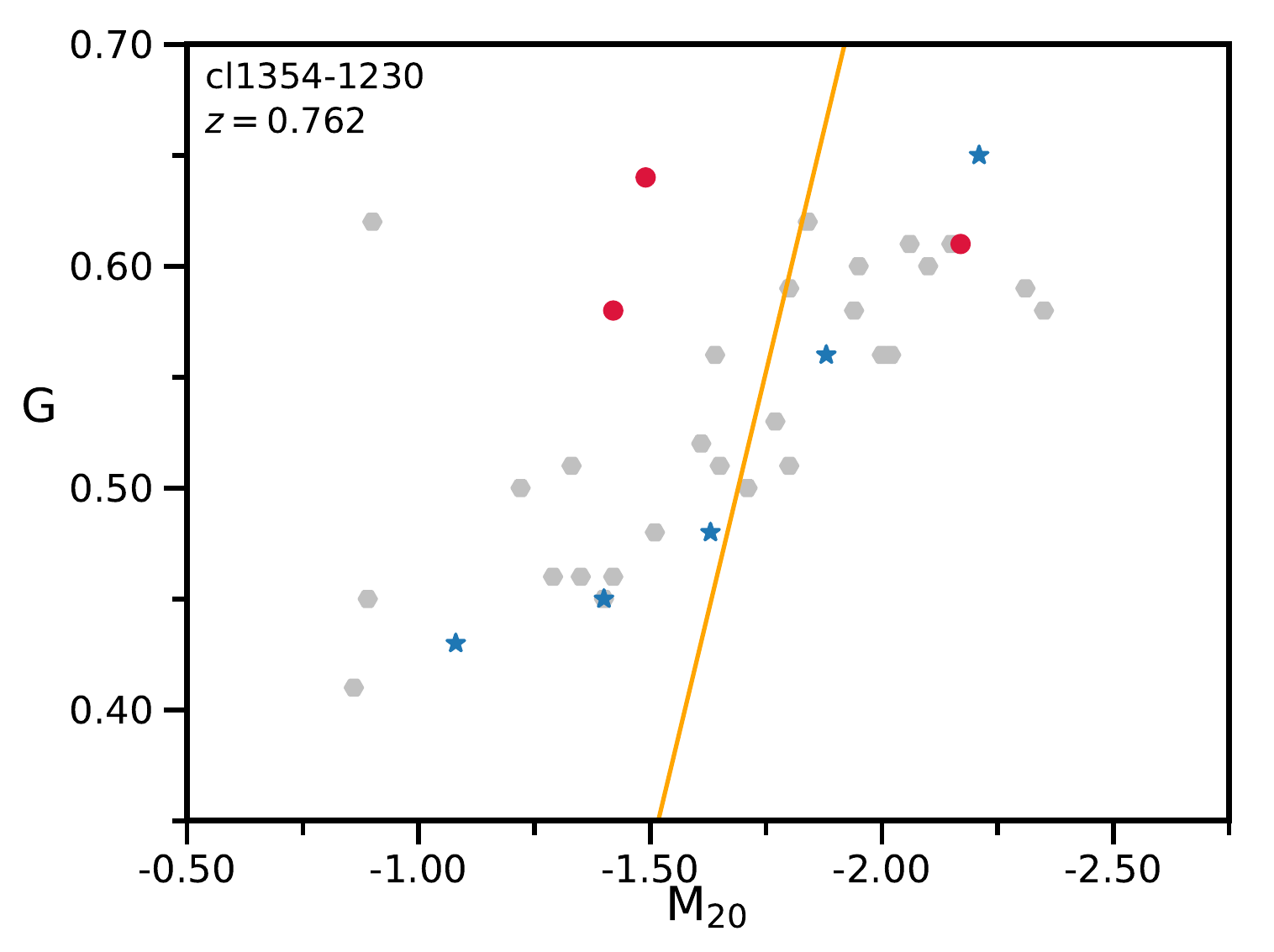}\\
\includegraphics[width=.33\textwidth]{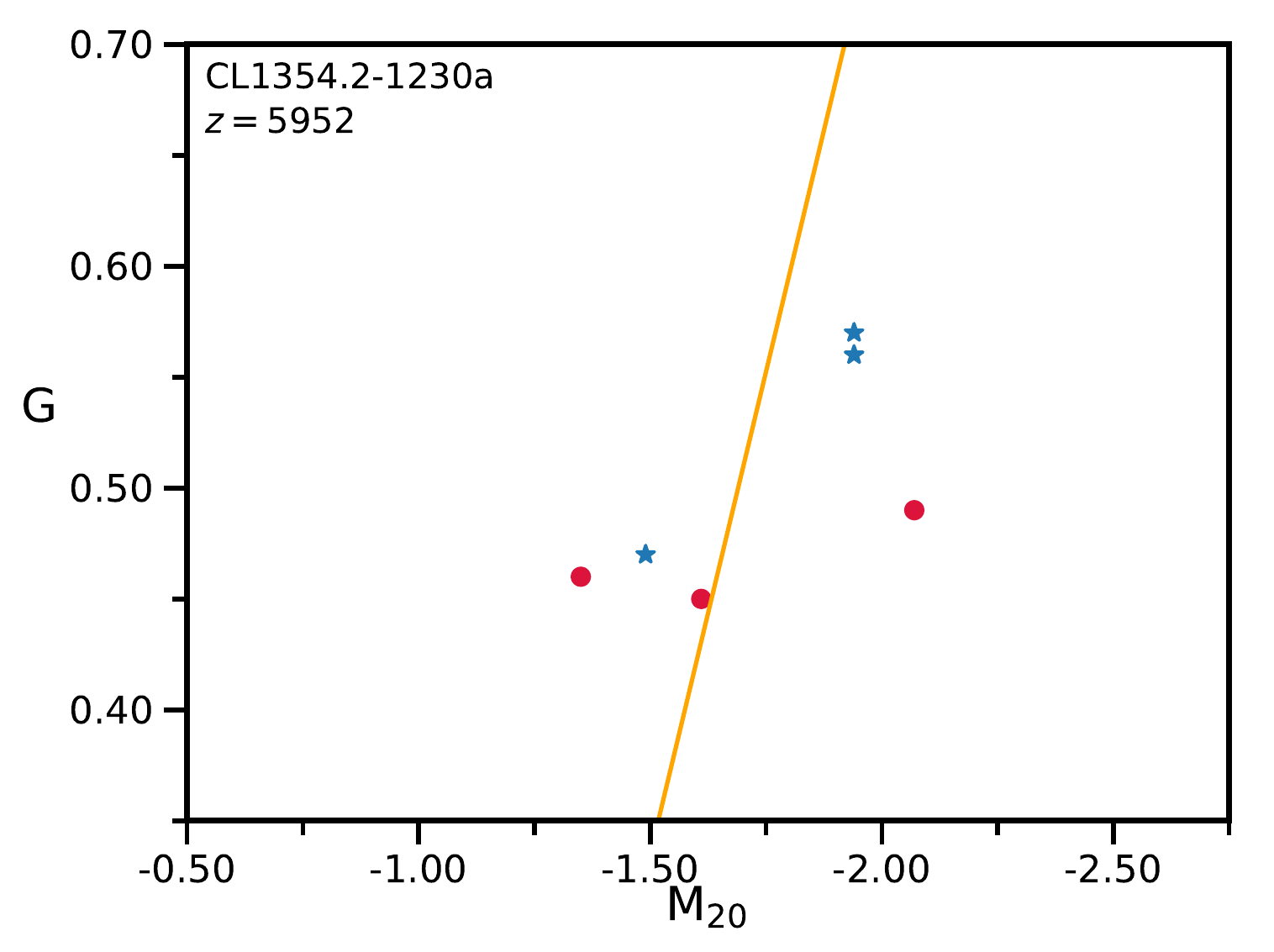}\hfill
\includegraphics[width=.33\textwidth]{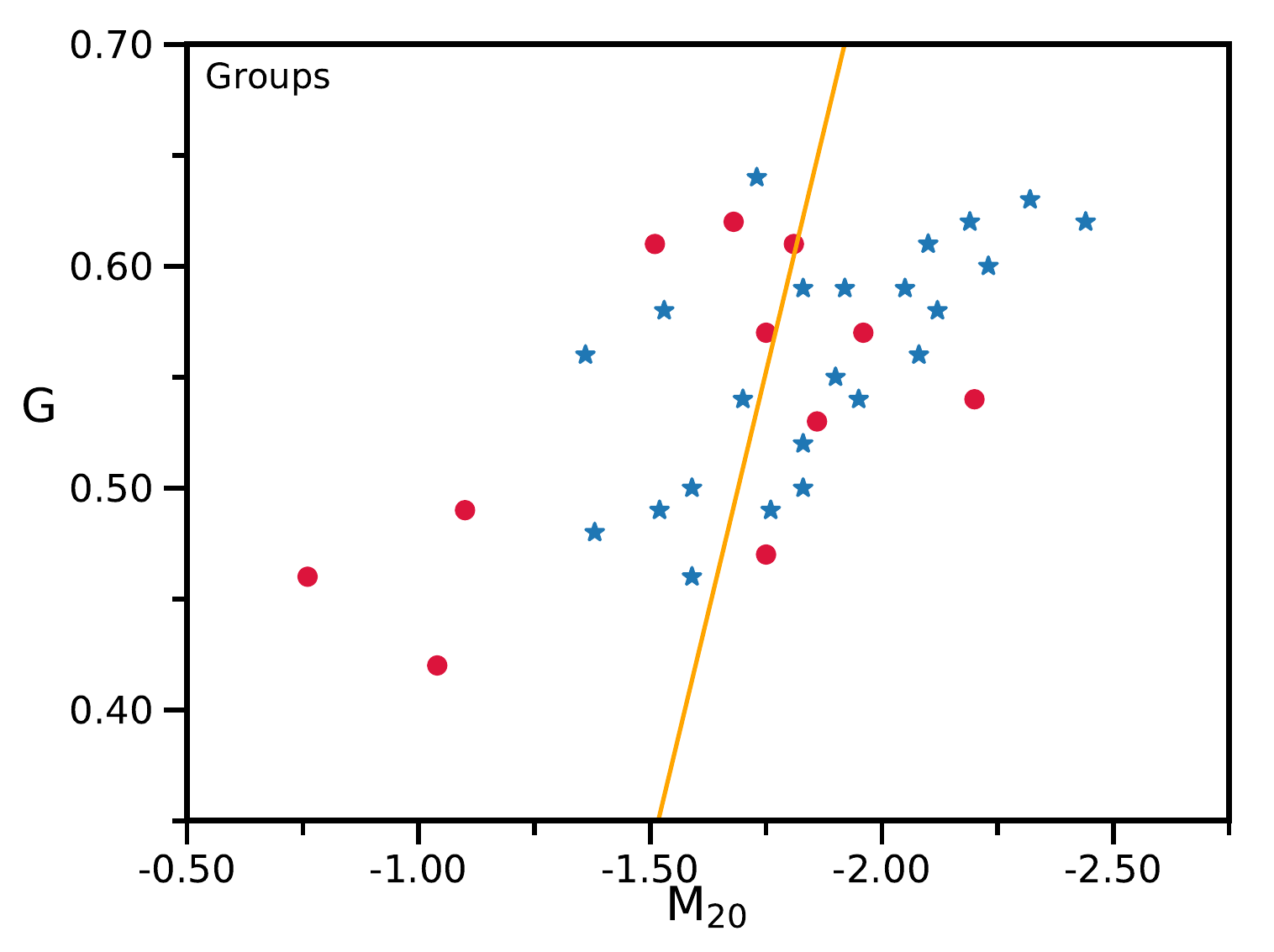}
\caption{$G-M_{20}$ plots for all of our individual clusters and an aggregate plot of our groups. Red data points are visually classified TIM in our spectroscopic sample, blue data points are undisturbed galaxies in our spectroscopic sample. Gray circles are our members that only have photometric redshifts and hence lack visual classification. The line is our calibrated TIM decision boundary, see \S\ref{Sec:MorphClass} for its derivation. As discussed in \S\ref{Sec:MorphClass}, CL1227.9-1138a has been excluded from any following analysis as it only contains two members that satisfy our selection criteria.}
\label{Fig:GM20-SpecSample}
\end{figure*}

For our analysis, we rely on both a visual classification and an automated classifier of galaxy morphology, in hopes of combining the particular strengths of both methods. The $G$ and $M_{20}$ values we measured for this sample together with their visual classes (TIM and undisturbed) are shown in Figure~\ref{Fig:GM20}. Our results reveal that the selection line used by (\cite{Lotz08}; L08) to separate merging galaxies from nonmergers is missing a substantial fraction of our visually classified mergers. L08 uses a lower stellar mass cut compared to ours and their line is optimized to avoid selecting low mass, high gas fraction irregular galaxies that are not undergoing an interaction or merger event. Our higher mass cut ensures that our analysis is not contaminated by such galaxies. To address this issue we decided to utilize the visual classifications to calibrate our merger selection criteria with the $G-M_{20}$ method. We derived a selection line with the premise of maximizing the number of mergers above and maximizing the number of nonmergers below it. We first define purity $\rho$ as
\begin{equation}
 \rho = \frac{N_{above}^{VisTIM}}{N_{total}^{VisTIM}} + \frac{N_{below}^{VisUnd}}{N_{total}^{VisUnd}},
\end{equation}
where $VisTIM$ and $VisUnd$ is used for objects visually classified as tidal interactions and mergers, and undisturbed, respectively, above and below to denote above and below the selection line. We optimized our line by requiring maximum purity, which we obtained by varying values of y-intercept and slope of the selection line. In Figure~\ref{Fig:Purity} we show the purity value as a function of slope and y-intercept of the line obtained from our spectroscopic sample. The line with maximum purity is used in our $G$ versus $M_{20}$ plots, and for all subsequent calculations of $f_{\rm TIM}$. We show the $G-M_{20}$ plots with this line for our spectroscopic cluster and group members (left panel), and our field sample (right panel) in Figure~\ref{Fig:GM20}. Using a plot where both these samples were plotted together, we find that that among the galaxies that remain above this selection line, 60\% are tidal interactions and mergers (TIM). As we describe in more detail below, we are concerned with identifying TIM with well established observability timescales. Therefore, for the purpose of this analysis we regard the objects visually classified as TIM that reside below our line as part of our undisturbed population. Hence we define the fraction of tidal interactions and mergers with well identified observability timescales, or $f_{\rm TIM}$, for samples with visual classification as
\begin{equation}
 f_{\rm TIM} = \frac{N_{above}^{VisTIM}}{N_{total}},
\end{equation}
where $N_{above}^{VisTIM}$ is the number of objects visually classified as tidal interactions and mergers above our line, and $N_{total}$ is the total number of objects in the sample. By using a sample with visual classifications, we explicitly correct for the contamination by symmetric galaxies above our line. We applied a correction factor $C = N_{above}^{VisTIM} / N_{above}$ calculated using our visually classified spectroscopic sample to the $G-M_{20}$ TIM fraction of samples we do not have visual classifications for, to account for the visually symmetric galaxies that would be identified as TIM by the $G-M_{20}$ technique. The TIM fraction for our photometric sample, for which a visual classification has not been performed, is hence calculated as
\begin{equation}
 f_{\rm TIM}^{p} = \frac{N_{above}^{p}}{N_{total}^{p}}\times C,
\end{equation}
where the superscript $p$ is to symbolize that this $f_{\rm TIM}$ calculation has been used for our photometric sample only. As also mentioned above, we find $C = 0.60$ from the $G-M_{20}$ distribution of our entire spectroscopic sample.

In a future paper we will couple the observability timescale of the mergers with a study of the stellar populations in our galaxies to determine the relative timing of morphological transformation and star formation quenching. In this study we therefore have deliberately chosen to only optimize our selection based on obtaining a clean sample of mergers above our dividing line, as those galaxies will have the most well constrained observability timescales, unlike \textquotedblleft true" mergers below our line.  In other words, our goal is not to measure a total merger fraction, but rather to isolate a sample of mergers with a well identified observability timescale.

In Appendix C we explore the discriminatory power of both $G$ and $M_{20}$ and find that the disturbed and undisturbed populations are significantly separated in both parameters. Our line and the distribution in $G-M_{20}$ space of our spectroscopic and photometric cluster members, and our aggregate group members is presented in Figure~\ref{Fig:GM20-SpecSample}.

We conclude this section with a final remark. The best purity value we obtained from our code was a single value, corresponding to a y-intercept of -0.97 and slope of -0.87. As is evident from Figure~\ref{Fig:Purity}, there are many other outcomes close to our purity value corresponding to different y-intercept and slope values. In order to test the robustness of our results, we drew $10^{5}$ random combinations of (y-intercept, slope)$_{\rm test}$. For each combination we also drew a random purity $1 \leq \rho_{\rm test}\leq \rho_{\rm max}$, where $\rho_{\rm max}$ is the maximum purity over all y-intercepts and slopes. If $\rho_{\rm test}$ was less than or equal to the purity corresponding to (y-intercept, slope)$_{ \rm test}$, we kept the (y-intercept, slope)$_{\rm test}$ pair. Otherwise we discarded it and drew another (y-intercept,slope)$_{\rm test}$. This resulted in $\sim 10^4$ sets of (y-intercept,slope)$_{\rm test}$. We find that the same visual TIM galaxies are isolated by most of the accepted lines. We demonstrate this by plotting a random subset of the accepted lines on the $G-M_{20}$ space of our entire spec sample in Appendix C. We then calculated the TIM fraction per global environment at every accepted (y-intercept, slope) pair to assess the impact of different lines on our analysis in \S\ref{SubSec:Environment}. All of our results presented in \S\ref{SubSec:Environment} computed using the best purity line are within the $68\%$ confidence interval of the distribution in $f_{\rm TIM}$ we derive using this procedure. Furthermore, at every (y-intercept, slope)$_{\rm test}$, we performed a two sample KS Test comparing the distribution of $G-M_{20}$ TIM and undisturbed objects (where the $G-M_{20}$ TIM and undisturbed populations are picked relative to (y-intercept, slope)$_{test}$ each time) in $\Delta V/\sigma$ and $R_{proj}/R_{200}$. The KS \textit{p}-values we report in \S\ref{SubSec:PhaseSpace} for the best purity line are close in value to the peak of the distribution in each case. Therefore we decided to use the line corresponding to our best purity value for the entire analysis presented in this paper.

\section{Results}
\label{Sec:Results}

After getting the fraction of tidal interactions and mergers with well identified observability timescales ($f_{\rm TIM}$) for our spectroscopic and photometric catalogs, we looked at the dependence of $f_{\rm TIM}$ on redshift, cluster velocity dispersion, global environment, and local environment. We present our findings from each of these in the subsequent subsections. Most errors have been obtained through bootstrapping respective catalogs, except for spectroscopic errors attributed to the merger fractions of CL1054.7-1245 and CL1138.2-1133a which didn't have any visually classified TIM galaxies above our selection line. We calculated errors for these clusters using the binomial error formulas as given in \cite{Gehrels86}. For all other structures, we confirmed that the error we obtain from bootstrapping of respective samples is equal to the error we obtain from the same binomial error formula. We present a table showing $f_{\rm TIM}$ values in Table~\ref{Table:ftimresults}.

\begin{deluxetable}{ccc}
\tablecolumns{3}
\tablewidth{0pc}
\tablecaption{$f_{\rm TIM}$ Results}
\tablehead{
\colhead{Structure Name} & \colhead{$f_{\rm TIM}^{\rm phot+spec}$} & \colhead{$f_{\rm TIM}^{\rm spec}$}}

\startdata
CL1040.7-1155 & $0.24^{+0.12}_{-0.12}$ & $0.40^{+0.2}_{-0.1}$ \\
CL1054.4-1146 & $0.17^{+0.05}_{-0.05}$ & $0.20^{+0.9}_{-0.075}$ \\
CL1054.7-1245 & $0.04^{+0.035}_{-0.04}$ & $0.0^{+0.2}_{-0.0}$ \\
CL1138.2-1133 & - & $0.23^{+0.07}_{-0.08}$ \\ 
CL1138.2-1133a & - & $0.0^{+0.2}_{-0.0}$ \\ 
CL1216.8-1201 & $0.23^{+0.05}_{-0.05}$ & $0.25^{+0.08}_{-0.06}$  \\ 
CL1227.9-1138 & $0.20^{+0.10}_{-0.10}$ & $0.32^{+0.10}_{-0.14}$ \\
CL1232.5-1250 & $0.10^{+0.03}_{-0.04}$ & $0.03^{+0.03}_{-0.03}$ \\
CL1354.2-1230 & $0.37^{+0.08}_{-0.07}$ & $0.25^{+0.125}_{-0.125}$ \\
CL1354.2-1230a & - & $0.34^{+0.16}_{-0.17}$\\
\hline
Field & $0.18^{+0.02}_{-0.02}$ & $0.12^{+0.04}_{-0.05}$ \\
Groups &  -  &   $0.20^{+0.06}_{-0.05}$ \\
Cluster: $R>0.5 \times R_{200}$ & $0.25^{+0.02}_{-0.02}$ & $0.23^{+0.06}_{-0.05}$  \\
Cluster: $R<0.5 \times R_{200}$ & $0.14^{+0.02}_{-0.03}$ &  $0.16^{+0.04}_{-0.04}$ \\
Cluster: $R < 0.15\times R_{200}$ & $0.16^{+0.04}_{-0.03}$ & $0.24^{+0.06}_{-0.06}$ \\
\enddata
\tablecomments{Column 1: Structure Name. Column 2: TIM fraction in the phot+spec sample. Column 3: TIM fraction using the spectroscopic sample only. }
\label{Table:ftimresults}
\end{deluxetable}

\subsection{$f_{\rm TIM}$ versus Redshift}
\label{SubSec:MF_vs_z}

\begin{figure*}
\epsscale{1}
\plottwo{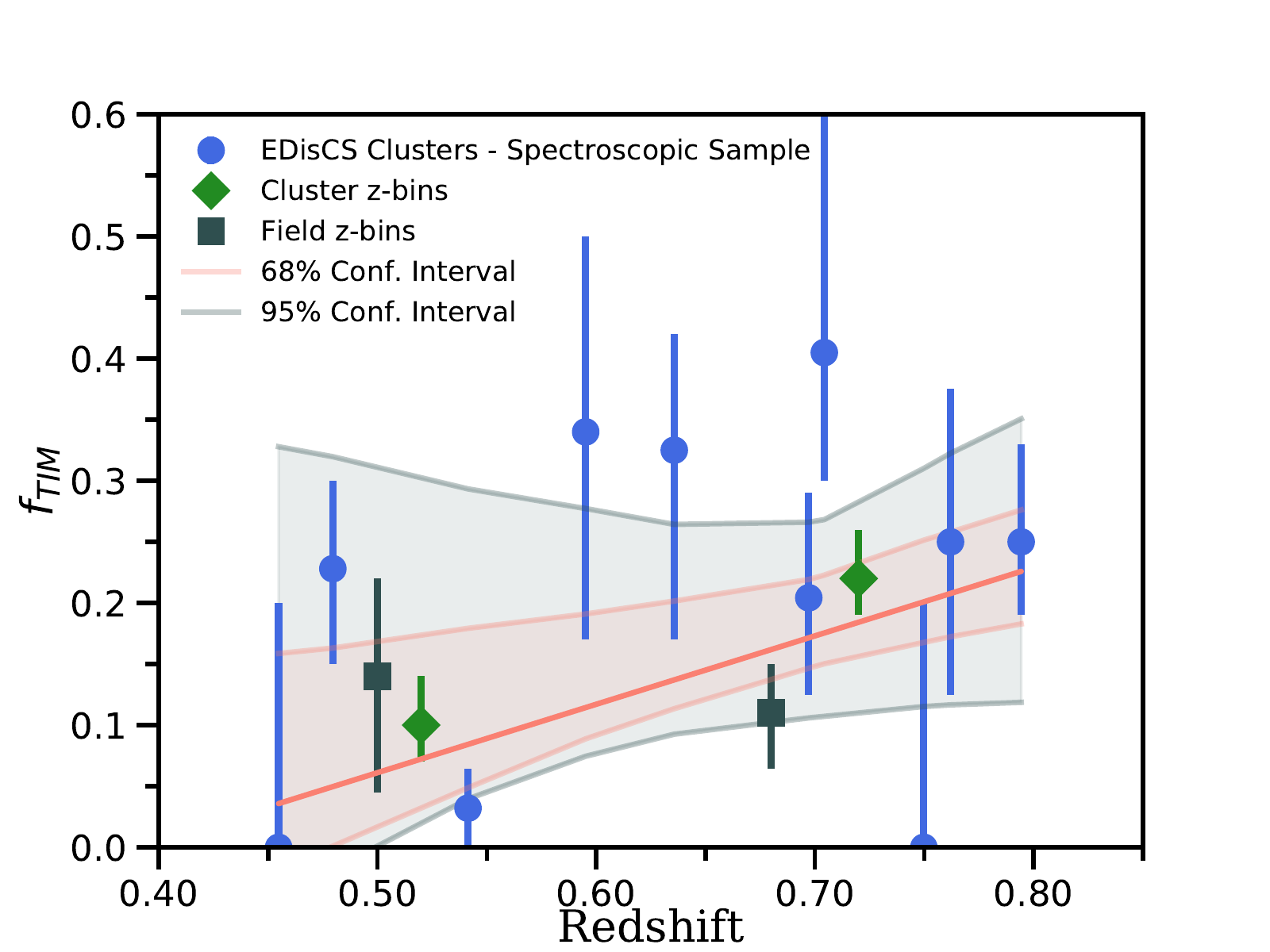}{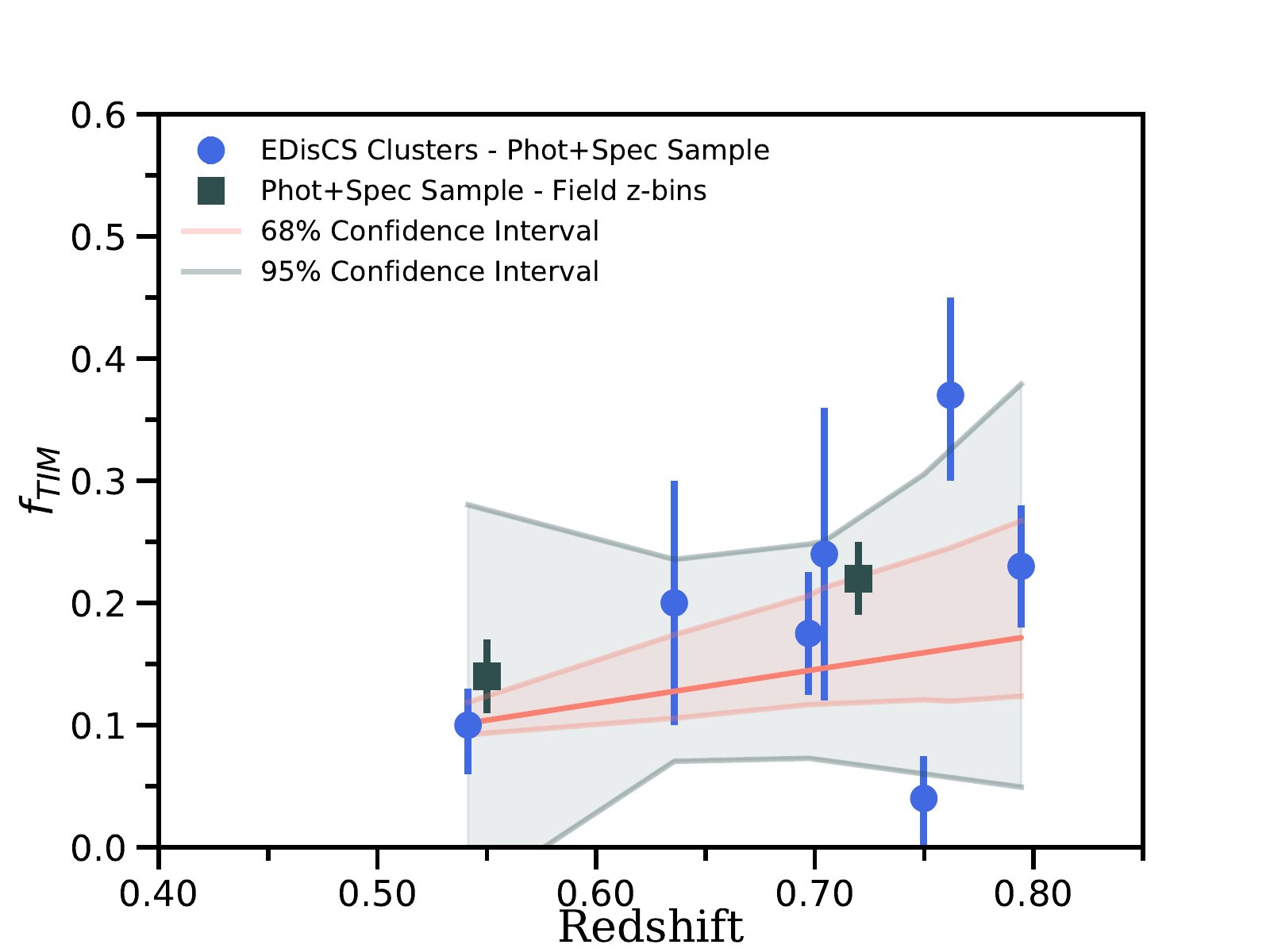}
\caption{\textit{Left panel --} Evolution of $f_{\rm TIM}$ for our spectroscopic sample. Blue circles are the clusters in our spectroscopic sample, green diamonds are galaxies from these clusters in redshift bins of $0.45<z<0.6$ and $0.6<z<0.8$. Dark gray squares are spectroscopically confirmed field galaxies in bins of $0.4<z<0.6$ and $0.6<z<0.8$. \textit{Right panel --}  Evolution of $f_{\rm TIM}$ for our phot+spec sample. Blue circles are the clusters, and dark gray squares are field galaxies in bins of $0.4<z<0.6$ and $0.6<z<0.8$. We obtain the red fitted line via a weighted linear regression algorithm for both panels. In both panels, error bars in $f_{\rm TIM}$ are the $68\%$ confidence limits obtained through a bootstrapping of the $G-M_{20}$ catalogs of respective clusters. Finally, for both panels, the pink and light gray lines above and below the fit are the $68\%$ and $95\%$ confidence limits of the fit respectively. The best fit line in both plots show an increasing $f_{\rm TIM}$ with redshift. However, we cannot rule out no evolution $f_{\rm TIM}$ at more than 68\% confidence for either sample. The Spearman rank \textit{p}-values, at $0.42$ for the clusters in the left panel, and $0.29$ for the clusters in the right panel, further point to our results being consistent with no evolution of $f_{\rm TIM}$ with $z$.}
\label{Fig:MF_vs_redshift}
\end{figure*}

Our findings for how $f_{\rm TIM}$ varies with redshift are shown in Figure~\ref{Fig:MF_vs_redshift}. The left plot shows results from our spectroscopic sample. It displays each cluster from this sample we used for our analysis, members from these clusters binned in equal redshift intervals, and field galaxies binned in two redshift bins containing roughly equal numbers of galaxies. The right plot shows results from our phot+spec sample. We obtained a weighted fit of the cluster data for both plots, which we present with the confidence intervals on the fit.

While the best fit line in both panels show an increasing $f_{\rm TIM}$ with redshift, we cannot rule out a non-evolving $f_{\rm TIM}$ at more than 68\% confidence for either sample. This is reinforced by the results of a Spearman rank test, which gives a \textit{p}-value of $0.42$ for the clusters in spectroscopic sample (blue data points in Figure~\ref{Fig:MF_vs_redshift}, left panel), and $0.29$ for the clusters in the phot+spec sample (blue data points in Figure~\ref{Fig:MF_vs_redshift}, right panel), indicating that there is a $42$ and $29\%$ chance respectively that a random sample would show as strong a correlation as ours. Thus our results are consistent with no evolution of $f_{\rm TIM}$ with redshift. We finalize this section by stating that our results rule out a line with a slope greater than 1.23 $\Delta f_{\rm TIM} /\Delta z$ for the spec, and a line with a slope 1.36 $\Delta f_{\rm TIM} /\Delta z$ for our phot+spec sample at a $99.5\%$ confidence level. Thus, we can also rule out at high confidence very strong evolution in $f_{\rm TIM}$. Furthermore, our results rule out lines with slopes less than -1.65 $\Delta f_{\rm TIM} /\Delta z$ for the spec, and less than -1.96 $\Delta f_{\rm TIM} /\Delta z$ for our phot+spec sample with $99.5\%$ confidence.

\subsection{$f_{\rm TIM}$ versus Velocity Dispersion}
\label{SubSec:MF_vs_sigma}

\begin{figure*}
\epsscale{1}
\plottwo{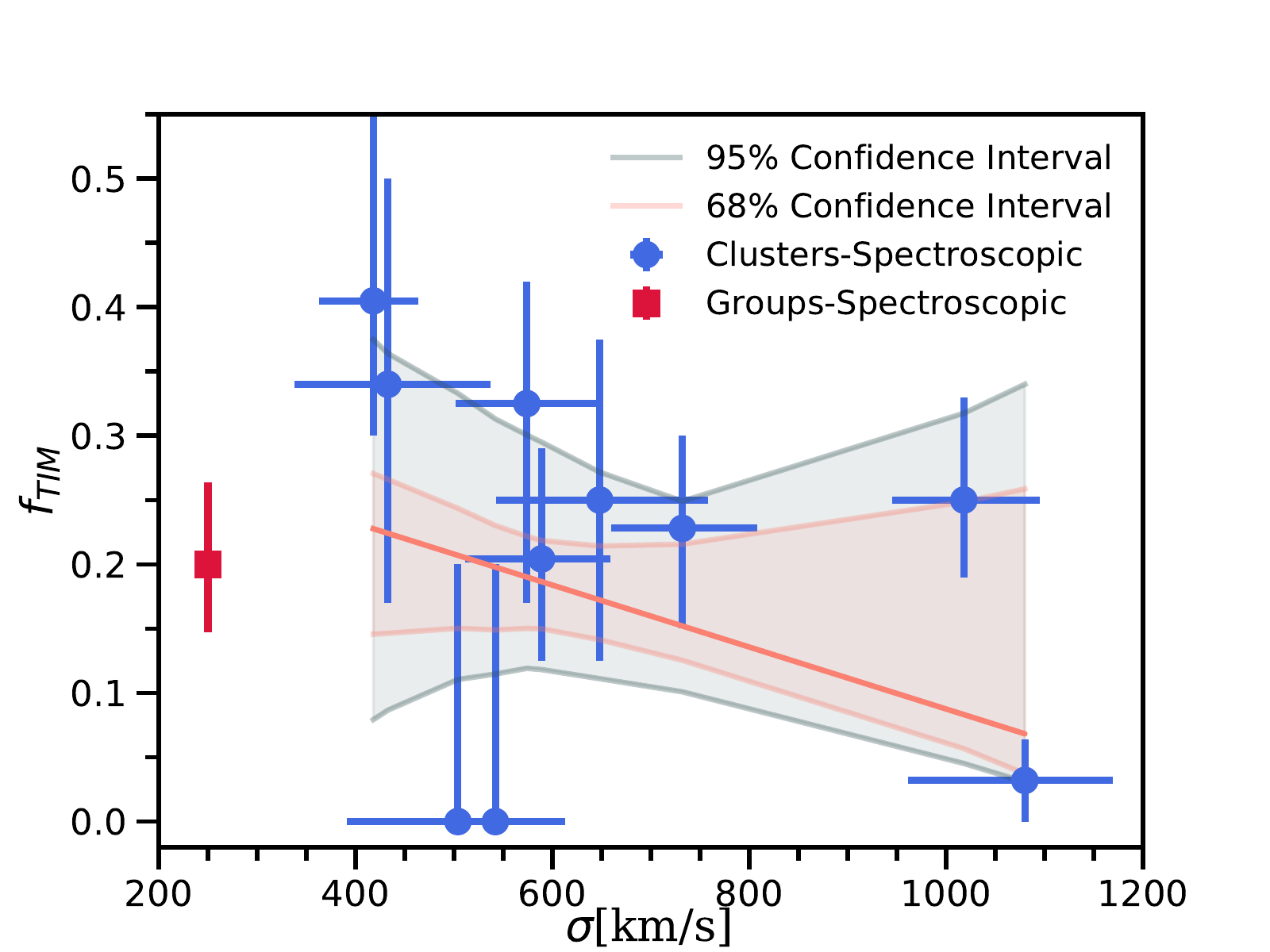}{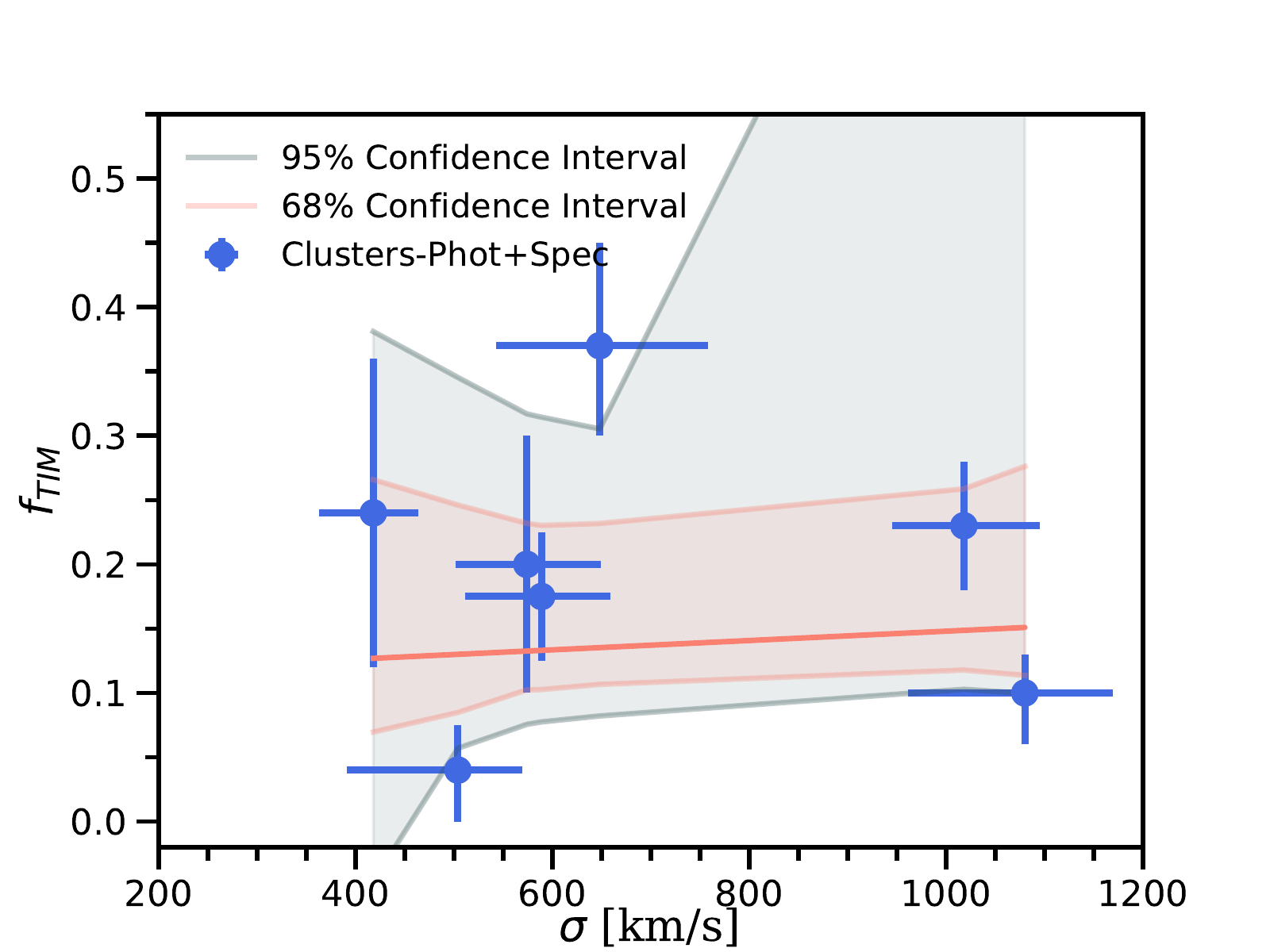}
\caption {\textit{Left panel --} $f_{\rm TIM}$ versus velocity dispersion results for our spectroscopic sample. Blue circles are the clusters in the spectroscopic sample. The red square data point is the $f_{\rm TIM}$ value of our aggregate group sample shown at the mean $\sigma$ of our groups. \textit{Right panel --} $f_{\rm TIM}$ versus velocity dispersion results for our phot+spec sample. Blue circles are the clusters in our phot+spec sample. We do not present a group result for our photometric sample, as explained in \S\ref{Sec:Sample}. For both panels the error bars are the $68\%$ confidence limits obtained through a bootstrapping of the $G-M_{20}$ catalogs of the respective clusters or groups. We obtain the red fitted line via a weighted linear regression algorithm for both panels. The pink and light gray lines above and below the fit are the $68\%$ and $95\%$ confidence limits of the fit respectively. These reveal that our data are completely consistent with no dependence on velocity dispersion. The Spearman rank \textit{p}-values of the left and right panels are $0.37$ and $0.93$, respectively, in support of this conclusion.}
\label{Fig:MF_vs_sigma}
\end{figure*}

We present our findings for how $f_{\rm TIM}$ varies with velocity dispersion in Figure~\ref{Fig:MF_vs_sigma}. The left panel of the figure shows results from our spectroscopic sample and the right panel from our phot+spec sample. In the plot for our spectroscopic sample we display the result for groups at a $\sigma$ value that is the average of the individual group $\sigma$ values. The right panel does not present a data point for groups, as discussed in \S\ref{Sec:Sample}. Similar to Figure~\ref{Fig:MF_vs_redshift}, we also present a weighted best fit to the cluster data in both panels, and the $68\%$ and $95\%$ confidence limits of the fit. In both panels we see that our results are fully consistent with no correlation of $f_{\rm TIM}$ with $\sigma$. The Spearman rank test results also point out to a probable no correlation with $\sigma$. We obtain a Spearman rank \textit{p}-value of $0.37$ for our spectroscopic sample (left panel, Figure~\ref{Fig:MF_vs_sigma}), and $0.93$ for our phot+spec sample (right panel, Figure~\ref{Fig:MF_vs_sigma}), indicating that there is a $37$ and $93\%$ chance respectively that a random uncorrelated sample would show as strong a correlation as ours. We therefore conclude that there is no significant trend of $f_{\rm TIM}$ with velocity dispersion for either sample. Finally, we find that our results rule out lines with slopes greater than a slope of $9\times10^{-4}$ $\Delta f_{\rm TIM} /\Delta \sigma$ for our spectroscopic sample, and a slope of $2\times10^{-3}$ $\Delta f_{\rm TIM} /\Delta \sigma$ for our phot+spec sample at a $99.5\%$ confidence level. Similarly, we are able to rule out at a $99.5\%$ confidence level lines with slopes less than $-6\times10^{-4}$ $\Delta f_{\rm TIM} /\Delta \sigma$ for the spec, and $-7\times10^{-4}$ $\Delta f_{\rm TIM} /\Delta \sigma$ for our phot+spec sample.

\subsection{$f_{\rm TIM}$ in Different Environments}
\label{SubSec:Environment}

\begin{figure}
 \plotone{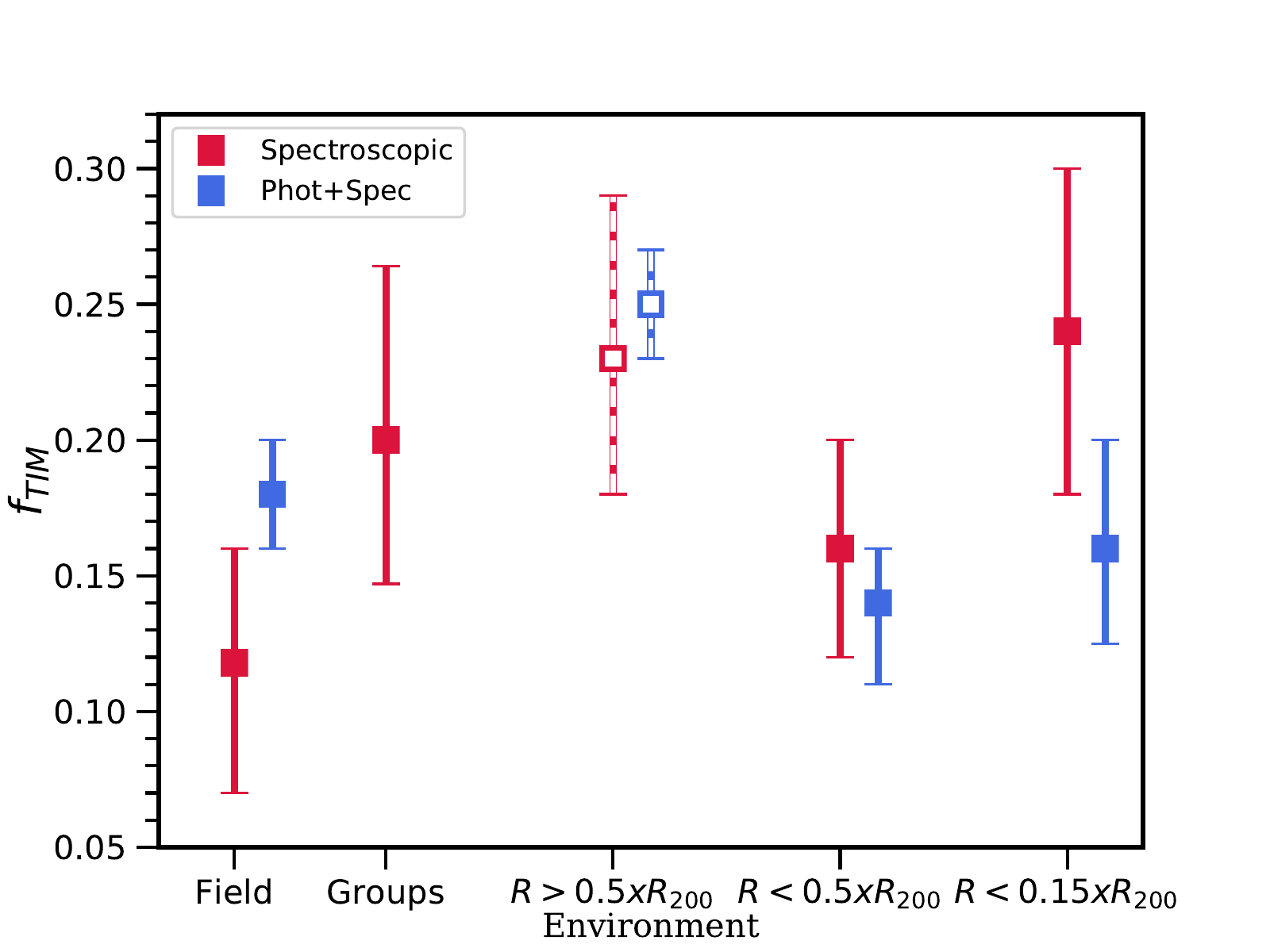}
 \caption{$f_{\rm TIM}$ of our spectroscopic and phot+spec samples at different environments. Red and blue markers represent our spectroscopic and phot+spec samples respectively. Error bars are obtained by bootstrapping catalogs per each composite data point. The group result only uses spectroscopic sample as discussed in \S\ref{Sec:Sample}. We split our cluster members into three regions according to clustercentric radius, namely $R < 0.5\times R_{200}$, $R > 0.5\times R_{200}$, and $R < 0.15\times R_{200}$. Our sample is not equally represented in $R > 0.5\times R_{200}$, therefore we present our results for that region as open squares with dashed error bars. The plot shows that $f_{\rm TIM}$ has suggestive peaks at groups, and at radii in clusters larger than $0.5\times R_{200}$.}
 \label{Fig:Environment}
\end{figure}

In Figure~\ref{Fig:Environment} we show how our $f_{\rm TIM}$ values vary across environment. This figure shows the $f_{\rm TIM}$ results using our spectroscopic and phot+spec samples for field galaxies, groups (using the spectroscopic sample only, see \S\ref{Sec:Sample} for an explanation of why only the spectroscopic sample has been used for groups), and our cluster result in three radial bins, $R < 0.5\times R_{200}$, $R > 0.5\times R_{200}$, and and $R < 0.15\times R_{200}$. We remark that while the random uncertainties are smaller for our phot+spec sample, the spectroscopic sample has lower systematic uncertainties due to the more precise determination of membership. Our results show that $f_{\rm TIM}$ has a peak at $R > 0.5\times R_{200}$ in clusters for our phot+spec sample. We find that $f_{\rm TIM}$ has peaks at groups, at $R > 0.5\times R_{200}$ and $R < 0.15\times R_{200}$ in clusters for our spectroscopic sample, though these peaks are weaker and are of low significance. Here we note that for some of our clusters, our data does not extend to the full $R_{200}$, so our clusters are not equally represented in the cluster outskirt result (see Figure~\ref{Fig:AllClust}). For example, CL1232.5-1250, one of our most massive and lowest $f_{\rm TIM}$ clusters does not have HST coverage past $ 0.5\times R_{200}$. Since the result at this radius is inevitably affected by this unequal representation, we present it with a caveat and plot our findings with different markers in Figure~\ref{Fig:Environment}. We also note that we excluded CL1354.2-1230a and CL1138.2-1133a from the cluster outskirts and core results, as discussed in \S\ref{Sec:Sample}.

\subsection{Phase Space Analysis}
\label{SubSec:PhaseSpace}

\begin{figure}
\plotone{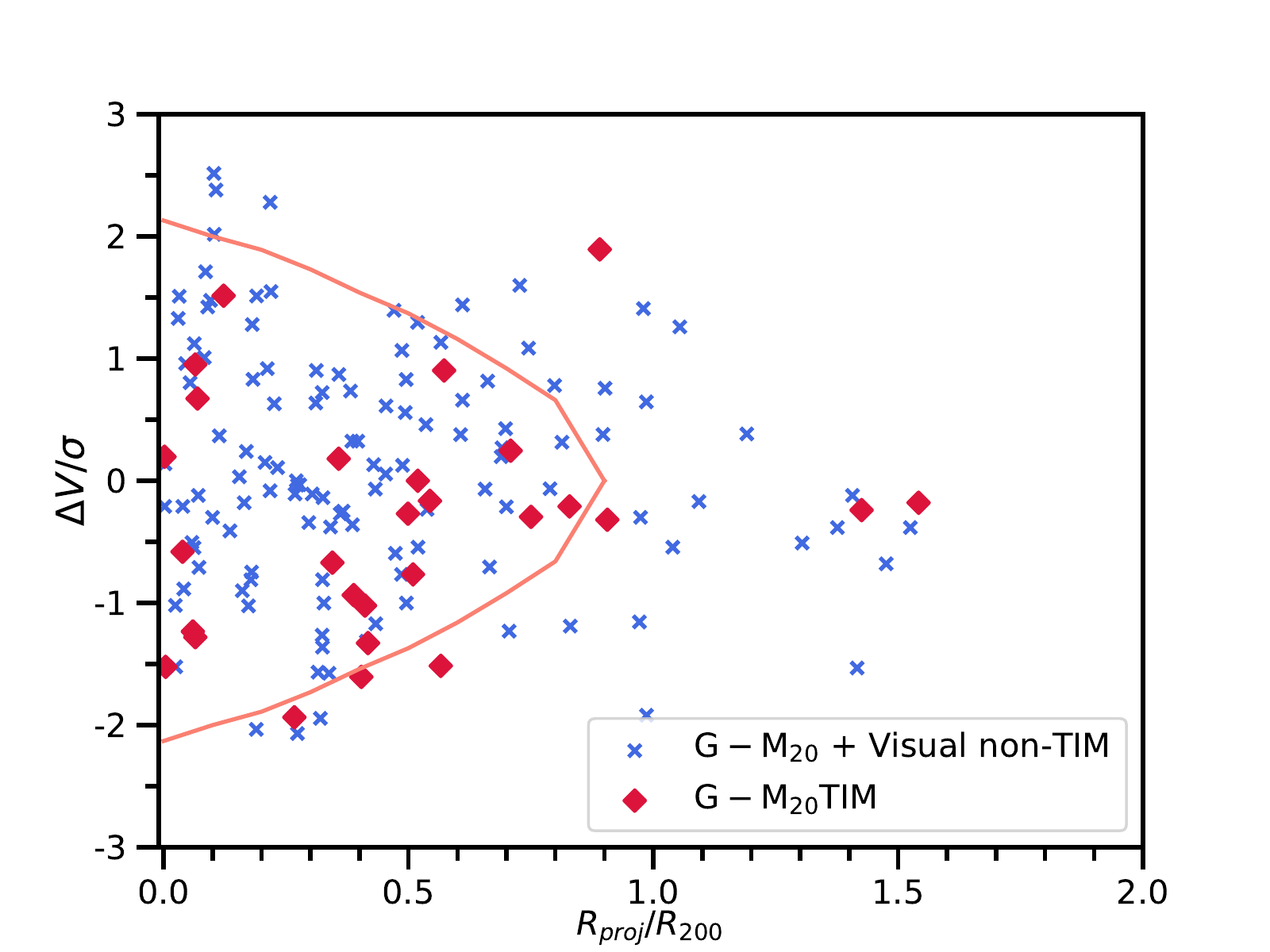}
\caption{\textit{Left panel --} Phase space analysis of our spectroscopic sample. Red circles are cluster members from our spectroscopic sample visually classified as TIM that also reside above our TIM selection line ($G-M_{20}$ TIM), blue crosses are galaxies visually classified as undisturbed, or visually classified TIM that reside below our line. The orange solid line from \cite{Mahajan11} indicates the region where the majority of virialized galaxies lie. No significant trend is apparent in the phase space. We further investigate this in Figure~\ref{Fig:CumulativeHists}.}
\label{Fig:PhaseSpace}
\end{figure}

\begin{figure*}
\plottwo{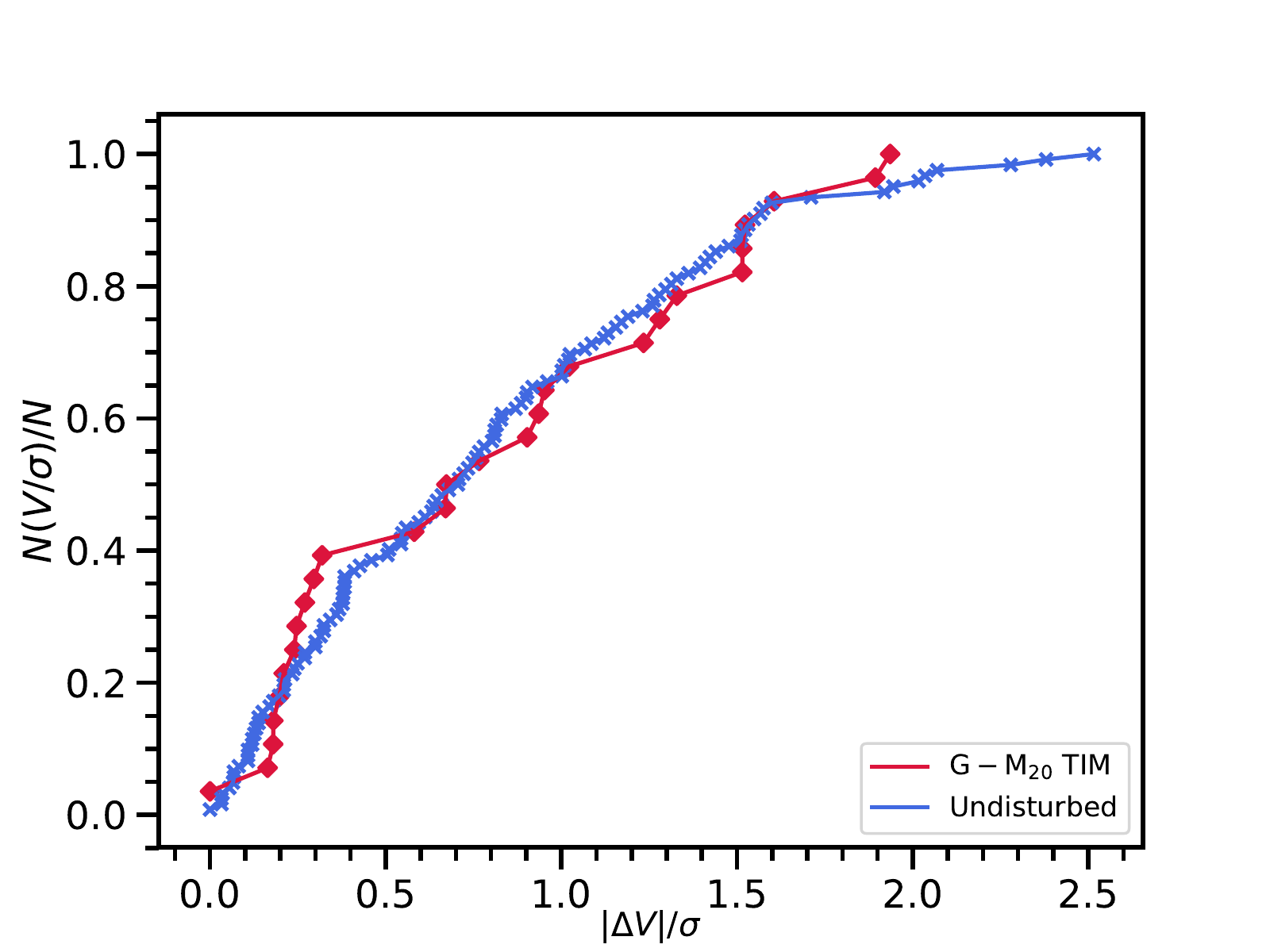}{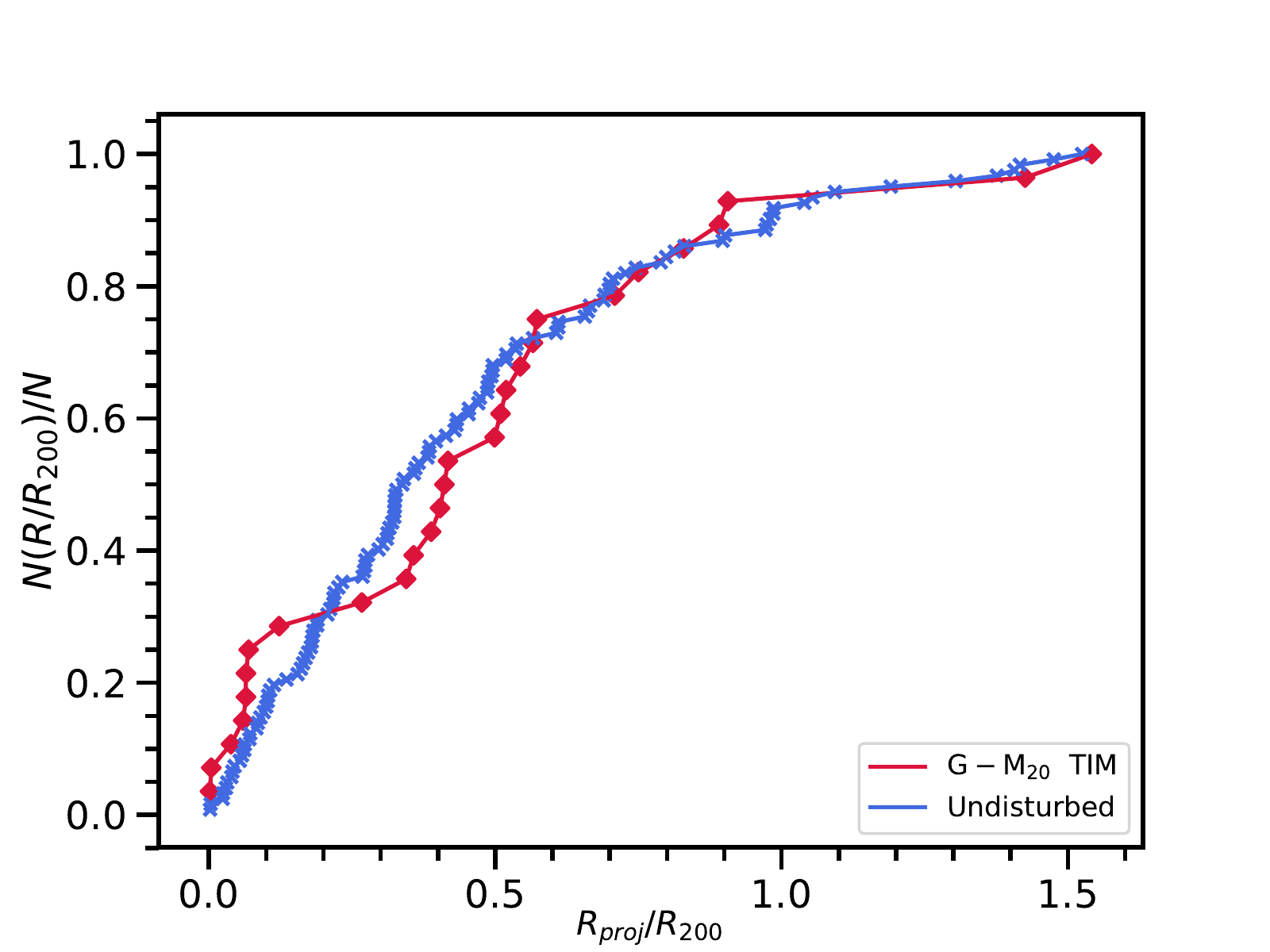}
\caption{\textit{Left panel --} The cumulative histogram of $|\Delta V|/\sigma$ for the sample we used in our phase space plot, Figure~\ref{Fig:PhaseSpace}. The colors represent the same populations as in the phase space plot, red for $G-M_{20}$ TIM, and blue for undisturbed galaxies. \textit{Right panel --} The cumulative histogram of $R_{proj}/R_{200}$ for the sample we used in our phase space plot, Figure~\ref{Fig:PhaseSpace}. Red and blue colors represent the same populations as in the left panel. KS test results show that there is a 6\% and 37\% probability that our samples are drawn from the same distribution when their $\Delta V/\sigma$ and $R_{proj}/R_{200}$ values are compared, respectively.}
\label{Fig:CumulativeHists}
\end{figure*}

We performed a phase space analysis using cluster members from our spectroscopic sample to observe whether TIM and undisturbed galaxies show any trends. We limit the our analysis to our clusters as they are the only systems with sufficient member counts for a precise determination of $\sigma$ and $R_{200}$. From our clusters we additionally excluded CL1354.2-1230a and CL1138.2-1133a for this analysis, as discussed in \S\ref{Sec:Sample}. The distribution of our sample in the phase space is shown in Figure~\ref{Fig:PhaseSpace}. The plot displays the phase space distribution of our undisturbed and $G-M_{20}$ TIM galaxies. The solid orange line is from \cite{Mahajan11}, and it signifies the region where most virialized galaxies reside. Since the majority of our galaxies are inside the virialized region, it is difficult to draw conclusions with respect to the virialized nature of the subpopulations.

We present cumulative histograms of $|\Delta V|/\sigma$ and $R_{proj}/R_{200}$ for our two classes in Figure~\ref{Fig:CumulativeHists}. We further investigate the environmental dependence of our sample by performing the Kolmogorov-Smirnov (KS test) for our undisturbed and $G-M_{20}$ TIM samples, comparing their $\Delta V/\sigma$ and $R_{proj}/R_{200}$ values. The KS test shows that there is a 6\% probability that our TIM and undisturbed galaxies have been drawn from the same parent population in their $\Delta V/\sigma$ distribution. For the $R_{proj}/R_{200}$ values the KS test finds that the probability is 37\%. These results are comparable in statistical significance to results from our analysis of $f_{\rm TIM}$ in clusters (in Figure~\ref{Fig:Environment}), where we found that the sample of cluster members with higher radii have a moderately higher $f_{\rm TIM}$ value compared to the members closer to the cluster core.

\subsection{Local Density Analysis}
\label{SubSec:LocDen}

\begin{figure*}
\plottwo{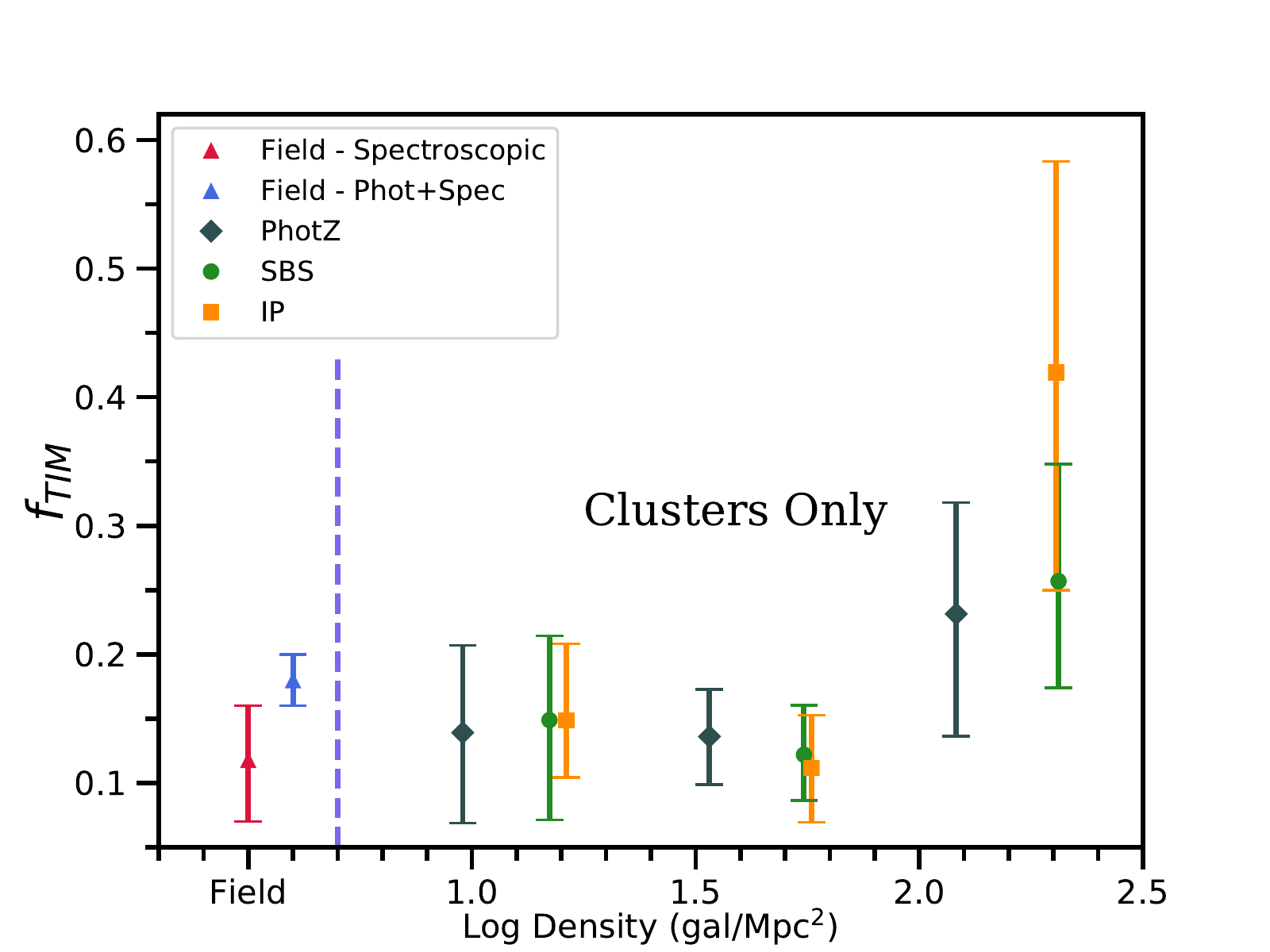}{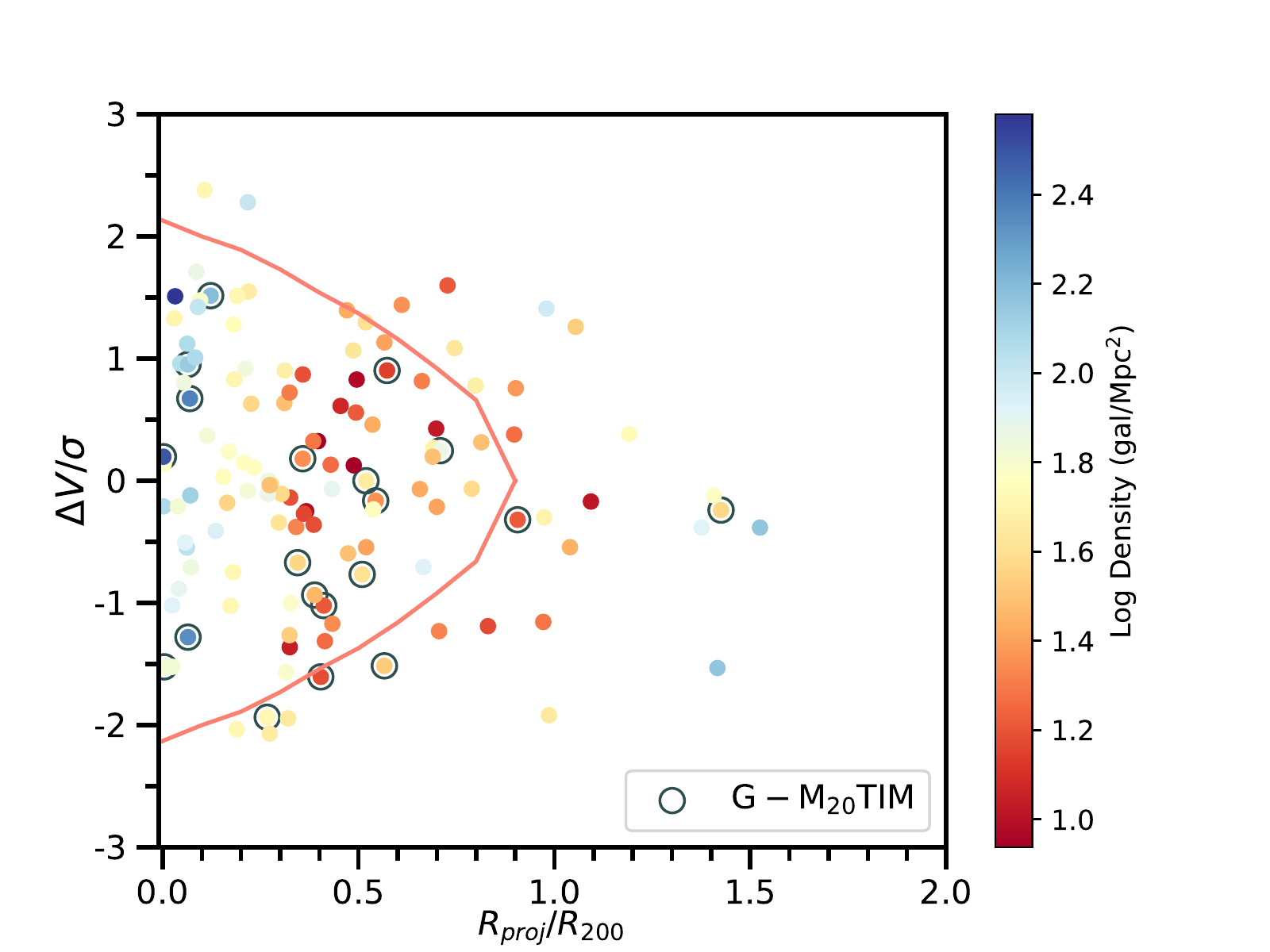}
\caption{\textit{Left panel --} The local density versus $f_{\rm TIM}$ plot for our spectroscopic cluster members, using the three different measures derived in \cite{Poggianti08}. The local densities are given as the logarithm of the number of galaxies per $\rm Mpc^{2}$.  We calculated the $f_{\rm TIM}$ within three equal size bins for each measure, and the markers are displayed at the centers of their respective bins. Green circles show the result with the SBS local density measure, orange squares for the IP measure, and gray diamonds for the PhotZ measure. Please see text in \S\ref{SubSec:LocDen} and \cite{Poggianti08} for the details. We also display the $f_{\rm TIM}$ results for our field samples, from Figure~\ref{Fig:Environment}, to the left of our local density results. We note that we did not measure local density for our field samples, and the field results here are presented at an arbitrary point on the $\rm Log\, Density$ axis (we indicate this region of the plot with the vertical purple line). Our results show a mild boost in $f_{\rm TIM}$ at the highest density bins. \textit{Right panel --} The phase space plot of the IP local density measure. The color bar represents the local density in the IP measure per spectroscopic cluster member, red colors for lower density and blue for higher. The orange line is as described in \S\ref{SubSec:PhaseSpace}. We note the diversity in local density values within the $0.5\times R_{200}$ of the cluster environment, ranging from the lowest values all the way to the highest.}
\label{Fig:LocDenPS}
\end{figure*}

In \S\ref{SubSec:Environment} we presented our findings for how the $f_{\rm TIM}$ varies according to global environment. Here we present the results of the local density analysis of our sample, using values derived by \cite{Poggianti08}. In their paper they measured the local density via a projected tenth nearest neighbor analysis for the spectroscopic cluster members of the EDisCS sample. As described below, it is not possible to measure accurate local densities for group and field galaxies in our sample and so we restrict ourselves to the local density measurements for cluster members. \cite{Poggianti08} made use of the EDisCS photometric catalogs to derive the local density per spectroscopic cluster member using three different methods. The first method uses every galaxy in the photometric catalogs with the sample corrected using a statistical background subtraction (SBS). The other two methods use different ways of determining photometric membership, one requiring the integrated probability that the galaxy is within $\pm0.1z$ of the cluster to be above a certain threshold for two different photometric redshift codes (hyperz, \cite{Bolzonella00}, and the code of \cite{Rudnick01}, and the other accepting a galaxy as a cluster member if its best photometric estimate using the hyperz code \citep{Bolzonella00} is within $\pm0.1z$ of the cluster redshift. We label these measures as IP and PhotZ here respectively. We remark that the IP method is the accepted method of determining photometric redshift-based membership in EDisCS \citep{Pello09}, and that the photometric redshifts we use and present in this work comes from this method. We refer the interested reader to \cite{Poggianti08}, and subsequently to \cite{Pello09} and \cite{Rudnick09}, for the details of each method. We note that \cite{Poggianti08} excluded some galaxies from their analysis for which reliable local densities could not be measured. For galaxies close to the edges of the field, the circular region containing the ten nearest neighbors extends off the image, hence these objects were taken out of the analysis. We therefore end up with a local density measure for 134 out of the 163 spectroscopic cluster members we use in this paper. There are excluded objects from each of our 10 fields, with no bias towards rejecting more from a particular field.

We present the $f_{\rm TIM}$ versus local density plot in Figure~\ref{Fig:LocDenPS}, left panel, for all three measures of local density. For comparison purposes we also included our field $f_{\rm TIM}$ results from \S\ref{SubSec:Environment} for both of our samples to this plot, at an arbitrary point on the local density axis. Even though we do not measure their local densities, our field samples are reasonable choices to represent low local densities, as they exclude all group and cluster members. We find a mild boost in $f_{\rm TIM}$ at the highest density bin with the most significant increase seen for the IP density measure. This tentative enhancement is in agreement with the potentially higher $f_{\rm TIM}$ result at $R < 0.15\times R_{200}$, in Figure~\ref{Fig:Environment}. We attempt to better understand the trends in $f_{\rm TIM}$ in the IP measure by looking at the distribution of local density values in the phase space, we show our results in the right panel of Figure~\ref{Fig:LocDenPS}. The plot reveals the diversity in the local density values at $R < 0.5\times R_{200}$. The distribution does not show a monotonic decrease in local density with increasing radius. This nonmonotonic behavior could explain why we observe a flat relation in $f_{\rm TIM}$ in low and intermediate local densities in our clusters, as opposed to the tentative enhancement we find in clusters at $R > 0.5\times R_{200}$, in Figure~\ref{Fig:Environment}.

We have a small number of spectroscopic members per group and our photometric redshifts are not adequate to select a high completeness sample of group members (see \S\ref{Sec:Sample}). Therefore any local density estimate in our groups would suffer from significant systematic uncertainties, and we therefore choose not to compute local densities for our groups. \cite{Cooper05} finds that a contiguous and relatively high sampling rate is essential for accurate local density measurements. Our field sample lacks this high sampling rate in the spectroscopic sample and supplementing it with photometric redshifts would induce significant systematic errors. Hence we do not calculate the local density for our field sample. We only present the $f_{\rm TIM}$ results from \S\ref{SubSec:Environment} on Figure~\ref{Fig:LocDenPS}.

\section{Discussion}
\label{Sec:Discussion}

Our results imply that $f_{\rm TIM}$ does not depend strongly on redshift. The weighted best fits in our $f_{\rm TIM}$ versus redshift plots (Figure~\ref{Fig:MF_vs_redshift}) reveal a tentative correlation for both our spectroscopic and photometric samples, but we cannot rule out the null hypothesis above 68\% confidence in either case. We also find no correlation between $f_{\rm TIM}$ and cluster velocity dispersion (Figure~\ref{Fig:MF_vs_sigma}). This result goes against the simplistic expectation that merger fraction should be higher for dense systems with lower velocity dispersion.

When we separate our galaxies into environmental classifications based on their position in the cluster, or their inclusion in clusters, groups, or the field (Figure~\ref{Fig:Environment}), we find that $f_{\rm TIM}$ shows its most significant peak at $R > 0.5\times R_{200}$ in clusters for our phot+spec sample. Our spec sample shows tentative peaks of low significance at groups, at $R > 0.5\times R_{200}$ and $R < 0.15\times R_{200}$ within the cluster environment. We relate this tentative enhancement within the innermost parts of the clusters to trends in local density later in this section. The high uncertainties in the group $f_{\rm TIM}$ due to low sample size inhibits us from being able to more definitively conclude that groups have higher $f_{\rm TIM}$. However, assuming this result holds let us consider its origin. Our groups have lower velocity dispersions than our clusters, yet have relatively high galaxy density, making them especially conducive for galaxy mergers and interactions to occur. Our results are therefore potentially in support of the preprocessing scheme, where groups serve as a preprocessing stage for the evolution of cluster galaxies \citep{Zabludoff98, Fujita04, Cortese06, Dressler13, Abramson13, Vijayaraghavan13, Man16}. Likewise, the outer regions of our clusters have lower galaxy-galaxy velocities and therefore may also host regions with an enhanced merger and interaction probability. We note that our sample is not equally represented outside of $0.5\times R_{200}$, which may have an effect on the peak we see in $f_{\rm TIM}$ at cluster outer regions. We clearly need more TIM measurements in different environments to conclusively determine how $f_{\rm TIM}$ depends on detailed environment. 

When we analyze the local environment of our spectroscopic cluster members (Figure~\ref{Fig:LocDenPS}), we see that $f_{\rm TIM}$ remains constant over the majority of the range of the cluster environment, with only a tentative enhancement in the highest density regions. The potential elevation of $f_{\rm TIM}$ at the highest local densities is driven mostly by the elevated $f_{\rm TIM}$ in the very centers of the clusters at $R < 0.15\times R_{200}$, seen in our spectroscopic cluster members. The marginal enhancement we see in $f_{\rm TIM}$ at $R > 0.5\times R_{200}$ is likely not reflected in the $f_{\rm TIM}$ versus local density plot because of the non-monotonic relation of local density and radius (Figure~\ref{Fig:LocDenPS}). As the phase space diagram of one of our local density measures displays (Figure~\ref{Fig:LocDenPS}), there is a high diversity in density values around $R = 0.5\times R_{200}$. As discussed in \S\ref{SubSec:LocDen}, we are unable compute local densities reliably for the field or group galaxies, which limits our ability to understand how $f_{\rm TIM}$ behaves at intermediate and low densities outside of the cluster.

The potentially elevated $f_{\rm TIM}$ values in the outskirts of the cluster and in groups are broadly consistent with a picture in which galaxies are morphologically transformed before their passage through the cluster core, and perhaps even before their entry into the cluster. Thus, the morphology-density relation might not be driven by processes specific to clusters. As far as our results for the core regions of clusters, it is not clear what drives the marginal elevation in $f_{\rm TIM}$ at $R < 0.15\times R_{200}$, although it is possible that the much higher densities make conditions favorable for high-speed tidal interactions without actually increasing the merger and interaction rate. Given the marginal signal we cannot make any more definitive statements at this time. Similar to our results, \cite{Adams12} find that the fraction of tidally disturbed galaxies drops within $0.5\times R_{200}$. Within the considerable uncertainties of our $f_{\rm TIM}$ measurements, this agrees with our result that $f_{\rm TIM}$ drops within $0.5\times R_{200}$, but may be inconsistent with the slight increase in  $f_{\rm TIM}$ that we see at the very highest densities and smallest ($R < 0.15\times R_{200}$) clustercentric radii. If this discrepancy turns out to be real, it could be due to the different tidal interaction and merger classification techniques, or because their clusters are older and more dynamically developed, and therefore better at removing the faint tidal features that they measure.

Our results for global environment is broadly consistent with the conclusion we draw from an analysis of the phase space of our spectroscopic cluster members. When considering the cumulative distribution of $|\Delta V|/\sigma$ (Figure~\ref{Fig:CumulativeHists}) we find only a 6\% KS probability that TIM and undisturbed galaxies are drawn from the same distribution. The cumulative distribution of radii shows a higher KS probability, of 37\% in this case, that the two samples are drawn from the same population. Nonetheless, these two phase space results are consistent with the modest differences in $f_{\rm TIM}$ seen in the environment plot. In our analysis we have assumed that the merger observability timescale is the same in all environments and at all redshifts in our study. For example, we have not accounted for the potentially shorter lifetimes of some TIM signatures, e.g. tidal tails, via interaction with the cluster tidal environment. Accounting for this particular effect would serve to enhance the $f_{\rm TIM}$ in cluster cores compared to our measured value. We will explore the implications for this phase space distribution in a future paper that constrains the visibility timescale of our $G-M_{20}$ merger classification and compare it to the infall histories of our clusters as derived by simulations.

\section{Summary and Conclusion}
\label{Sec:Summary}

In this paper we presented our analysis of tidal interactions and mergers (TIM) in the EDisCS cluster, group, and field galaxies. For our analysis we make use of a visual identification of morphological signatures indicative of tidal interactions and mergers, performed on every galaxy in our sample that has a spectroscopic redshift. We then calibrated a line selecting TIM with high purity in the $G-M_{20}$ space using this visual classification. We showed that both $G$ and $M_{20}$ are effective at identifying visually disturbed galaxies. For our galaxies with photometric redshifts, for which a visual classification was not performed, we used a correction factor derived using the visual classification of our spectroscopic sample. We then derived $f_{\rm TIM}$, the fraction of TIM objects with well identified observability timescales utilizing the selection line we calibrated, and analyzed its dependence on redshift, velocity dispersion, and both global and local environment. We also analyzed the projected radius-velocity dispersion phase space distribution of our spectroscopic cluster sample. Our conclusions are as follows.

\begin{enumerate}
 \item We find tentative evidence that $f_{\rm TIM}$ increases with increasing redshift. However, we cannot rule out at more than 68\% confidence that there is no evolution in redshift for either of our samples. Our results do rule out very strong evolution of $f_{\rm TIM}$ with high confidence. Our results rule out any line with slopes outside of $[-1.65, 1.23]$ for the spectroscopic sample, and outside of $[-1.96, 1.36]$ for the phot+spec sample with $99.5\%$ confidence.
 
 \item $f_{\rm TIM}$ shows no trend with velocity dispersion for either sample.
 
 \item $f_{\rm TIM}$ has a potentially higher value in our groups and our cluster outskirts, compared to the field and cluster cores. We tentatively conclude that $f_{\rm TIM}$ is enhanced in these environments.
 
 \item Our results are also statistically consistent with the cluster core playing no strong role in enhancing $f_{\rm TIM}$.  However, given the limited precision of our $f_{\rm TIM}$ values, we also cannot strongly rule out a more significant trend with clustercentric radius.
 
 \item We perform a phase space analysis of our cluster members, an environment where we can measure $R_{200}$ and $\sigma$ reliably, and find a marginally significant difference in the velocity distributions of the TIM and undisturbed galaxies. This supports our tentative identification of the outskirts of clusters as potentially being the site of an enhanced fraction of tidal interactions and mergers. However, it is also worth noting that the radial distribution of TIM and undisturbed galaxies does not differ significantly. Clearly we need more clusters with $f_{\rm TIM}$ estimates to make stronger constraints.
 
 \item Except for an elevated $f_{\rm TIM}$ value of low significance at the highest density bin for one of our local density measures, our results show no trend between $f_{\rm TIM}$ and local density within the cluster environment.
\end{enumerate}

While our limited number of galaxies prevents us from drawing more robust conclusions, this analysis lays the groundwork for future studies that will make stronger constraints. For example, this analysis can be readily applied to any data set with excellent image quality and precision redshifts. Space-based missions like Euclid and WFIRST will be the prime candidates thanks to their high resolution and grism-based redshifts. LSST will also have very precise photometric redshifts and good image quality and this technique should be possible for lower redshift samples where the ground-based seeing results in sufficient physical resolution. Finally deconvolution methods such as in \cite{Cantale16a} can be applied to ground-based imaging surveys making it possible to carry out analysis on these surveys to much larger distances.

G.R. acknowledges funding support from HST program HST-GO-12590.011-A, HST-AR-12152.01-A, HST-AR- 14310.001; NSF AST grants 1211358 and 1517815; and the NSF under award no. EPS-0903806, with matching support from the state of Kansas through Kansas Technology Enterprise Corporation. G.R. would also like to acknowledge the support of an Alexander von Humboldt Foundation fellowship for experienced researchers and the excellent hospitality of the Max-Planck-Institute for Astronomy, the University of Hamburg Observatory, the Max-Planck-Institute for Extraterrestrial Physics, the International Space Sciences Institute, and the European Southern Observatory, where some of this research was conducted.

\section{Appendix A}
\label{Sec:GM20-FalsePositives}

\begin{figure*}
\centering
\includegraphics[width=.33\textwidth]{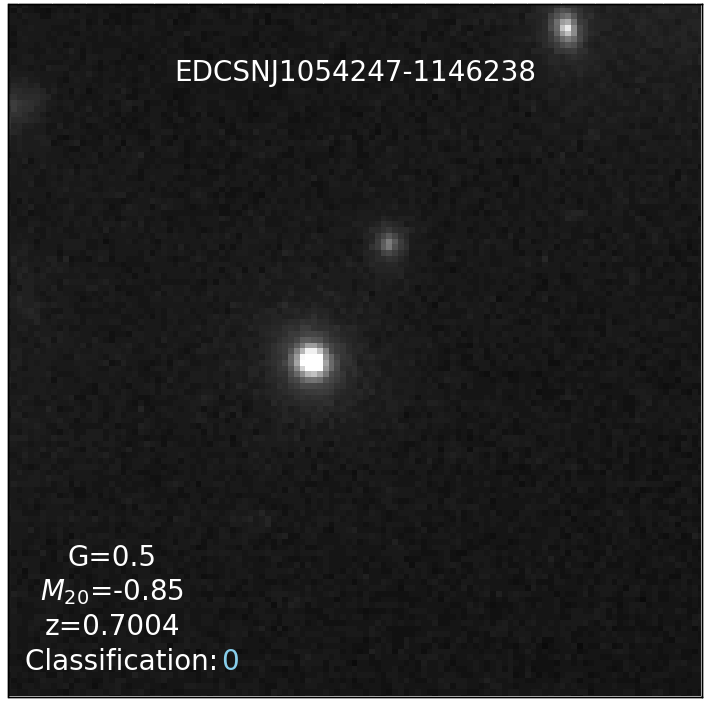}\hfill
\includegraphics[width=.33\textwidth]{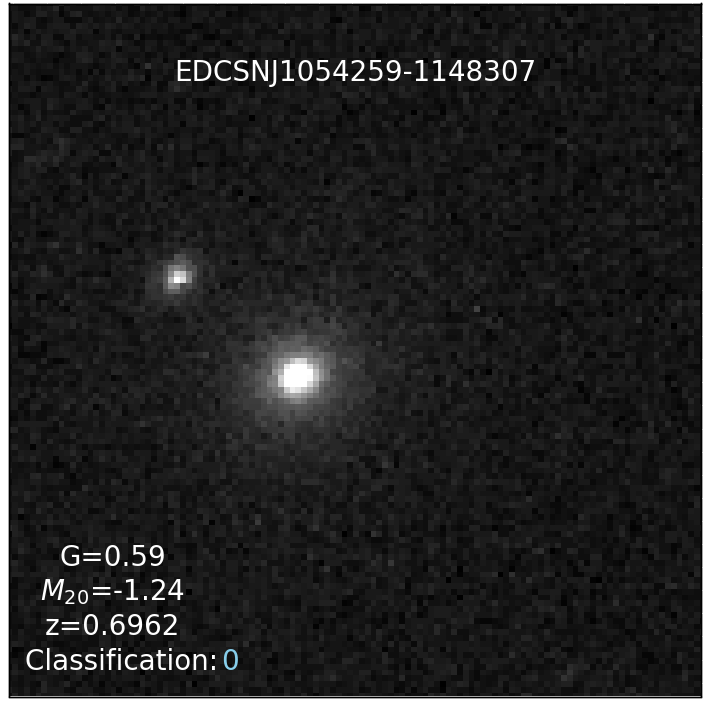}\hfill
\includegraphics[width=.33\textwidth]{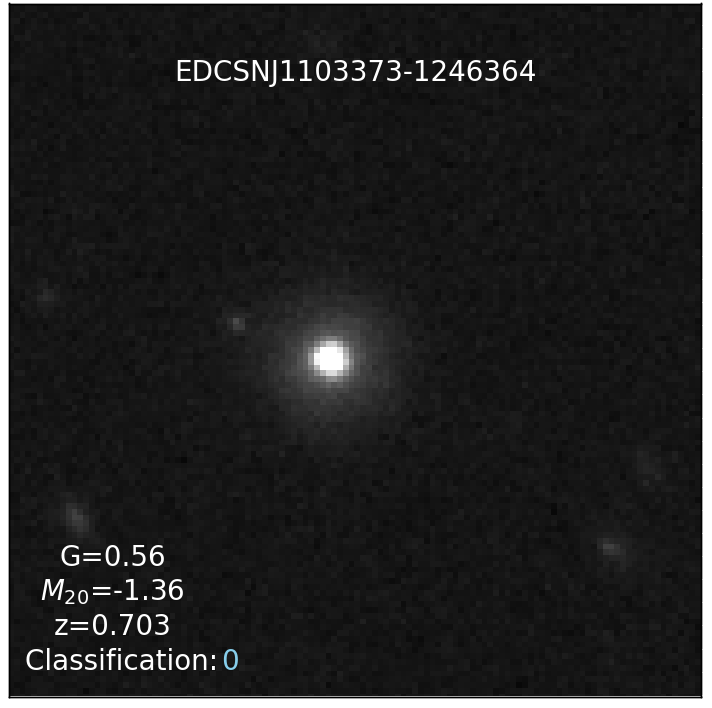}
\caption{Galaxies visually classified as undisturbed but lie above our merger selection line, or false-positives. Visual classes and colors the same as in Figure~\ref{Fig:Classes}. All images show galaxies with a neighboring object. These objects cause a variance in the flux distribution and therefore increase the $M_{20}$ value. This in turn pushes the object above our line.}
\label{Fig:FalsePositives}
\end{figure*}

\begin{figure*}
\centering
\includegraphics[width=.33\textwidth]{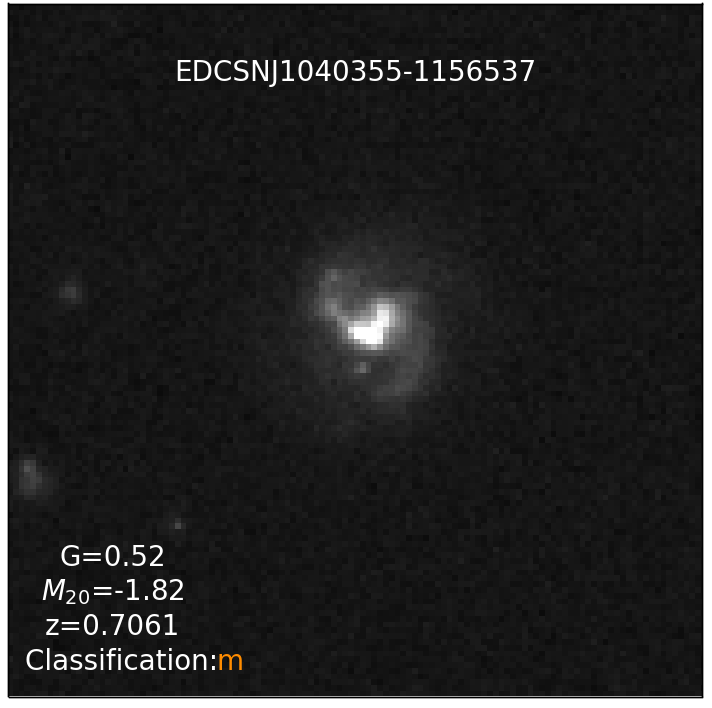}\hfill
\includegraphics[width=.33\textwidth]{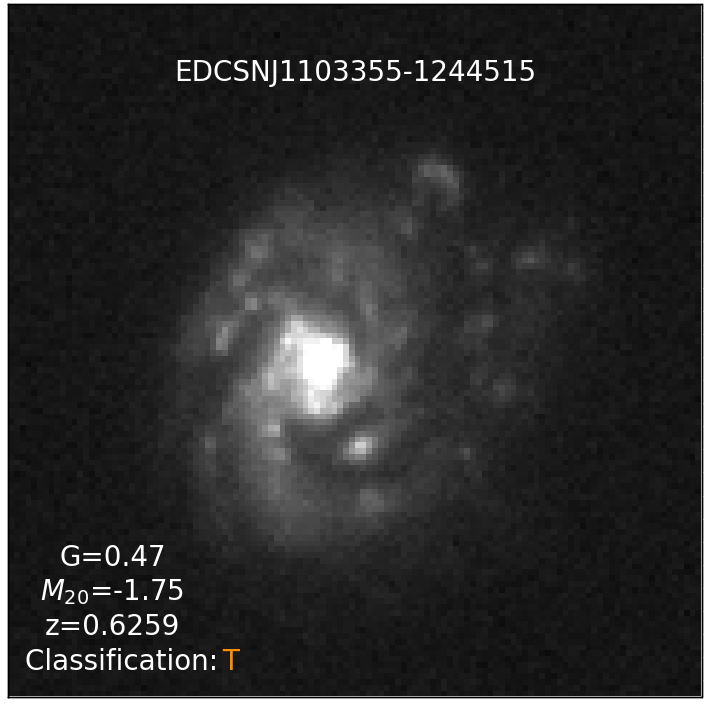}\hfill
\includegraphics[width=.33\textwidth]{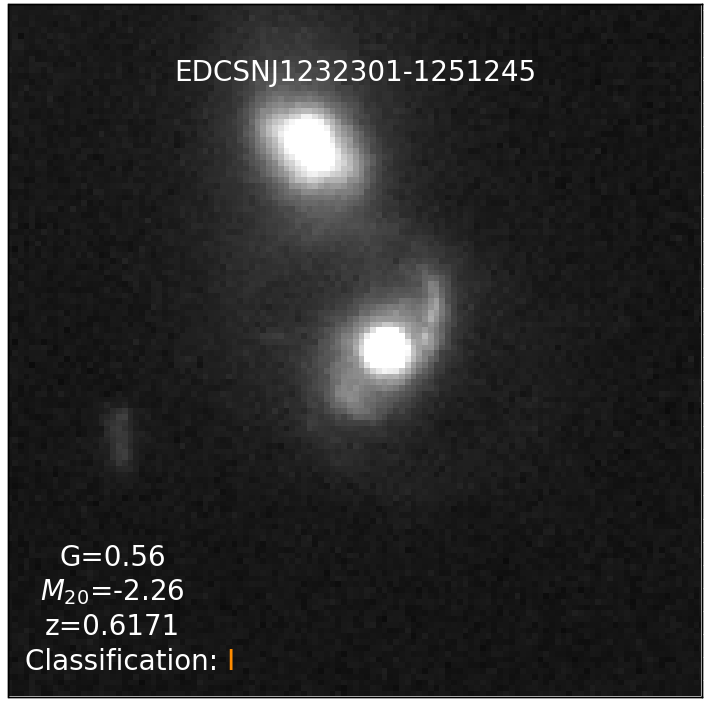}
\caption{Galaxies visually classified as TIM but lie below our merger selection line, or false-negatives. Visual classes and colors the same as in Figure~\ref{Fig:Classes}. As discussed in the text, these galaxies might be at a stage when they avoid classification as mergers by the $G-M_{20}$ method.}
\label{Fig:FalseNegatives}
\end{figure*}

Every automated method of merger detection suffers from incorrect classifications. We give some examples of such detections from our sample in this appendix. We present in Figure~\ref{Fig:FalsePositives} some of the galaxies that \cite{Kelkar17} visually classified as undisturbed, but are picked as TIM by the $G-M_{20}$ method. These are the undisturbed galaxies that remain above our selection line, or the false-positives of the $G-M_{20}$ detection. We also present some of the galaxies that \cite{Kelkar17} visually classified as having merger signatures, but remain below our selection line, in Figure~\ref{Fig:FalseNegatives}. So these form the false-negatives of the $G-M_{20}$ detection. Galaxies undergoing mergers will move on the $G-M_{20}$ space as their morphologies get altered by the merger event. They will be detected as mergers by the $G-M_{20}$ method only during a certain period of the merging process \citep{Lotz10}. It should be noted that stages too early and too late in the merger process are prone to avoid detection by automated methods, and are also challenging to identify by visual methods. The first panel in Figure~\ref{Fig:FalseNegatives} might be an example to a late stage event, that avoided detection the $G-M_{20}$ and hence resided below our line.

\section{Appendix B}
\label{appendix:Galaxies}

Here we present our spectroscopic and photometric sample in more detail. Table 3 contains every spectroscopic cluster and group member that passes all our quality cuts. Groups according to our definition have $\sigma < 400 \, \mathrm{km\, s^{-1}}$. We remind that we applied a stellar mass completeness cut of $log (M_{*}/M_{\odot}) > 10.4$ to obtain this sample. For other details, we refer the reader to \S\ref{Sec:Sample}. The table shows the galaxy ID, name of the cluster/group the galaxy is a member of, galaxy spectroscopic redshift, galaxy stellar mass, $G$ and $M_{20}$ values, and its visual classification. Table 4 contains every photometric member that passes all our quality cuts. Stellar mass completeness cut is the same as the spectroscopic sample. The table shows the galaxy ID, name of the cluster/group the galaxy is a member of, galaxy stellar mass, and its $G$ and $M_{20}$ values.
\clearpage
\startlongtable
\begin{deluxetable*}{cccccccc}
\tablecolumns{7}
\tablewidth{0pc}
\tablecaption{Spectroscopic Cluster \& Group Members} \label{Tab:SpecSample}
\tablehead{
\colhead{Galaxy ID} & \colhead{Cluster Name} & \colhead{Redshift}  & \colhead{$\log_{10}(M_{*}/M_{\odot})$}  & \colhead{G}  & \colhead{$M_{20}$}  & \colhead{Vis. Class}}
\startdata
\multicolumn{7}{c}{\bf{Cluster Members}}\\
  EDCSNJ1040415-1156559 & cl1040.7-1155 & 0.7007 & 10.49 & 0.52 & -1.51 & I\\
  EDCSNJ1040410-1155590 & cl1040.7-1155 & 0.7079 & 10.87 & 0.37 & -1.11 & m\\
  EDCSNJ1040409-1156282 & cl1040.7-1155 & 0.6997 & 10.72 & 0.46 & -1.51 & I\\
  EDCSNJ1040407-1156015 & cl1040.7-1155 & 0.703 & 11.15 & 0.58 & -1.94 & 0\\
  EDCSNJ1040402-1155587 & cl1040.7-1155 & 0.7031 & 10.92 & 0.61 & -2.16 & 0\\
  EDCSNJ1040396-1155183 & cl1040.7-1155 & 0.7046 & 10.93 & 0.57 & -2.17 & 0\\
  EDCSNJ1040369-1157141 & cl1040.7-1155 & 0.7052 & 11.04 & 0.61 & -2.24 & 0\\
  EDCSNJ1040356-1156026 & cl1040.7-1155 & 0.7081 & 11.25 & 0.53 & -1.98 & 0\\
  EDCSNJ1040355-1156537 & cl1040.7-1155 & 0.7061 & 10.97 & 0.52 & -1.82 & m\\
  EDCSNJ1040346-1155511 & cl1040.7-1155 & 0.7088 & 10.42 & 0.52 & -1.47 & i\\
  EDCSNJ1054323-1147213 & cl1054.4-1146 & 0.7019 & 10.99 & 0.56 & -1.95 & i\\
  EDCSNJ1054316-1147400 & cl1054.4-1146 & 0.6908 & 10.97 & 0.54 & -2.02 & 0\\
  EDCSNJ1054311-1149250 & cl1054.4-1146 & 0.6966 & 10.67 & 0.48 & -1.47 & i\\
  EDCSNJ1054309-1147095 & cl1054.4-1146 & 0.6998 & 11.01 & 0.57 & -2.14 & 0\\
  EDCSNJ1054303-1149132 & cl1054.4-1146 & 0.6964 & 11.63 & 0.62 & -1.51 & i\\
  EDCSNJ1054296-1147123 & cl1054.4-1146 & 0.6981 & 11.36 & 0.63 & -2.32 & 0\\
  EDCSNJ1054296-1145499 & cl1054.4-1146 & 0.6994 & 10.5 & 0.57 & -1.99 & 0\\
  EDCSNJ1054292-1149179 & cl1054.4-1146 & 0.6968 & 10.47 & 0.6 & -1.91 & 0\\
  EDCSNJ1054264-1147207 & cl1054.4-1146 & 0.6963 & 10.71 & 0.39 & -1.25 & i\\
  EDCSNJ1054263-1148407 & cl1054.4-1146 & 0.7014 & 10.92 & 0.54 & -1.81 & 0\\
  EDCSNJ1054259-1148307 & cl1054.4-1146 & 0.6962 & 10.91 & 0.59 & -1.24 & 0\\
  EDCSNJ1054255-1146331 & cl1054.4-1146 & 0.6942 & 10.58 & 0.56 & -2.15 & 0\\
  EDCSNJ1054255-1146441 & cl1054.4-1146 & 0.7048 & 11.04 & 0.58 & -2.33 & i\\
  EDCSNJ1054254-1145547 & cl1054.4-1146 & 0.6977 & 11.55 & 0.54 & -2.1 & 0\\
  EDCSNJ1054251-1145360 & cl1054.4-1146 & 0.6945 & 10.61 & 0.53 & -1.85 & 0\\
  EDCSNJ1054250-1146238 & cl1054.4-1146 & 0.6968 & 11.42 & 0.62 & -2.29 & 0\\
  EDCSNJ1054247-1146238 & cl1054.4-1146 & 0.7004 & 10.72 & 0.5 & -0.85 & 0\\
  EDCSNJ1054244-1146194 & cl1054.4-1146 & 0.6965 & 11.5 & 0.61 & -2.44 & 0\\
  EDCSNJ1054242-1146564 & cl1054.4-1146 & 0.6903 & 10.68 & 0.59 & -2.36 & 0\\
  EDCSNJ1054237-1146107 & cl1054.4-1146 & 0.6962 & 10.58 & 0.54 & -1.83 & 0\\
  EDCSNJ1054233-1146024 & cl1054.4-1146 & 0.698 & 10.6 & 0.53 & -1.76 & 0\\
  EDCSNJ1054209-1145141 & cl1054.4-1146 & 0.7020 & 11.28 & 0.6 & -1.95 & M\\
  EDCSNJ1054198-1146337 & cl1054.4-1146 & 0.6972 & 10.55 & 0.54 & -1.42 & M\\
  EDCSNJ1054182-1147240 & cl1054.4-1146 & 0.6965 & 10.98 & 0.51 & -1.49 & m\\
  EDCSNJ1054478-1244244 & cl1054.7-1245 & 0.7517 & 10.53 & 0.53 & -1.91 & 0\\
  EDCSNJ1054471-1246412 & cl1054.7-1245 & 0.7522 & 10.7 & 0.6 & -2.0 & 0\\
  EDCSNJ1054445-1246173 & cl1054.7-1245 & 0.7498 & 10.77 & 0.6 & -1.97 & 0\\
  EDCSNJ1054440-1246390 & cl1054.7-1245 & 0.7496 & 10.72 & 0.55 & -1.94 & 0\\
  EDCSNJ1054439-1245556 & cl1054.7-1245 & 0.7531 & 10.84 & 0.6 & -2.22 & 0\\
  EDCSNJ1054438-1245409 & cl1054.7-1245 & 0.7568 & 11.12 & 0.58 & -2.03 & 0\\
  EDCSNJ1054437-1246028 & cl1054.7-1245 & 0.7572 & 10.45 & 0.57 & -2.07 & 0\\
  EDCSNJ1054436-1244202 & cl1054.7-1245 & 0.7463 & 10.91 & 0.58 & -2.08 & 0\\
  EDCSNJ1054435-1246152 & cl1054.7-1245 & 0.7525 & 11.09 & 0.57 & -1.91 & 0\\
  EDCSNJ1054433-1245534 & cl1054.7-1245 & 0.7468 & 10.65 & 0.57 & -1.91 & 0\\
  EDCSNJ1054409-1246529 & cl1054.7-1245 & 0.7496 & 10.91 & 0.54 & -1.83 & 0\\
  EDCSNJ1054407-1247385 & cl1054.7-1245 & 0.7482 & 10.88 & 0.6 & -2.19 & 0\\
  EDCSNJ1054404-1248083 & cl1054.7-1245 & 0.7483 & 10.59 & 0.45 & -1.52 & 0\\
  EDCSNJ1054398-1246055 & cl1054.7-1245 & 0.7482 & 11.23 & 0.56 & -2.19 & I\\
  EDCSNJ1054396-1248241 & cl1054.7-1245 & 0.7478 & 10.94 & 0.63 & -2.01 & 0\\
  EDCSNJ1054356-1245264 & cl1054.7-1245 & 0.7493 & 11.19 & 0.61 & -2.22 & 0\\
  EDCSNJ1138130-1132345 & cl1138.2-1133 & 0.4791 & 10.72 & 0.55 & -1.87 & 0\\
  EDCSNJ1138116-1134448 & cl1138.2-1133 & 0.4571 & 10.77 & 0.58 & -2.06 & 0\\
  EDCSNJ1138113-1132017 & cl1138.2-1133 & 0.4748 & 10.73 & 0.58 & -0.83 & I\\
  EDCSNJ1138110-1133411 & cl1138.2-1133 & 0.4825 & 10.59 & 0.53 & -1.97 & 0\\
  EDCSNJ1138109-1134170 & cl1138.2-1133 & 0.4759 & 10.53 & 0.57 & -1.81 & 0\\
  EDCSNJ1138107-1133431 & cl1138.2-1133 & 0.4764 & 10.68 & 0.59 & -2.01 & 0\\
  EDCSNJ1138106-1133312 & cl1138.2-1133 & 0.4775 & 10.67 & 0.58 & -1.0 & I\\
  EDCSNJ1138104-1133319 & cl1138.2-1133 & 0.4844 & 10.53 & 0.57 & -2.29 & 0\\
  EDCSNJ1138102-1133379 & cl1138.2-1133 & 0.4801 & 11.14 & 0.6 & -2.1 & 0\\
  EDCSNJ1138086-1136549 & cl1138.2-1133 & 0.4519 & 10.83 & 0.55 & -1.95 & m\\
  EDCSNJ1138078-1133592 & cl1138.2-1133 & 0.4769 & 10.52 & 0.44 & -1.08 & 0\\
  EDCSNJ1138069-1134314 & cl1138.2-1133 & 0.4819 & 10.81 & 0.6 & -2.11 & 0\\
  EDCSNJ1138069-1132044 & cl1138.2-1133 & 0.4798 & 10.48 & 0.65 & -2.3 & 0\\
  EDCSNJ1138068-1132285 & cl1138.2-1133 & 0.4787 & 10.59 & 0.5 & -1.77 & 0\\
  EDCSNJ1138065-1136018 & cl1138.2-1133 & 0.4561 & 10.44 & 0.53 & -1.85 & 0\\
  EDCSNJ1138056-1136287 & cl1138.2-1133 & 0.4561 & 10.58 & 0.53 & -1.84 & 0\\
  EDCSNJ1138031-1134278 & cl1138.2-1133 & 0.4549 & 10.69 & 0.59 & -1.87 & 0\\
  EDCSNJ1138024-1136024 & cl1138.2-1133 & 0.4585 & 10.59 & 0.53 & -2.05 & 0\\
  EDCSNJ1138022-1135459 & cl1138.2-1133 & 0.4541 & 10.77 & 0.58 & -2.2 & 0\\
  EDCSNJ1216522-1200595 & cl1216.8-1201 & 0.7882 & 10.71 & 0.6 & -1.91 & 0\\
  EDCSNJ1216504-1200480 & cl1216.8-1201 & 0.7886 & 11.07 & 0.43 & -1.1 & M\\
  EDCSNJ1216498-1201358 & cl1216.8-1201 & 0.7882 & 10.72 & 0.56 & -2.08 & 0\\
  EDCSNJ1216490-1201531 & cl1216.8-1201 & 0.7998 & 10.66 & 0.41 & -1.32 & 0\\
  EDCSNJ1216490-1200091 & cl1216.8-1201 & 0.7863 & 10.88 & 0.54 & -2.05 & 0\\
  EDCSNJ1216480-1200220 & cl1216.8-1201 & 0.7859 & 10.86 & 0.54 & -2.04 & 0\\
  EDCSNJ1216470-1159267 & cl1216.8-1201 & 0.7971 & 10.82 & 0.6 & -2.02 & 0\\
  EDCSNJ1216468-1202226 & cl1216.8-1201 & 0.7987 & 11.04 & 0.55 & -2.23 & 0\\
  EDCSNJ1216464-1203257 & cl1216.8-1201 & 0.7966 & 10.56 & 0.58 & -2.15 & 0\\
  EDCSNJ1216462-1202253 & cl1216.8-1201 & 0.7866 & 10.74 & 0.53 & -2.0 & 0\\
  EDCSNJ1216462-1200073 & cl1216.8-1201 & 0.7847 & 10.66 & 0.59 & -2.0 & 0\\
  EDCSNJ1216456-1158383 & cl1216.8-1201 & 0.7925 & 10.98 & 0.45 & -1.26 & I\\
  EDCSNJ1216454-1200017 & cl1216.8-1201 & 0.7996 & 11.04 & 0.61 & -2.15 & 0\\
  EDCSNJ1216452-1203134 & cl1216.8-1201 & 0.7933 & 10.41 & 0.52 & -1.13 & M\\
  EDCSNJ1216453-1201176 & cl1216.8-1201 & 0.7955 & 11.76 & 0.48 & -0.7 & I\\
  EDCSNJ1216451-1158493 & cl1216.8-1201 & 0.7969 & 11.04 & 0.54 & -1.72 & 0\\
  EDCSNJ1216449-1202036 & cl1216.8-1201 & 0.7938 & 10.79 & 0.41 & -1.41 & 0\\
  EDCSNJ1216449-1201203 & cl1216.8-1201 & 0.8035 & 11.52 & 0.57 & -1.98 & 0\\
  EDCSNJ1216448-1201309 & cl1216.8-1201 & 0.7984 & 11.32 & 0.53 & -1.28 & I\\
  EDCSNJ1216447-1201282 & cl1216.8-1201 & 0.7865 & 10.82 & 0.45 & -0.73 & M\\
  EDCSNJ1216446-1201089 & cl1216.8-1201 & 0.8001 & 10.92 & 0.61 & -1.11 & i\\
  EDCSNJ1216443-1201429 & cl1216.8-1201 & 0.7918 & 11.36 & 0.56 & -0.85 & 0\\
  EDCSNJ1216438-1200536 & cl1216.8-1201 & 0.7945 & 11.36 & 0.56 & -1.76 & 0\\
  EDCSNJ1216429-1159536 & cl1216.8-1201 & 0.7951 & 10.82 & 0.61 & -1.92 & 0\\
  EDCSNJ1216428-1203395 & cl1216.8-1201 & 0.7955 & 11.42 & 0.52 & -2.09 & 0\\
  EDCSNJ1216420-1201509 & cl1216.8-1201 & 0.7941 & 11.46 & 0.6 & -2.3 & i\\
  EDCSNJ1216419-1202440 & cl1216.8-1201 & 0.8028 & 10.59 & 0.48 & -1.2 & 0\\
  EDCSNJ1216417-1203054 & cl1216.8-1201 & 0.8012 & 10.78 & 0.56 & -1.85 & 0\\
  EDCSNJ1216401-1202352 & cl1216.8-1201 & 0.8022 & 11.15 & 0.62 & -2.15 & 0\\
  EDCSNJ1216387-1203120 & cl1216.8-1201 & 0.7958 & 11.36 & 0.46 & -1.06 & T\\
  EDCSNJ1216387-1201503 & cl1216.8-1201 & 0.8008 & 11.07 & 0.59 & -2.02 & 0\\
  EDCSNJ1216382-1202517 & cl1216.8-1201 & 0.79 & 11.19 & 0.6 & -2.13 & 0\\
  EDCSNJ1216381-1203266 & cl1216.8-1201 & 0.7939 & 11.26 & 0.57 & -2.28 & 0\\
  EDCSNJ1216364-1200087 & cl1216.8-1201 & 0.7868 & 10.9 & 0.61 & -2.06 & 0\\
  EDCSNJ1216359-1200294 & cl1216.8-1201 & 0.7929 & 10.93 & 0.6 & -1.85 & 0\\
  EDCSNJ1228025-1135219 & cl1227.9-1138 & 0.638 & 10.82 & 0.47 & -1.72 & 0\\
  EDCSNJ1228001-1136095 & cl1227.9-1138 & 0.6325 & 10.42 & 0.39 & -0.9 & t\\
  EDCSNJ1227581-1135364 & cl1227.9-1138 & 0.6383 & 11.0 & 0.6 & -2.16 & 0\\
  EDCSNJ1227572-1135552 & cl1227.9-1138 & 0.6336 & 10.8 & 0.48 & -1.1 & M\\
  EDCSNJ1227571-1136178 & cl1227.9-1138 & 0.6333 & 10.77 & 0.51 & -1.8 & 0\\
  EDCSNJ1227566-1136545 & cl1227.9-1138 & 0.6391 & 10.66 & 0.58 & -2.29 & 0\\
  EDCSNJ1227551-1135584 & cl1227.9-1138 & 0.6333 & 10.61 & 0.51 & -0.94 & I\\
  EDCSNJ1227548-1137529 & cl1227.9-1138 & 0.6369 & 11.23 & 0.6 & -2.33 & 0\\
  EDCSNJ1227541-1138174 & cl1227.9-1138 & 0.6345 & 11.56 & 0.5 & -1.83 & I\\
  EDCSNJ1227537-1138210 & cl1227.9-1138 & 0.6309 & 10.56 & 0.56 & -1.91 & 0\\
  EDCSNJ1227533-1136527 & cl1227.9-1138 & 0.6347 & 10.69 & 0.51 & -1.48 & T\\
  EDCSNJ1227531-1138340 & cl1227.9-1138 & 0.6345 & 10.48 & 0.59 & -1.97 & 0\\
  EDCSNJ1232384-1251324 & cl1232.5-1250 & 0.5349 & 10.56 & 0.57 & -2.13 & 0\\
  EDCSNJ1232370-1248239 & cl1232.5-1250 & 0.5401 & 10.74 & 0.58 & -2.06 & 0\\
  EDCSNJ1232370-1248495 & cl1232.5-1250 & 0.5381 & 10.47 & 0.6 & -1.97 & 0\\
  EDCSNJ1232365-1251264 & cl1232.5-1250 & 0.5393 & 10.71 & 0.59 & -2.29 & 0\\
  EDCSNJ1232347-1249462 & cl1232.5-1250 & 0.5408 & 10.51 & 0.57 & -1.95 & 0\\
  EDCSNJ1232343-1249265 & cl1232.5-1250 & 0.5395 & 10.77 & 0.6 & -2.15 & 0\\
  EDCSNJ1232341-1252213 & cl1232.5-1250 & 0.5394 & 10.65 & 0.64 & -2.23 & 0\\
  EDCSNJ1232340-1249138 & cl1232.5-1250 & 0.5306 & 10.79 & 0.61 & -2.25 & 0\\
  EDCSNJ1232327-1249057 & cl1232.5-1250 & 0.5327 & 10.57 & 0.42 & -1.56 & 0\\
  EDCSNJ1232323-1251267 & cl1232.5-1250 & 0.5498 & 11.03 & 0.48 & -2.09 & 0\\
  EDCSNJ1232318-1249049 & cl1232.5-1250 & 0.5408 & 10.5 & 0.62 & -2.15 & 0\\
  EDCSNJ1232317-1249275 & cl1232.5-1250 & 0.542 & 11.14 & 0.56 & -1.94 & 0\\
  EDCSNJ1232311-1251061 & cl1232.5-1250 & 0.5526 & 10.48 & 0.55 & -2.27 & 0\\
  EDCSNJ1232309-1249408 & cl1232.5-1250 & 0.5485 & 11.34 & 0.59 & -2.22 & 0\\
  EDCSNJ1232303-1251441 & cl1232.5-1250 & 0.55 & 10.57 & 0.58 & -1.65 & 0\\
  EDCSNJ1232299-1251034 & cl1232.5-1250 & 0.5493 & 10.41 & 0.58 & -2.05 & 0\\
  EDCSNJ1232297-1250080 & cl1232.5-1250 & 0.5496 & 10.67 & 0.49 & -2.2 & 0\\
  EDCSNJ1232297-1249120 & cl1232.5-1250 & 0.5412 & 10.58 & 0.65 & -2.32 & 0\\
  EDCSNJ1232296-1250119 & cl1232.5-1250 & 0.5509 & 10.97 & 0.55 & -2.05 & 0\\
  EDCSNJ1232287-1252369 & cl1232.5-1250 & 0.5432 & 11.24 & 0.57 & -2.11 & 0\\
  EDCSNJ1232288-1250490 & cl1232.5-1250 & 0.547 & 10.82 & 0.58 & -2.08 & 0\\
  EDCSNJ1232280-1252528 & cl1232.5-1250 & 0.5448 & 10.47 & 0.51 & -1.49 & 0\\
  EDCSNJ1232281-1249480 & cl1232.5-1250 & 0.5301 & 10.57 & 0.58 & -1.82 & 0\\
  EDCSNJ1232280-1249353 & cl1232.5-1250 & 0.5449 & 11.21 & 0.52 & -1.37 & 0\\
  EDCSNJ1232275-1248540 & cl1232.5-1250 & 0.5424 & 11.03 & 0.48 & -1.34 & M\\
  EDCSNJ1232273-1251080 & cl1232.5-1250 & 0.5369 & 10.61 & 0.59 & -2.12 & T\\
  EDCSNJ1232271-1253013 & cl1232.5-1250 & 0.5445 & 10.88 & 0.61 & -2.21 & 0\\
  EDCSNJ1232271-1250195 & cl1232.5-1250 & 0.5404 & 10.91 & 0.65 & -2.37 & 0\\
  EDCSNJ1232250-1251551 & cl1232.5-1250 & 0.5399 & 10.73 & 0.61 & -2.22 & i\\
  EDCSNJ1232228-1251168 & cl1232.5-1250 & 0.5432 & 10.74 & 0.62 & -2.26 & 0\\
  EDCSNJ1232204-1249547 & cl1232.5-1250 & 0.546 & 11.21 & 0.53 & -2.02 & 0\\
  EDCSNJ1354175-1230391 & cl1354.2-1230 & 0.7632 & 10.47 & 0.43 & -1.08 & 0\\
  EDCSNJ1354164-1231599 & cl1354.2-1230 & 0.5937 & 11.24 & 0.47 & -1.49 & 0\\
  EDCSNJ1354159-1232272 & cl1354.2-1230 & 0.5929 & 10.46 & 0.57 & -1.94 & 0\\
  EDCSNJ1354144-1231514 & cl1354.2-1230 & 0.5946 & 10.4 & 0.46 & -1.35 & m\\
  EDCSNJ1354118-1232499 & cl1354.2-1230 & 0.5946 & 10.44 & 0.45 & -1.61 & i\\
  EDCSNJ1354114-1230452 & cl1354.2-1230 & 0.5947 & 11.16 & 0.49 & -2.07 & i\\
  EDCSNJ1354106-1230499 & cl1354.2-1230 & 0.7634 & 11.03 & 0.61 & -2.17 & i\\
  EDCSNJ1354102-1230527 & cl1354.2-1230 & 0.7593 & 11.34 & 0.45 & -1.4 & $0^{*}$\\
  EDCSNJ1354101-1231041 & cl1354.2-1230 & 0.7612 & 10.97 & 0.56 & -1.88 & 0\\
  EDCSNJ1354098-1231098 & cl1354.2-1230 & 0.7573 & 10.64 & 0.64 & -1.49 & i\\
  EDCSNJ1354098-1231015 & cl1354.2-1230 & 0.7562 & 11.58 & 0.58 & -1.42 & i\\
  EDCSNJ1354097-1230579 & cl1354.2-1230 & 0.7562 & 11.27 & 0.65 & -2.21 & 0\\
  EDCSNJ1354026-1230127 & cl1354.2-1230 & 0.5942 & 10.5 & 0.56 & -1.94 & 0\\
  EDCSNJ1354025-1232300 & cl1354.2-1230 & 0.7576 & 10.99 & 0.48 & -1.63 & 0\\
\hline
\multicolumn{7}{c}{\bf{Group Members}}\\
  EDCSNJ1037548-1245113 & cl1037.9-1243 & 0.5789 & 11.09 & 0.59 & -2.05 & 0\\
  EDCSNJ1037535-1244006 & cl1037.9-1243 & 0.5775 & 11.06 & 0.46 & -0.76 & M\\
  EDCSNJ1037535-1241538 & cl1037.9-1243 & 0.5789 & 10.88 & 0.58 & -1.53 & 0\\
  EDCSNJ1037531-1243551 & cl1037.9-1243 & 0.5788 & 10.48 & 0.5 & -1.59 & 0\\
  EDCSNJ1037527-1243456 & cl1037.9-1243 & 0.5807 & 10.8 & 0.61 & -1.81 & m\\
  EDCSNJ1037525-1243541 & cl1037.9-1243 & 0.5772 & 10.84 & 0.59 & -1.83 & 0\\
  EDCSNJ1037521-1243392 & cl1037.9-1243 & 0.5799 & 11.05 & 0.42 & -1.04 & M\\
  EDCSNJ1040471-1153262 & cl1040.7-1155 & 0.7792 & 10.53 & 0.49 & -1.1 & M\\
  EDCSNJ1040420-1155525 & cl1040.7-1155 & 0.6308 & 10.95 & 0.49 & -1.76 & 0\\
  EDCSNJ1040409-1157230 & cl1040.7-1155 & 0.6316 & 10.67 & 0.46 & -1.59 & 0\\
  EDCSNJ1040343-1155414 & cl1040.7-1155 & 0.7807 & 11.48 & 0.63 & -2.32 & 0\\
  EDCSNJ1054308-1147557 & cl1054.4-1146 & 0.615 & 10.68 & 0.48 & -1.38 & 0\\
  EDCSNJ1054297-1148146 & cl1054.4-1146 & 0.6143 & 11.03 & 0.62 & -1.68 & $T^{*}$\\
  EDCSNJ1054249-1147556 & cl1054.4-1146 & 0.6139 & 10.93 & 0.61 & -1.51 & I\\
  EDCSNJ1054197-1145282 & cl1054.4-1146 & 0.6127 & 11.05 & 0.49 & -1.52 & $0^{*}$\\
  EDCSNJ1054525-1244189 & cl1054.7-1245 & 0.7283 & 11.17 & 0.57 & -1.75 & T\\
  EDCSNJ1054466-1247161 & cl1054.7-1245 & 0.7302 & 10.43 & 0.64 & -1.73 & 0\\
  EDCSNJ1054457-1246373 & cl1054.7-1245 & 0.7302 & 10.86 & 0.5 & -1.83 & 0\\
  EDCSNJ1054451-1247336 & cl1054.7-1245 & 0.7305 & 10.89 & 0.57 & -1.96 & i\\
  EDCSNJ1054450-1244089 & cl1054.7-1245 & 0.7305 & 10.86 & 0.53 & -1.86 & i\\
  EDCSNJ1054387-1243048 & cl1054.7-1245 & 0.7314 & 10.78 & 0.58 & -2.12 & 0\\
  EDCSNJ1054350-1243344 & cl1054.7-1245 & 0.7293 & 10.74 & 0.55 & -1.9 & 0\\
  EDCSNJ1103438-1247251 & cl1103.7-1245 & 0.6238 & 10.52 & 0.61 & -2.1 & 0\\
  EDCSNJ1103413-1244379 & cl1103.7-1245 & 0.7038 & 11.17 & 0.6 & -2.23 & 0\\
  EDCSNJ1103401-1244377 & cl1103.7-1245 & 0.7032 & 10.6 & 0.52 & -1.83 & 0\\
  EDCSNJ1103386-1247210 & cl1103.7-1245 & 0.6276 & 11.18 & 0.54 & -2.2 & I\\
  EDCSNJ1103373-1246364 & cl1103.7-1245 & 0.703 & 10.54 & 0.56 & -1.36 & 0\\
  EDCSNJ1103372-1245215 & cl1103.7-1245 & 0.6251 & 10.93 & 0.62 & -2.19 & 0\\
  EDCSNJ1103365-1244223 & cl1103.7-1245 & 0.7031 & 11.79 & 0.56 & -2.08 & 0\\
  EDCSNJ1103363-1246220 & cl1103.7-1245 & 0.6288 & 11.11 & 0.54 & -1.95 & 0\\
  EDCSNJ1103357-1246398 & cl1103.7-1245 & 0.6278 & 10.74 & 0.59 & -1.92 & 0\\
  EDCSNJ1103355-1244515 & cl1103.7-1245 & 0.6259 & 10.95 & 0.47 & -1.75 & T\\
  EDCSNJ1103349-1246462 & cl1103.7-1245 & 0.6257 & 11.34 & 0.62 & -2.44 & 0\\
  EDCSNJ1103339-1243415 & cl1103.7-1245 & 0.7004 & 10.53 & 0.54 & -1.7 & 0\\
\enddata
\tablecomments{Column 1: Galaxy ID. Column 2: Field name. Column 3: Galaxy spectroscopic redshift. Column 4: Log galaxy stellar mass in solar masses. Column 5: Galaxy $G$ value. Column 6: Galaxy $M_{20}$ value. Column 7: Visual classifier. M: Major merger. m: Minor merger. I: Strong interaction. i: Weak interaction. T: Strong tidal features. t: Mild tidal features. 0: undisturbed. For details on the classification scheme see \S\ref{Sec:MorphClass}. Star superscript implies we changed the visual classification by \cite{Kelkar17}.}
\end{deluxetable*}

\startlongtable
\begin{deluxetable*}{cccccc}  
\tablecolumns{5}
\tablewidth{0pc}
\tablecaption{Photometric Cluster Members} \label{Tab:PhotSample}
\tablehead{
\colhead{Galaxy ID} & \colhead{Cluster Name}  & \colhead{$\log_{10}(M_{*}/M_{\odot})$}  & \colhead{G}  & \colhead{$M_{20}$}}
\startdata
\multicolumn{5}{c}{\bf{Cluster Members}}\\
  EDCSNJ1040506-1154108 & cl1040.7-1155 & 11.32 & 0.6 & -2.27\\
  EDCSNJ1040495-1153125 & cl1040.7-1155 & 10.7 & 0.61 & -2.37\\
  EDCSNJ1040488-1155078 & cl1040.7-1155 & 10.76 & 0.54 & -1.9\\
  EDCSNJ1040486-1156217 & cl1040.7-1155 & 11.26 & 0.46 & -1.23\\
  EDCSNJ1040473-1154038 & cl1040.7-1155 & 11.16 & 0.48 & -1.92\\
  EDCSNJ1040426-1157532 & cl1040.7-1155 & 11.03 & 0.53 & -2.07\\
  EDCSNJ1040383-1153176 & cl1040.7-1155 & 10.67 & 0.59 & -1.99\\
  EDCSNJ1040382-1153506 & cl1040.7-1155 & 10.79 & 0.52 & -1.31\\
  EDCSNJ1040381-1153518 & cl1040.7-1155 & 10.83 & 0.46 & -1.09\\
  EDCSNJ1040380-1157000 & cl1040.7-1155 & 10.89 & 0.46 & -1.67\\
  EDCSNJ1040374-1154010 & cl1040.7-1155 & 10.63 & 0.53 & -1.12\\
  EDCSNJ1040361-1156054 & cl1040.7-1155 & 10.7 & 0.41 & -1.28\\
  EDCSNJ1040337-1157231 & cl1040.7-1155 & 11.09 & 0.57 & -1.77\\
  EDCSNJ1040328-1152599 & cl1040.7-1155 & 10.87 & 0.58 & -1.98\\
  EDCSNJ1054345-1146503 & cl1054.4-1146 & 10.83 & 0.61 & -1.62\\
  EDCSNJ1054343-1146541 & cl1054.4-1146 & 10.67 & 0.57 & -2.06\\
  EDCSNJ1054338-1147230 & cl1054.4-1146 & 11.49 & 0.62 & -2.3\\
  EDCSNJ1054338-1145541 & cl1054.4-1146 & 11.2 & 0.63 & -2.18\\
  EDCSNJ1054335-1148197 & cl1054.4-1146 & 10.61 & 0.44 & -0.95\\
  EDCSNJ1054332-1147414 & cl1054.4-1146 & 10.56 & 0.57 & -2.07\\
  EDCSNJ1054331-1147379 & cl1054.4-1146 & 10.93 & 0.59 & -2.26\\
  EDCSNJ1054330-1147315 & cl1054.4-1146 & 11.05 & 0.63 & -1.68\\
  EDCSNJ1054320-1149211 & cl1054.4-1146 & 11.22 & 0.58 & -1.92\\
  EDCSNJ1054315-1147019 & cl1054.4-1146 & 10.91 & 0.58 & -1.04\\
  EDCSNJ1054308-1146114 & cl1054.4-1146 & 10.94 & 0.62 & -2.26\\
  EDCSNJ1054307-1146375 & cl1054.4-1146 & 10.46 & 0.57 & -1.79\\
  EDCSNJ1054304-1149226 & cl1054.4-1146 & 10.78 & 0.58 & -2.15\\
  EDCSNJ1054287-1146574 & cl1054.4-1146 & 10.73 & 0.63 & -2.06\\
  EDCSNJ1054284-1146500 & cl1054.4-1146 & 10.89 & 0.57 & -1.02\\
  EDCSNJ1054278-1146280 & cl1054.4-1146 & 10.57 & 0.6 & -2.07\\
  EDCSNJ1054272-1145430 & cl1054.4-1146 & 10.48 & 0.48 & -1.46\\
  EDCSNJ1054270-1146240 & cl1054.4-1146 & 10.66 & 0.51 & -1.36\\
  EDCSNJ1054266-1146566 & cl1054.4-1146 & 10.86 & 0.49 & -1.88\\
  EDCSNJ1054265-1146316 & cl1054.4-1146 & 10.83 & 0.57 & -1.09\\
  EDCSNJ1054257-1147149 & cl1054.4-1146 & 10.67 & 0.58 & -2.27\\
  EDCSNJ1054256-1147235 & cl1054.4-1146 & 11.0 & 0.63 & -2.3\\
  EDCSNJ1054254-1148048 & cl1054.4-1146 & 10.41 & 0.62 & -1.9\\
  EDCSNJ1054254-1147523 & cl1054.4-1146 & 11.24 & 0.33 & -1.04\\
  EDCSNJ1054254-1148135 & cl1054.4-1146 & 10.88 & 0.54 & -1.75\\
  EDCSNJ1054254-1146005 & cl1054.4-1146 & 10.63 & 0.57 & -1.76\\
  EDCSNJ1054248-1148509 & cl1054.4-1146 & 10.43 & 0.46 & -1.46\\
  EDCSNJ1054245-1146139 & cl1054.4-1146 & 10.59 & 0.61 & -1.88\\
  EDCSNJ1054243-1146168 & cl1054.4-1146 & 11.03 & 0.61 & -2.09\\
  EDCSNJ1054243-1145565 & cl1054.4-1146 & 10.59 & 0.42 & -0.83\\
  EDCSNJ1054241-1146407 & cl1054.4-1146 & 10.98 & 0.64 & -2.02\\
  EDCSNJ1054241-1145283 & cl1054.4-1146 & 10.44 & 0.46 & -1.5\\
  EDCSNJ1054241-1146427 & cl1054.4-1146 & 11.02 & 0.57 & -2.14\\
  EDCSNJ1054240-1147297 & cl1054.4-1146 & 10.72 & 0.58 & -2.02\\
  EDCSNJ1054239-1144031 & cl1054.4-1146 & 10.43 & 0.54 & -1.89\\
  EDCSNJ1054235-1146205 & cl1054.4-1146 & 10.46 & 0.57 & -1.97\\
  EDCSNJ1054224-1146208 & cl1054.4-1146 & 11.09 & 0.62 & -2.36\\
  EDCSNJ1054217-1147249 & cl1054.4-1146 & 10.65 & 0.49 & -1.65\\
  EDCSNJ1054213-1146186 & cl1054.4-1146 & 10.94 & 0.59 & -2.33\\
  EDCSNJ1054211-1146162 & cl1054.4-1146 & 10.47 & 0.51 & -1.85\\
  EDCSNJ1054199-1146282 & cl1054.4-1146 & 10.62 & 0.46 & -1.47\\
  EDCSNJ1054183-1149011 & cl1054.4-1146 & 10.94 & 0.62 & -2.03\\
  EDCSNJ1054180-1146217 & cl1054.4-1146 & 10.8 & 0.48 & -1.0\\
  EDCSNJ1054177-1146083 & cl1054.4-1146 & 10.6 & 0.6 & -1.16\\
  EDCSNJ1054169-1148162 & cl1054.4-1146 & 10.45 & 0.42 & -1.6\\
  EDCSNJ1054158-1148203 & cl1054.4-1146 & 10.72 & 0.46 & -1.63\\
  EDCSNJ1054151-1144080 & cl1054.4-1146 & 10.61 & 0.54 & -1.07\\
  EDCSNJ1054528-1245171 & cl1054.7-1245 & 10.47 & 0.48 & -1.06\\
  EDCSNJ1054528-1244126 & cl1054.7-1245 & 10.96 & 0.58 & -2.0\\
  EDCSNJ1054522-1244173 & cl1054.7-1245 & 11.01 & 0.62 & -2.43\\
  EDCSNJ1054520-1244178 & cl1054.7-1245 & 10.72 & 0.57 & -1.74\\
  EDCSNJ1054504-1243398 & cl1054.7-1245 & 10.48 & 0.48 & -0.94\\
  EDCSNJ1054494-1244376 & cl1054.7-1245 & 10.4 & 0.42 & -1.47\\
  EDCSNJ1054487-1245119 & cl1054.7-1245 & 10.84 & 0.49 & -1.7\\
  EDCSNJ1054479-1246592 & cl1054.7-1245 & 10.95 & 0.46 & -1.33\\
  EDCSNJ1054478-1246292 & cl1054.7-1245 & 10.83 & 0.49 & -1.1\\
  EDCSNJ1054477-1245080 & cl1054.7-1245 & 10.68 & 0.57 & -1.68\\
  EDCSNJ1054476-1246405 & cl1054.7-1245 & 10.41 & 0.46 & -1.6\\
  EDCSNJ1054474-1245580 & cl1054.7-1245 & 10.59 & 0.44 & -1.32\\
  EDCSNJ1054471-1246276 & cl1054.7-1245 & 10.86 & 0.42 & -1.19\\
  EDCSNJ1054466-1247248 & cl1054.7-1245 & 11.11 & 0.49 & -1.93\\
  EDCSNJ1054459-1246290 & cl1054.7-1245 & 10.75 & 0.6 & -2.03\\
  EDCSNJ1054450-1247318 & cl1054.7-1245 & 10.64 & 0.62 & -1.87\\
  EDCSNJ1054446-1243367 & cl1054.7-1245 & 10.81 & 0.51 & -1.75\\
  EDCSNJ1054443-1245198 & cl1054.7-1245 & 11.29 & 0.45 & -2.21\\
  EDCSNJ1054442-1246441 & cl1054.7-1245 & 11.35 & 0.53 & -2.12\\
  EDCSNJ1054437-1246270 & cl1054.7-1245 & 10.88 & 0.56 & -2.09\\
  EDCSNJ1054432-1245541 & cl1054.7-1245 & 10.75 & 0.55 & -1.92\\
  EDCSNJ1054432-1245241 & cl1054.7-1245 & 10.53 & 0.57 & -1.96\\
  EDCSNJ1054427-1246359 & cl1054.7-1245 & 10.54 & 0.39 & -1.14\\
  EDCSNJ1054424-1246085 & cl1054.7-1245 & 11.1 & 0.6 & -2.35\\
  EDCSNJ1054424-1246157 & cl1054.7-1245 & 10.66 & 0.55 & -2.06\\
  EDCSNJ1054422-1244154 & cl1054.7-1245 & 10.47 & 0.5 & -1.83\\
  EDCSNJ1054418-1246350 & cl1054.7-1245 & 10.58 & 0.54 & -1.91\\
  EDCSNJ1054417-1246282 & cl1054.7-1245 & 10.66 & 0.55 & -1.93\\
  EDCSNJ1054413-1245341 & cl1054.7-1245 & 10.57 & 0.58 & -1.93\\
  EDCSNJ1054408-1245594 & cl1054.7-1245 & 10.74 & 0.58 & -2.18\\
  EDCSNJ1054404-1246478 & cl1054.7-1245 & 10.49 & 0.6 & -1.95\\
  EDCSNJ1054402-1246022 & cl1054.7-1245 & 10.76 & 0.42 & -0.9\\
  EDCSNJ1054395-1248181 & cl1054.7-1245 & 10.78 & 0.45 & -1.38\\
  EDCSNJ1054383-1247373 & cl1054.7-1245 & 10.48 & 0.61 & -1.95\\
  EDCSNJ1054378-1246245 & cl1054.7-1245 & 10.47 & 0.36 & -1.19\\
  EDCSNJ1054377-1247394 & cl1054.7-1245 & 10.63 & 0.54 & -2.1\\
  EDCSNJ1054363-1247075 & cl1054.7-1245 & 10.64 & 0.56 & -1.83\\
  EDCSNJ1054361-1246580 & cl1054.7-1245 & 10.8 & 0.53 & -2.11\\
  EDCSNJ1054361-1244568 & cl1054.7-1245 & 10.66 & 0.58 & -2.01\\
  EDCSNJ1054335-1247110 & cl1054.7-1245 & 10.59 & 0.58 & -1.94\\
  EDCSNJ1054334-1245246 & cl1054.7-1245 & 10.98 & 0.52 & -2.04\\
  EDCSNJ1216554-1200183 & cl1216.8-1201 & 10.65 & 0.58 & -1.86\\
  EDCSNJ1216546-1201460 & cl1216.8-1201 & 10.71 & 0.51 & -0.87\\
  EDCSNJ1216544-1201328 & cl1216.8-1201 & 11.1 & 0.49 & -1.53\\
  EDCSNJ1216541-1203104 & cl1216.8-1201 & 10.91 & 0.57 & -1.89\\
  EDCSNJ1216542-1159077 & cl1216.8-1201 & 10.53 & 0.61 & -1.96\\
  EDCSNJ1216540-1159240 & cl1216.8-1201 & 10.74 & 0.42 & -0.98\\
  EDCSNJ1216532-1201359 & cl1216.8-1201 & 11.14 & 0.52 & -1.78\\
  EDCSNJ1216530-1201504 & cl1216.8-1201 & 10.51 & 0.36 & -0.92\\
  EDCSNJ1216531-1158378 & cl1216.8-1201 & 10.86 & 0.52 & -1.98\\
  EDCSNJ1216525-1158523 & cl1216.8-1201 & 10.41 & 0.52 & -1.05\\
  EDCSNJ1216522-1158170 & cl1216.8-1201 & 10.89 & 0.51 & -1.3\\
  EDCSNJ1216512-1201331 & cl1216.8-1201 & 10.48 & 0.58 & -1.8\\
  EDCSNJ1216509-1202177 & cl1216.8-1201 & 10.41 & 0.58 & -1.86\\
  EDCSNJ1216508-1201063 & cl1216.8-1201 & 10.51 & 0.45 & -1.51\\
  EDCSNJ1216506-1200064 & cl1216.8-1201 & 10.43 & 0.53 & -1.67\\
  EDCSNJ1216502-1159425 & cl1216.8-1201 & 10.76 & 0.48 & -1.8\\
  EDCSNJ1216498-1201392 & cl1216.8-1201 & 11.2 & 0.57 & -2.02\\
  EDCSNJ1216497-1201117 & cl1216.8-1201 & 10.89 & 0.52 & -2.11\\
  EDCSNJ1216492-1202036 & cl1216.8-1201 & 10.61 & 0.62 & -1.0\\
  EDCSNJ1216490-1201426 & cl1216.8-1201 & 11.21 & 0.62 & -2.12\\
  EDCSNJ1216489-1201239 & cl1216.8-1201 & 11.14 & 0.56 & -1.96\\
  EDCSNJ1216470-1201216 & cl1216.8-1201 & 10.93 & 0.54 & -1.05\\
  EDCSNJ1216469-1201494 & cl1216.8-1201 & 10.46 & 0.59 & -2.04\\
  EDCSNJ1216469-1201241 & cl1216.8-1201 & 11.14 & 0.6 & -2.05\\
  EDCSNJ1216465-1201574 & cl1216.8-1201 & 10.78 & 0.59 & -2.08\\
  EDCSNJ1216457-1158368 & cl1216.8-1201 & 10.54 & 0.5 & -1.62\\
  EDCSNJ1216452-1202262 & cl1216.8-1201 & 10.45 & 0.56 & -1.93\\
  EDCSNJ1216451-1202531 & cl1216.8-1201 & 10.52 & 0.5 & -1.69\\
  EDCSNJ1216447-1201234 & cl1216.8-1201 & 10.76 & 0.58 & -2.09\\
  EDCSNJ1216447-1201434 & cl1216.8-1201 & 11.26 & 0.53 & -2.13\\
  EDCSNJ1216446-1201139 & cl1216.8-1201 & 10.58 & 0.57 & -2.0\\
  EDCSNJ1216445-1201132 & cl1216.8-1201 & 11.12 & 0.42 & -1.38\\
  EDCSNJ1216443-1201201 & cl1216.8-1201 & 11.03 & 0.53 & -1.99\\
  EDCSNJ1216441-1201553 & cl1216.8-1201 & 10.51 & 0.43 & -0.79\\
  EDCSNJ1216434-1201434 & cl1216.8-1201 & 10.68 & 0.52 & -1.73\\
  EDCSNJ1216431-1203334 & cl1216.8-1201 & 10.81 & 0.53 & -1.21\\
  EDCSNJ1216429-1200591 & cl1216.8-1201 & 10.57 & 0.55 & -1.9\\
  EDCSNJ1216423-1201576 & cl1216.8-1201 & 10.99 & 0.62 & -1.92\\
  EDCSNJ1216418-1202044 & cl1216.8-1201 & 10.58 & 0.58 & -1.97\\
  EDCSNJ1216418-1201081 & cl1216.8-1201 & 11.25 & 0.61 & -2.28\\
  EDCSNJ1216414-1203332 & cl1216.8-1201 & 10.78 & 0.53 & -1.44\\
  EDCSNJ1216412-1201554 & cl1216.8-1201 & 10.46 & 0.58 & -1.85\\
  EDCSNJ1216410-1203293 & cl1216.8-1201 & 10.48 & 0.41 & -1.5\\
  EDCSNJ1216411-1159579 & cl1216.8-1201 & 10.65 & 0.51 & -1.89\\
  EDCSNJ1216408-1201433 & cl1216.8-1201 & 10.59 & 0.5 & -1.81\\
  EDCSNJ1216405-1200496 & cl1216.8-1201 & 10.83 & 0.49 & -1.35\\
  EDCSNJ1216393-1202262 & cl1216.8-1201 & 10.81 & 0.49 & -1.77\\
  EDCSNJ1216392-1201333 & cl1216.8-1201 & 10.52 & 0.56 & -1.71\\
  EDCSNJ1216391-1200154 & cl1216.8-1201 & 10.8 & 0.57 & -1.89\\
  EDCSNJ1216388-1200176 & cl1216.8-1201 & 10.67 & 0.5 & -1.47\\
  EDCSNJ1216387-1201386 & cl1216.8-1201 & 10.74 & 0.5 & -1.34\\
  EDCSNJ1216386-1202099 & cl1216.8-1201 & 10.58 & 0.57 & -2.02\\
  EDCSNJ1216385-1203051 & cl1216.8-1201 & 10.78 & 0.47 & -1.35\\
  EDCSNJ1216383-1202205 & cl1216.8-1201 & 10.63 & 0.54 & -1.79\\
  EDCSNJ1216381-1202515 & cl1216.8-1201 & 11.23 & 0.6 & -2.16\\
  EDCSNJ1216380-1202393 & cl1216.8-1201 & 10.56 & 0.62 & -2.16\\
  EDCSNJ1216379-1201545 & cl1216.8-1201 & 10.43 & 0.54 & -1.76\\
  EDCSNJ1216368-1200357 & cl1216.8-1201 & 10.59 & 0.58 & -1.78\\
  EDCSNJ1216367-1202298 & cl1216.8-1201 & 11.42 & 0.59 & -1.06\\
  EDCSNJ1216366-1202317 & cl1216.8-1201 & 10.94 & 0.52 & -1.36\\
  EDCSNJ1216366-1202253 & cl1216.8-1201 & 11.11 & 0.48 & -1.7\\
  EDCSNJ1216365-1159452 & cl1216.8-1201 & 10.67 & 0.58 & -2.06\\
  EDCSNJ1216364-1203174 & cl1216.8-1201 & 10.65 & 0.51 & -1.66\\
  EDCSNJ1216361-1200431 & cl1216.8-1201 & 11.0 & 0.56 & -1.64\\
  EDCSNJ1216358-1203011 & cl1216.8-1201 & 10.71 & 0.57 & -2.06\\
  EDCSNJ1216358-1201415 & cl1216.8-1201 & 10.48 & 0.57 & -1.98\\
  EDCSNJ1228031-1136039 & cl1227.9-1138 & 11.52 & 0.64 & -2.44\\
  EDCSNJ1228025-1140247 & cl1227.9-1138 & 10.84 & 0.59 & -2.24\\
  EDCSNJ1228022-1135468 & cl1227.9-1138 & 10.65 & 0.55 & -1.66\\
  EDCSNJ1228021-1140299 & cl1227.9-1138 & 10.49 & 0.56 & -1.8\\
  EDCSNJ1228013-1138450 & cl1227.9-1138 & 11.3 & 0.54 & -1.18\\
  EDCSNJ1228007-1140469 & cl1227.9-1138 & 11.64 & 0.53 & -1.24\\
  EDCSNJ1228003-1137041 & cl1227.9-1138 & 10.84 & 0.53 & -2.09\\
  EDCSNJ1227596-1138024 & cl1227.9-1138 & 10.52 & 0.55 & -2.02\\
  EDCSNJ1227589-1138408 & cl1227.9-1138 & 10.7 & 0.58 & -2.05\\
  EDCSNJ1227586-1136295 & cl1227.9-1138 & 10.48 & 0.49 & -1.75\\
  EDCSNJ1227586-1139362 & cl1227.9-1138 & 11.1 & 0.46 & -2.2\\
  EDCSNJ1227585-1140265 & cl1227.9-1138 & 11.29 & 0.52 & -1.55\\
  EDCSNJ1227570-1135193 & cl1227.9-1138 & 10.81 & 0.56 & -1.97\\
  EDCSNJ1227569-1136423 & cl1227.9-1138 & 11.17 & 0.6 & -2.36\\
  EDCSNJ1227554-1137391 & cl1227.9-1138 & 10.49 & 0.59 & -2.28\\
  EDCSNJ1227553-1136118 & cl1227.9-1138 & 10.7 & 0.48 & -1.46\\
  EDCSNJ1227551-1136202 & cl1227.9-1138 & 10.88 & 0.66 & -1.52\\
  EDCSNJ1227550-1135278 & cl1227.9-1138 & 11.09 & 0.56 & -2.14\\
  EDCSNJ1227550-1137464 & cl1227.9-1138 & 11.43 & 0.62 & -2.34\\
  EDCSNJ1227548-1138463 & cl1227.9-1138 & 11.16 & 0.59 & -2.18\\
  EDCSNJ1227546-1138212 & cl1227.9-1138 & 10.51 & 0.57 & -2.06\\
  EDCSNJ1227545-1139383 & cl1227.9-1138 & 11.32 & 0.54 & -2.11\\
  EDCSNJ1227542-1138246 & cl1227.9-1138 & 10.95 & 0.61 & -2.13\\
  EDCSNJ1227538-1139470 & cl1227.9-1138 & 10.51 & 0.47 & -1.8\\
  EDCSNJ1227538-1138257 & cl1227.9-1138 & 10.99 & 0.56 & -2.03\\
  EDCSNJ1227530-1138474 & cl1227.9-1138 & 11.2 & 0.62 & -0.78\\
  EDCSNJ1227527-1139218 & cl1227.9-1138 & 11.07 & 0.58 & -2.14\\
  EDCSNJ1227524-1135155 & cl1227.9-1138 & 10.67 & 0.58 & -2.38\\
  EDCSNJ1227521-1139587 & cl1227.9-1138 & 10.51 & 0.43 & -1.18\\
  EDCSNJ1227510-1137559 & cl1227.9-1138 & 10.56 & 0.53 & -0.86\\
  EDCSNJ1227506-1135282 & cl1227.9-1138 & 10.68 & 0.54 & -1.73\\
  EDCSNJ1227505-1136072 & cl1227.9-1138 & 11.19 & 0.48 & -1.95\\
  EDCSNJ1227504-1135224 & cl1227.9-1138 & 10.42 & 0.57 & -1.97\\
  EDCSNJ1227503-1140297 & cl1227.9-1138 & 10.65 & 0.52 & -1.95\\
  EDCSNJ1227493-1139524 & cl1227.9-1138 & 10.94 & 0.54 & -1.87\\
  EDCSNJ1227488-1137593 & cl1227.9-1138 & 10.88 & 0.61 & -2.06\\
  EDCSNJ1227486-1135281 & cl1227.9-1138 & 10.61 & 0.6 & -2.03\\
  EDCSNJ1227486-1135342 & cl1227.9-1138 & 10.49 & 0.47 & -1.25\\
  EDCSNJ1227482-1140258 & cl1227.9-1138 & 11.17 & 0.6 & -2.28\\
  EDCSNJ1227465-1139168 & cl1227.9-1138 & 11.18 & 0.62 & -2.5\\
  EDCSNJ1227452-1138369 & cl1227.9-1138 & 10.75 & 0.57 & -2.02\\
  EDCSNJ1227444-1138305 & cl1227.9-1138 & 11.19 & 0.58 & -2.1\\
  EDCSNJ1232401-1248452 & cl1232.5-1250 & 11.15 & 0.53 & -2.06\\
  EDCSNJ1232398-1250269 & cl1232.5-1250 & 10.46 & 0.55 & -1.75\\
  EDCSNJ1232394-1248165 & cl1232.5-1250 & 10.98 & 0.59 & -1.91\\
  EDCSNJ1232391-1249025 & cl1232.5-1250 & 10.76 & 0.49 & -2.03\\
  EDCSNJ1232391-1248278 & cl1232.5-1250 & 10.69 & 0.55 & -2.0\\
  EDCSNJ1232390-1250300 & cl1232.5-1250 & 10.76 & 0.62 & -2.25\\
  EDCSNJ1232387-1248459 & cl1232.5-1250 & 10.64 & 0.41 & -1.27\\
  EDCSNJ1232384-1251509 & cl1232.5-1250 & 10.58 & 0.62 & -2.29\\
  EDCSNJ1232386-1248154 & cl1232.5-1250 & 10.89 & 0.6 & -2.23\\
  EDCSNJ1232376-1248384 & cl1232.5-1250 & 11.23 & 0.51 & -2.13\\
  EDCSNJ1232371-1250322 & cl1232.5-1250 & 10.87 & 0.65 & -2.5\\
  EDCSNJ1232369-1248246 & cl1232.5-1250 & 10.4 & 0.47 & -1.17\\
  EDCSNJ1232364-1250394 & cl1232.5-1250 & 11.13 & 0.51 & -1.08\\
  EDCSNJ1232362-1250098 & cl1232.5-1250 & 10.64 & 0.58 & -2.05\\
  EDCSNJ1232358-1250099 & cl1232.5-1250 & 10.67 & 0.58 & -2.25\\
  EDCSNJ1232357-1251214 & cl1232.5-1250 & 11.35 & 0.62 & -2.24\\
  EDCSNJ1232349-1252505 & cl1232.5-1250 & 10.55 & 0.52 & -1.99\\
  EDCSNJ1232347-1252164 & cl1232.5-1250 & 10.61 & 0.46 & -1.5\\
  EDCSNJ1232343-1249594 & cl1232.5-1250 & 10.99 & 0.48 & -1.58\\
  EDCSNJ1232340-1248326 & cl1232.5-1250 & 10.68 & 0.59 & -2.19\\
  EDCSNJ1232339-1250106 & cl1232.5-1250 & 10.94 & 0.62 & -2.26\\
  EDCSNJ1232336-1250207 & cl1232.5-1250 & 10.79 & 0.61 & -2.18\\
  EDCSNJ1232334-1250578 & cl1232.5-1250 & 10.63 & 0.58 & -1.97\\
  EDCSNJ1232335-1250052 & cl1232.5-1250 & 10.86 & 0.61 & -2.29\\
  EDCSNJ1232333-1252436 & cl1232.5-1250 & 10.46 & 0.47 & -1.63\\
  EDCSNJ1232325-1250105 & cl1232.5-1250 & 11.02 & 0.6 & -2.43\\
  EDCSNJ1232325-1251214 & cl1232.5-1250 & 10.6 & 0.5 & -1.27\\
  EDCSNJ1232321-1249489 & cl1232.5-1250 & 10.47 & 0.39 & -1.52\\
  EDCSNJ1232320-1250423 & cl1232.5-1250 & 10.47 & 0.49 & -1.23\\
  EDCSNJ1232319-1250383 & cl1232.5-1250 & 10.88 & 0.58 & -2.02\\
  EDCSNJ1232315-1250454 & cl1232.5-1250 & 10.96 & 0.56 & -2.22\\
  EDCSNJ1232313-1250327 & cl1232.5-1250 & 10.69 & 0.59 & -2.01\\
  EDCSNJ1232304-1251184 & cl1232.5-1250 & 11.53 & 0.61 & -2.22\\
  EDCSNJ1232302-1251229 & cl1232.5-1250 & 10.71 & 0.57 & -1.98\\
  EDCSNJ1232299-1250418 & cl1232.5-1250 & 10.6 & 0.6 & -1.91\\
  EDCSNJ1232290-1251407 & cl1232.5-1250 & 10.8 & 0.6 & -2.15\\
  EDCSNJ1232292-1248278 & cl1232.5-1250 & 10.68 & 0.58 & -1.93\\
  EDCSNJ1232290-1250437 & cl1232.5-1250 & 10.55 & 0.6 & -2.17\\
  EDCSNJ1232281-1248188 & cl1232.5-1250 & 10.79 & 0.6 & -2.1\\
  EDCSNJ1232272-1250593 & cl1232.5-1250 & 11.1 & 0.62 & -1.79\\
  EDCSNJ1232255-1250409 & cl1232.5-1250 & 10.41 & 0.55 & -2.04\\
  EDCSNJ1232252-1248313 & cl1232.5-1250 & 10.76 & 0.68 & -2.31\\
  EDCSNJ1232245-1252467 & cl1232.5-1250 & 10.44 & 0.6 & -2.23\\
  EDCSNJ1232243-1249307 & cl1232.5-1250 & 10.81 & 0.59 & -2.15\\
  EDCSNJ1232221-1251299 & cl1232.5-1250 & 10.41 & 0.58 & -1.94\\
  EDCSNJ1232219-1252098 & cl1232.5-1250 & 10.98 & 0.55 & -1.88\\
  EDCSNJ1232212-1248234 & cl1232.5-1250 & 10.93 & 0.6 & -2.32\\
  EDCSNJ1232208-1251077 & cl1232.5-1250 & 11.31 & 0.5 & -2.04\\
  EDCSNJ1232206-1252401 & cl1232.5-1250 & 10.56 & 0.63 & -2.01\\
  EDCSNJ1232206-1250553 & cl1232.5-1250 & 10.9 & 0.6 & -2.11\\
  EDCSNJ1232203-1251098 & cl1232.5-1250 & 10.5 & 0.6 & -2.09\\
  EDCSNJ1354192-1232556 & cl1354.2-1230 & 10.54 & 0.61 & -2.15\\
  EDCSNJ1354193-1229343 & cl1354.2-1230 & 11.05 & 0.62 & -1.84\\
  EDCSNJ1354185-1229217 & cl1354.2-1230 & 11.31 & 0.5 & -1.22\\
  EDCSNJ1354172-1230479 & cl1354.2-1230 & 10.87 & 0.48 & -1.51\\
  EDCSNJ1354171-1232073 & cl1354.2-1230 & 10.64 & 0.51 & -1.8\\
  EDCSNJ1354168-1230046 & cl1354.2-1230 & 10.42 & 0.46 & -1.29\\
  EDCSNJ1354164-1231544 & cl1354.2-1230 & 10.48 & 0.51 & -1.65\\
  EDCSNJ1354160-1229367 & cl1354.2-1230 & 10.63 & 0.5 & -1.71\\
  EDCSNJ1354149-1231202 & cl1354.2-1230 & 10.52 & 0.51 & -1.33\\
  EDCSNJ1354147-1231467 & cl1354.2-1230 & 10.65 & 0.59 & -2.31\\
  EDCSNJ1354140-1232426 & cl1354.2-1230 & 10.65 & 0.77 & -1.65\\
  EDCSNJ1354130-1230274 & cl1354.2-1230 & 10.59 & 0.6 & -2.1\\
  EDCSNJ1354126-1230338 & cl1354.2-1230 & 10.8 & 0.58 & -1.94\\
  EDCSNJ1354125-1233145 & cl1354.2-1230 & 10.7 & 0.56 & -2.02\\
  EDCSNJ1354122-1228350 & cl1354.2-1230 & 11.09 & 0.61 & -2.06\\
  EDCSNJ1354108-1233308 & cl1354.2-1230 & 10.44 & 0.45 & -0.89\\
  EDCSNJ1354103-1231039 & cl1354.2-1230 & 10.68 & 0.6 & -1.95\\
  EDCSNJ1354093-1229167 & cl1354.2-1230 & 10.59 & 0.46 & -1.42\\
  EDCSNJ1354081-1229334 & cl1354.2-1230 & 11.19 & 0.52 & -1.61\\
  EDCSNJ1354072-1231083 & cl1354.2-1230 & 11.0 & 0.58 & -2.35\\
  EDCSNJ1354070-1230595 & cl1354.2-1230 & 10.43 & 0.53 & -1.77\\
  EDCSNJ1354058-1232373 & cl1354.2-1230 & 11.0 & 0.62 & -0.9\\
  EDCSNJ1354039-1230317 & cl1354.2-1230 & 10.59 & 0.45 & -1.4\\
  EDCSNJ1354020-1233406 & cl1354.2-1230 & 10.7 & 0.56 & -1.64\\
  EDCSNJ1354014-1229441 & cl1354.2-1230 & 10.62 & 0.59 & -1.8\\
  EDCSNJ1354013-1231011 & cl1354.2-1230 & 10.43 & 0.46 & -1.35\\
  EDCSNJ1354011-1231288 & cl1354.2-1230 & 10.6 & 0.56 & -2.0\\
  EDCSNJ1354008-1231321 & cl1354.2-1230 & 11.04 & 0.41 & -0.86\\
 \enddata
 \tablecomments{Column 1: Galaxy ID. Column 2: Field name. Column 3: Log galaxy stellar mass in solar masses. Column 5: Galaxy $G$ value. Column 6: Galaxy $M_{20}$ value. A visual classification of structural disturbances has not been performed for this sample.}
\end{deluxetable*}

\section{Appendix C}
\label{Appendix:GM20-KSTest}

\begin{figure*}
\epsscale{1.0}
\plottwo{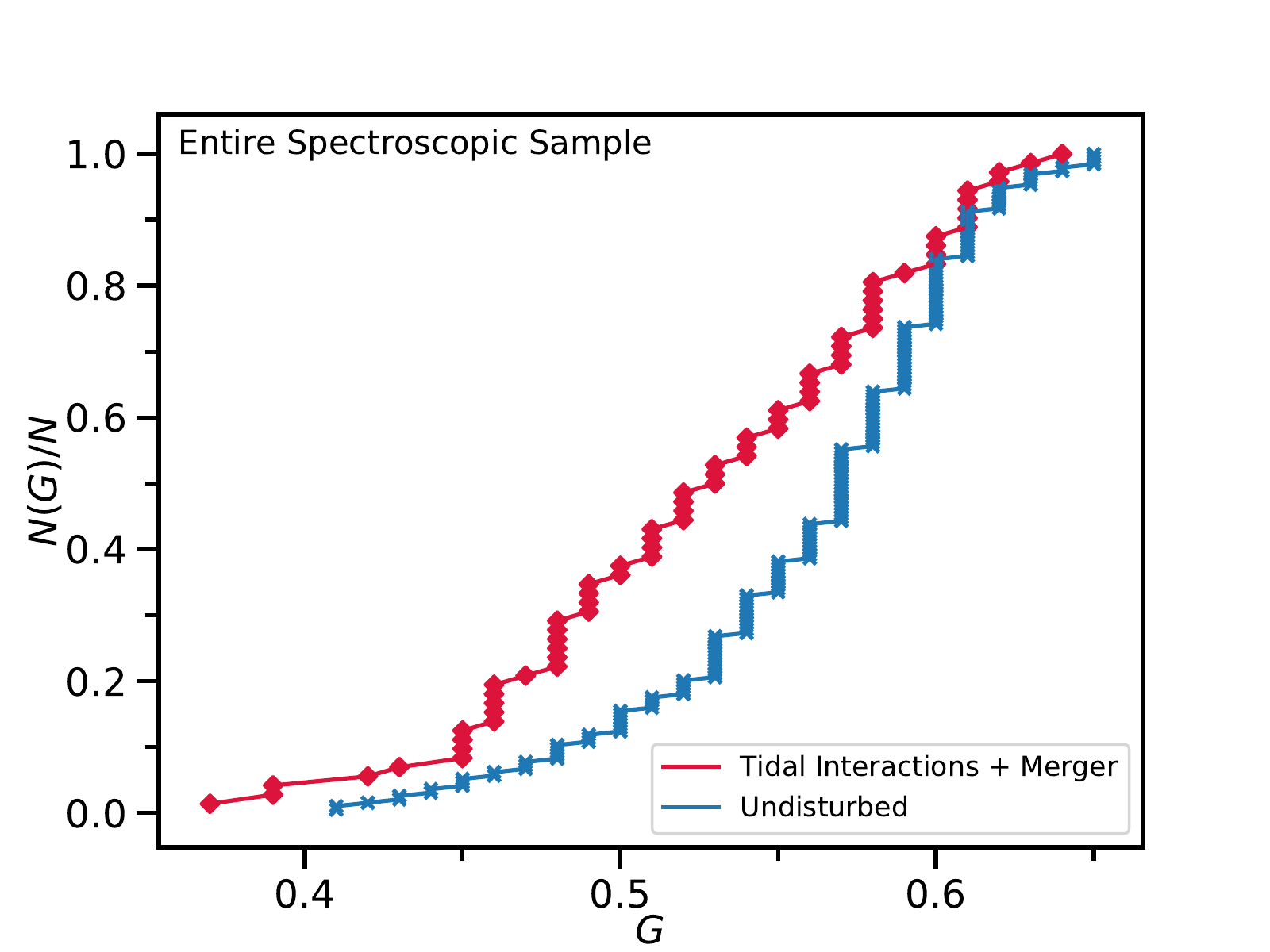}{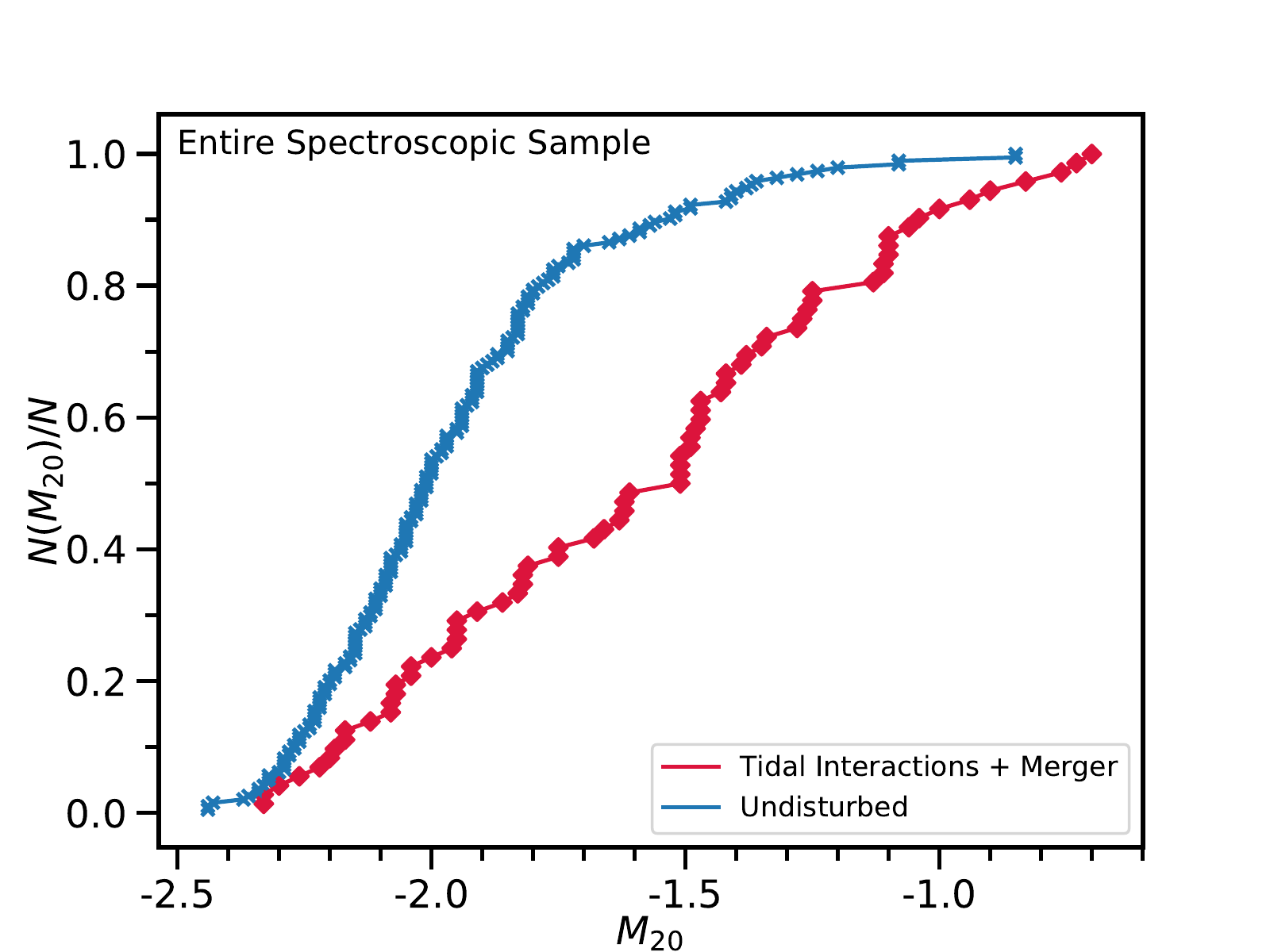}
\caption{Cumulative histograms for our spectroscopic sample which includes every galaxy from both panels of Figure~\ref{Fig:GM20}. In both panels the red line is for tidal interactions and mergers, and blue line is for our undisturbed galaxies. \textit{Left panel --} The cumulative histogram for $G$. \textit{Right panel --} The cumulative histogram for $M_{20}$. The KS test result is $0.0003$ for the left panel, and $10^{-10}$ for the right panel.}
\label{Fig:CumulativeGM20}
\end{figure*}

\begin{figure}
\plotone{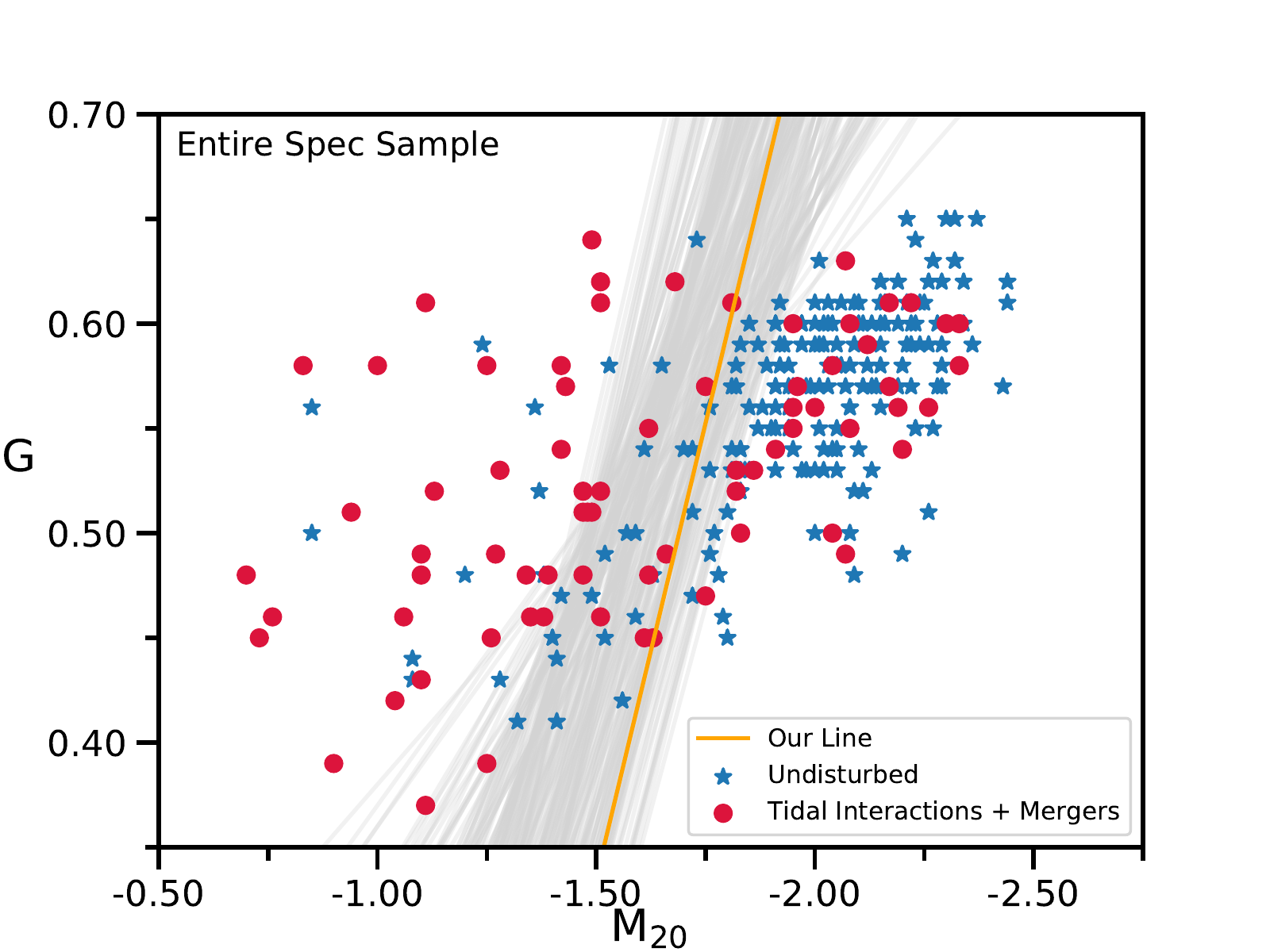}
\caption{\textit{Left panel --} Distribution of a random subsample of accepted lines (gray lines), shown on the $G-M_{20}$ plot of our entire spectroscopic sample. The orange line is the line with maximum purity ($\rho = 1.46$), and is the line we used for the analysis of this entire paper. To emphasize the region isolated by lines corresponding to high purity, we only plot the accepted lines with purity values larger than 1.35. Our test shows that such draws already dominate the distribution of accepted lines. Furthermore, for visual clarity, we also display only every twentieth accepted line.}
\label{Fig:AcceptedLines}
\end{figure}

We further investigate the distributions in $G$ and $M_{20}$ by plotting cumulative histograms of the TIM and undisturbed galaxies using our entire spectroscopic sample in Figure~\ref{Fig:CumulativeGM20}. We ran a Kolmogorov-Smirnov test (KS test) for both panels. We found that the KS \textit{p}-value over the $G$ parameter for the TIM and undisturbed galaxies is $0.0003$, showing that the probability that these classes are drawn from the same parent distribution in $G$ is $0.03\%$. The KS \textit{p}-value is significantly smaller for the $M_{20}$ parameter, which we found to be $10^{-10}$. Hence the probability that the TIM and undisturbed galaxies are drawn from the same parent distribution in $M_{20}$ is significantly less than $1\%$. These results indicate that the $M_{20}$ parameter is especially effective at separating TIM galaxies from undisturbed galaxies and that $G$, while still having discriminatory power, is less effective.

We finalize this section by investigating the distribution of lines accepted as a result of the test we describe at the end of \S\ref{Sec:MorphClass}. We show a subsample of such lines in Figure~\ref{Fig:AcceptedLines}. We chose to display only values with $\rho > 1.35$  to emphasize the region spanned by the higher purity lines. Our test preferentially accepts (y-intercept, slope) values with higher purities, so $\rho > 1.35$ draws already form the majority of the distribution of accepted lines. For visual clarity we plot every twentieth accepted line. We also display our line of maximum purity plotted on this distribution. Our results show that the same visual TIM galaxies remain above most of the accepted lines. Due to this result, combined with the results we presented at the end of \S\ref{Sec:MorphClass}, we chose to perform the entire analysis of this paper using the maximum purity line.

\end{document}